\pdfoutput=1
\documentclass[11pt]{article}

\usepackage{epsfig,latexsym,amsfonts,amsmath,amsthm,amssymb,amsbsy,multirow,slashed,wasysym,textcomp,cite,comment,subfigure,wrapfig,datetime,mathtools,cancel,mathrsfs}
\usepackage{datetime2}
\usepackage[font={footnotesize,sl},bf]{caption}

\usepackage{setspace}
\usepackage{tikz}
\usetikzlibrary{calc}
\usetikzlibrary{decorations.pathreplacing}
\usetikzlibrary{decorations.pathmorphing}
\usetikzlibrary{decorations.markings}
\usetikzlibrary{arrows.meta}
\usetikzlibrary{positioning}

\usepackage{xcolor}
\definecolor{oldmauve}{rgb}{0.4, 0.19, 0.28}
\definecolor{pansypurple}{rgb}{0.47, 0.09, 0.29}
\definecolor{burgundy}{rgb}{0.5, 0.0, 0.13}
\definecolor{carminepink}{rgb}{0.92, 0.3, 0.26}
\definecolor{blue(pigment)}{rgb}{0.2, 0.2, 0.6}
\definecolor{darkseagreen}{rgb}{0.56, 0.74, 0.56}
\definecolor{darkspringgreen}{rgb}{0.09, 0.45, 0.27}
\definecolor{ceruleanblue}{rgb}{0.16, 0.32, 0.75}

\definecolor{navyblue}{RGB}{0,0,128}

\numberwithin{equation}{section}

\def\bea{\begin{eqnarray}}
\def\eea{\end{eqnarray}}
\newcommand{\beq}{\begin{eqnarray}}
\newcommand{\eqq}{\end{eqnarray}}
 \newcommand{\badat}{\begin{alignedat}}
 \newcommand{\eadat}{\end{alignedat}}

\newcommand{\eal}[1]{\be \begin{aligned} #1 \end{aligned}\end{equation}} 
\newcommand{\eqn}[1]{\be #1 \end{equation}} 
\newcommand{\eqa}[1]{\bea  #1\end{eqnarray}}

\newcommand{\an}[1]{\left\langle#1\right\rangle}
\newcommand{\mo}{\mathcal{O}}

\newcommand{\mA}{\mathcal{A}}

\newcommand{\zb}{\bar{z}}
\newcommand{\Zb}{\bar{Z}}

\newcommand{\D}{\Delta}
\newcommand{\e}{\epsilon}
\newcommand{\om}{\omega}

\long\def\new#1\endnew{{\bf #1}}		
\long\def\del#1\enddel{}

\def\del{\partial}

\newcommand{\be}{\begin{eqnarray}}
\newcommand{\en}{\end{eqnarray}}

\numberwithin{equation}{section} 

\author{}
\numberwithin{equation}{section} 
\usepackage{hyperref}
\usepackage{orcidlink}

\begin{document}

\begin{titlepage}

  \thispagestyle{empty}

 \begin{flushright}
 \end{flushright}


  \begin{center}  
{\Large\textbf{Towards a Flat Space Carrollian Hologram from AdS$_4$/CFT$_3$}}

\vskip1cm
Arthur Lipstein$^{\dagger}$\footnote{\fontsize{8pt}{10pt}\selectfont\ \href{mailto:arthur.lipstein@durham.ac.uk }
{arthur.lipstein@durham.ac.uk}},
Romain Ruzziconi$^{\star}$\footnote{\fontsize{8pt}{10pt}\selectfont\ \href{mailto:Romain.Ruzziconi@maths.ox.ac.uk}{romain.ruzziconi@maths.ox.ac.uk}}, 
Akshay Yelleshpur Srikant$^{\star}$\footnote{\fontsize{8pt}{10pt}\selectfont \ \href{mailto:Akshay.YelleshpurSrikant@maths.ox.ac.uk}{yelleshpursr@maths.ox.ac.uk}}
\vskip0.5cm

\normalsize
\medskip

$^{\dagger}$\textit{Department of Mathematical Sciences, Durham University, \\ Stockton Road, DH1 3LE, Durham, United Kingdom } \\

\smallskip

$^{\star}$\textit{Mathematical Institute, University of Oxford, \\ Andrew Wiles Building, Radcliffe Observatory Quarter, \\
Woodstock Road, Oxford, OX2 6GG, UK}

\end{center}

\vskip0.5cm

\begin{abstract}

\end{abstract}
Finding a concrete example holography in four dimensional asymptotically flat space is an important open problem. A natural strategy is to take the 
flat space limit of the celebrated AdS$_4$/CFT$_3$ correspondence, which relates M-theory in AdS$_4 \times$S$^7$ to a certain superconformal Chern-Simons-matter theory known as the ABJM theory. In this limit, the boundary of AdS$_4$ becomes null infinity and the ABJM theory should exhibit an emergent superconformal Carrollian symmetry. We investigate this possibility by matching the Carrollian limit of ABJM correlators with four-dimensional supergravity amplitudes that arise from taking the flat space limit of AdS$_4 \times$S$^7$ and reducing along the S$^7$. We also present a general analysis of three-dimensional superconformal Carrollian symmetry.
\end{titlepage}
\setcounter{page}{2}

\setcounter{tocdepth}{2}
\tableofcontents

\section{Introduction}

The holographic principle constitutes one of the most successful paths toward a description of quantum gravity and has been extensively studied for spacetimes with a negative cosmological constant through the celebrated AdS/CFT correspondence. On the other hand, it is of great interest to extend this approach to more realistic backgrounds, notably four-dimensional spacetimes which are approximately flat or have positive curvature. Early works \cite{Susskind:1998vk,Polchinski:1999ry, Giddings:1999jq} attempted to implement a flat space limit of AdS/CFT to obtain a holographic description of spacetimes with vanishing cosmological constant. In recent years there has been an explosion of activity seeking to formulate flat space holography in terms of a two-dimensional CFT at null infinity, known as a celestial CFT \cite{deBoer:2003vf,He:2015zea,Pasterski:2016qvg,Cheung:2016iub,Pasterski:2017kqt,Strominger:2017zoo,Pasterski:2017ylz}, or a three-dimensional Carrollian CFT living on all of null infinity \cite{Arcioni:2003xx,Dappiaggi:2005ci,Barnich:2006av,Barnich:2010eb,Bagchi:2010zz,Barnich:2012xq,Barnich:2012rz,Bagchi:2012xr,Bagchi:2014iea,Bagchi:2015wna,Bagchi:2016bcd,Ciambelli:2018wre,Donnay:2022aba,Bagchi:2022emh,Donnay:2022wvx,Saha:2023hsl}, and these two approaches have been related in \cite{Donnay:2022aba,Bagchi:2022emh,Donnay:2022wvx}. Many deep lessons have been learned about the nature of flat space holography \cite{Strominger:2013jfa,Adamo:2014yya,He:2014laa,Kapec:2014opa,Strominger:2014pwa,Stieberger:2018edy,Stieberger:2018onx,Banerjee:2018gce,Adamo:2019ipt,Fotopoulos:2019vac,Banerjee:2020zlg,Pasterski:2020pdk,Pasterski:2021fjn,Freidel:2021dfs,Freidel:2021ytz,Adamo:2021zpw,Strominger:2021mtt,Adamo:2021lrv,Donnay:2021wrk,Donnay:2022hkf,Hu:2021lrx,Mago:2021wje,Freidel:2022bai,Freidel:2022skz,Ren:2022sws,Bhardwaj:2022anh,Bu:2022iak,Mason:2022hly,Melton:2023bjw,Pano:2023slc,Sleight:2023ojm,Fiorucci:2023lpb,Agrawal:2023zea,Choi:2024ygx,Geiller:2024bgf,Adamo:2024mqn,Cresto:2024fhd,Cresto:2024mne, Ball:2023qim, Bhardwaj:2024wld, Ball:2022bgg, Ball:2023sdz, Guevara:2024ixn, Kulp:2024scx}, and tremendous progress has been made in constructing explicit examples involving self-dual sectors of Yang-Mills theory and gravity \cite{Costello:2022wso,Costello:2023hmi, Costello:2022jpg, Bittleston:2024efo, Bittleston:2023bzp}. Connections have also been established with certain non-gravitational amplitudes \cite{Fernandez:2024tue, Fernandez:2024qnu, Costello:2023vyy, Dixon:2024tsb, Melton:2024akx, Stieberger:2023fju, Stieberger:2022zyk}. However, to date there is no concrete example of a flat space hologram for a UV complete theory which reduces to Einstein gravity at low energies. As a result, the holographic principle is still far less established in flat spacetime than in AdS. The goal of this paper is to take the first steps in deriving a concrete flat space Carrollian hologram from a canonical example of the AdS/CFT correspondence, notably the AdS$_4$/CFT$_3$ correspondence which relates M-theory on AdS$_4 \times$S$^7$ to a 3D superconformal Chern-Simons theory known as the ABJM theory \cite{Aharony:2008ug}. 

It has recently been understood from a bottom-up perspective that the flat space limit of AdS spacetime corresponds to a Carrollian limit in the dual boundary theory. The latter is a non-relativistic limit of the Poincaré algebra, formally defined by taking the speed of light to zero \cite{Levy1965}. This correspondence has been explored at the level of Einstein's equations in three \cite{Barnich:2012aw,Barnich:2014cwa} and four dimensions \cite{Ciambelli:2018wre,Compere:2019bua,Compere:2020lrt,Campoleoni:2023fug} using Bondi-type coordinates and has been extended to holographic correlators in \cite{Alday:2024yyj}, see also \cite{deGioia:2022fcn,deGioia:2023cbd,deGioia:2024yne,Bagchi:2023fbj,Bagchi:2023cen,Marotta:2024sce} for recent related works. Notably, \cite{Alday:2024yyj} provides a universal procedure for implementing the Carrollian limit of scalar correlators in 3D CFTs.

In this paper, we focus on two, three, and four-point correlation functions of protected scalar operators in the ABJM theory which are dual to Kaluza-Klein (KK) modes on the seven-sphere whose mode number is tied to the $R$-symmetry charge of the dual operators. Such correlation functions have been extensively studied using superconformal bootstrap techniques \cite{Rastelli:2017udc,Chester:2018aca,Alday:2020dtb,Alday:2021ymb,Alday:2022rly,Chester:2024esn,Chester:2024bij} and are typically written in Mellin space. At four points, we write the Mellin space expressions in terms a finite number of $\bar{D}$-functions in position space (along the lines of \cite{Alday:2021ymb, Binder:2021cif}), allowing us to directly apply
the techniques developed in \cite{Alday:2024yyj} to derive their Carrollian limit. This provides data for a putative 3D Carrollian theory living at null infinity. At the same time, we show how to derive the Carrollian correlators from a bulk perspective
by restricting 11D supergravity amplitudes in flat space to a four-dimensional hyperplane, with polarisation vectors pointing along the transverse directions. This directly gives the lowest charge correlators and we show how to obtain higher-charge correlators by appropriately defining the external states of the bulk scattering amplitudes. We also show that two and three-point Carrollian correlators can be derived by truncating the sum over KK modes in AdS$_4 \times$S$^7$, integrating out the seven sphere, and taking the flat space limit of the resulting effective field theory in AdS$_4$. In the flat space limit, the $S^7$ decompactifies, and the KK modes lead to a tower of massless scalar fields. In the dual theory, this corresponds to a tower of Carrollian primaries, each with a corresponding conformal dimension. We also discuss the kinematic properties of the Carrollian ABJM theory obtained in this limit and show an isomorphism between the superconformal Carrollian algebra in three dimensions and the super-Poincaré algebra in four dimensions. Although we mostly restrict to the supergravity approximation, which corresponds to the large $N$ limit in the boundary theory, the approach developed in this paper can be extended to higher orders in $1/N$ (see Appendix \ref{sec:higherdercamp}).

This paper is organised as follows. In section \ref{sec:Review} we review some background material such as the AdS$_4$/CFT$_3$ correspondence and methods for extracting the Carrollian limit of 3D CFT correlators. In section \ref{sec:ABJM correlators in position space}, we then review correlators of protected scalar operators in the ABJM theory, obtaining new expressions for four-point correlators in position space. In section \ref{sec:Carrollian limit of ABJM correlators}, we then compute the Carrollian limit of these correlators and in section \ref{sec:Bulk perspective in flat space} we derive these results from a bulk perspective. In section \ref{sec:Super conformal Carrollian correlators}, we derive the superconformal Carrollian algebra in three dimensions, demonstrate the relation to the bulk 4D super Poincare algebra, and use it to define super conformal Carrollian primaries. Finally in section \ref{sec:Conclusion} we present our conclusions and future directions. There are also several Appendices, where we review previous results on ABJM correlators in Mellin space (Appendix \ref{sec:ABJMmellinamp}), analyse the relation between the high energy limit in Mellin space and the Carrollian limit (Appendix \ref{sec:MCequivalence}), provide more details on how to derive two and three-point Carrollian correlators in the ABJM theory from bulk supergravity amplitudes (Appendix \ref{sec:3ptamps}), and consider higher-derivative corrections to supergravity (Appendix \ref{sec:higherdercamp}).

\section{Review}
\label{sec:Review}

In this section we will review some important concepts that we will make use of throughout the paper, notably the AdS$_4$/CFT$_3$ correspondence and the Carrollian limit of CFT correlators. 

\subsection{AdS$_4$/CFT$_3$ correspondence}
\label{sec:M-Theory/ABJM correspondence}

We are interested in the correspondence between M-theory on AdS${}_4 \times$S$^7/\mathbb{Z}_{k_{CS}}$ and ABJM theory on $\mathbb{R}^{2,1}$ \cite{Aharony:2008ug}. The AdS${}_4$ has radius $\ell$ and the $S^7$ has radius $2\ell$.  The ABJM theory is a superconformal Chern-Simons matter theory with gauge group $U(N)_{k_{CS}} \times U(N)_{-k_{CS}}$, where $k_{CS}$ is the Chern-Simons level and the matter fields are in the bi-adjoint representation of the gauge group. The theory has a Lagrangian description with $\mathcal{N}=6$ supersymmetry \cite{Benna:2008zy,Bandres:2008ry}, but for $k_{CS}=1,2$, the quantum theory has maximal $\mathcal{N}=8$ supersymmetry \cite{Kapustin:2010xq}. We will only consider these case where the Chern-Simons level $k_{CS} = 1$. 

The central charge $c_T$ is defined as the coefficient of the stress tensor two-point function. When $N \gg k_{CS}$, the relationship between $c_T$ and $N$ is \cite{Klebanov:1996un,Aharony:2008ug}
\begin{align}
    \label{eq:ctNrel}
c_T = \frac{64}{3\pi} \sqrt{2k_{CS}} N^{\frac{3}{2}} +
\mathcal{O}\left(N^{1/2}\right).
\end{align}
Moreover when $N \gg k_{CS}^5$, the bulk is described by supergravity on AdS$_4 \times$S$^7$ and we have 
\begin{align}
    \label{eq:Nlrel}
    \frac{\ell^6}{\ell_{11}^6} = \left(\frac{3\pi c_T k_{CS}}{2^{11}}\right)^{\frac{2}{3}} + \mathcal{O}\left(c_T^0\right) = \frac{N k_{CS}}{8} + \mathcal{O}\left(N^0\right),
\end{align}
where $\ell_{11}$ is the 11-dimensional Planck length. 

We will focus on correlators of scalar operators which are $1/2$-BPS (i.e. annihilated by half of the supersymmetry generators) and are dual to modes on the 7-sphere. These operators take the form $\mo_k^{I_1 \dots I_k}$, where $I_1, \dots , I_k$ are $SO(8)$ $R$-symmetry indices and $\mo_k^{I_1 \dots I_k}$ is symmetric trace-free. 
To make this property manifest, we will contract the indices with null vectors $t_I$ 
\begin{align}
\label{eq:contractedop}
    \mo_k\left(x, t\right) \equiv \mo_k^{I_1 \dots I_k}t_{I_1} \dots t_{I_k},
\end{align}
where the subscript $k$ denotes the R-charge. The scaling dimensions of these operators are protected and an operator with R-charge $k$ has conformal dimension $\D_k = \frac{k}{2}$. 
For the minimal value $k=2$, these operators belong to the stress tensor multiplet. 

The t'Hooft coupling is $\lambda=N/k_{CS}$ and the planar limit corresponds to taking $k_{CS}$ and $N$ to infinity while holding $\lambda$ fixed. In this limit, the theory becomes integrable (see \cite{Klose:2010ki} for a review). On the other hand, the enhancement of supersymmetry at $k_{CS}=1,2$ arises from non-perturbative effects involving monopole operators. In this regime, the operators $\mathcal{O}_k$ are quantum operators which are not constructed directly out of the fields in the Lagrangian and have no classical analogue \cite{Klebanov:2009sg,Lambert:2019khh}. As a result, their correlation functions have been mainly been computed using superconformal bootstrap methods \cite{Rastelli:2017udc,Chester:2018aca,Alday:2020dtb,Alday:2021ymb,Alday:2022rly,Chester:2024esn,Chester:2024bij}.

\subsection{Carrollian amplitudes} 
\label{sec:Carrollian amplitudes and Carrollian/flat space limit correspondence}

In this section, we briefly review salient results on Carrollian amplitudes in flat space, and their relation with holographic correlators in AdS. We will mainly follow \cite{Mason:2023mti,Alday:2024yyj} and refer to \cite{Donnay:2022aba,Bagchi:2022emh,Donnay:2022wvx,Salzer:2023jqv,Saha:2023abr,Nguyen:2023vfz,Nguyen:2023miw,Bagchi:2023fbj,Bagchi:2023cen,Liu:2024nfc,Stieberger:2024shv,Adamo:2024mqn,Ruzziconi:2024zkr,Jorstad:2024yzm,Ruzziconi:2024kzo,Marotta:2024sce,Kraus:2024gso,Kraus:2025wgi,Nguyen:2025sqk,Surubaru:2025fmg} for recent developments on this topic. Carrollian holography suggests that gravity in 4D asymptotically flat spacetime is dual to a 3D Carrollian CFT living at null infinity ($\mathscr{I})$. These theories exhibit conformal Carrollian or, equivalently \cite{Duval:2014uva}, BMS symmetries, as spacetime symmetries, and can be constructed from standard Lorentzian CFT by taking the Carrollian limit. Explicit examples of Carrollian field theories have been presented e.g. in \cite{Barnich:2012rz,Barducci:2018thr,Bagchi:2019xfx,Bagchi:2019clu,Grumiller:2020elf,Gomis:2020wxp,deBoer:2021jej,Henneaux:2021yzg,Gupta:2020dtl,Baiguera:2022lsw,Barnich:2022bni,Rivera-Betancour:2022lkc,Hao:2022xhq,Bagchi:2022eui,Miskovic:2023zfz,Ara:2024vbe,Banerjee:2024ldl} and their quantization has been discussed in \cite{deBoer:2023fnj,Chen:2021xkw,Chen:2023pqf,Chen:2024voz,Cotler:2024xhb,Sharma:2025rug,Poulias:2025eck}.

Let us first review how a massless scattering amplitude in Minkowski space can be recast as a correlator of local operators in a putative Carrollian CFT at $\mathscr{I}$. The momentum of a massless particle $j$ in Minkowski space can be parametrized by 
\begin{align}
    \label{eq:mompar}
    p_j =\frac{1}{\sqrt{2}} \e_j \om_j \left(1+z_j \zb_j, z_j+\zb_j, -i(z_j-\zb_j), 1-z_j \zb_j \right).
\end{align} 
Here $\epsilon_j = \pm 1$ labels an outgoing/incoming particle, $\omega_j$ is the energy and $(z_j,\bar z_j)$ coordinates on the celestial sphere. We will often find it useful to work in Klein space (spacetime with with $(2,2)$ signature), in which case the appropriate parametrization of the momentum is obtained by Wick rotating the third component.\footnote{In this case, $\epsilon_j = \pm 1$ labels the two Poincaré patches on the celestial torus \cite{Jorge-Diaz:2022dmy}.}
The Carrollian amplitude corresponding to the scattering of massless scalars is  \cite{Banerjee:2018gce,Banerjee:2018fgd,Donnay:2022aba,Bagchi:2022emh, Donnay:2022wvx,Mason:2023mti}
\begin{align}
    \mathcal{C}_n^{\Delta_1, \ldots , \Delta_n} \Big(\{ u_j, z_j, \bar{z}_j \}^{\epsilon_j} \Big)  = \int_0^{+\infty} \prod_{j=1}^n \frac{d\om_j}{2\pi} \, (-i \epsilon_j \omega_j)^{\Delta_j-1} e^{-i\e_j \om_j u_j}  \mathcal{A}_n \left(\{ \omega_j, z_j, \bar{z}_j \}^{\epsilon_j} \right) ,
\label{def Carrollian amplitude}
\end{align}
where $\mathcal{A}_n \left(\{ \omega_j, z_j, \bar{z}_j \}^{\epsilon_j} \right)$ is the momentum space amplitude. They can also be interpreted as correlators of Carrollian CFT primaries inserted at null infinity,
\begin{align}
   \mathcal{C}_n^{\Delta_1 , \ldots , \Delta_n} \Big(\{ u_j, z_j, \bar{z}_j \}^{\epsilon_j} \Big) \equiv   \left\langle \prod_{j=1}^n \Phi_{\Delta_j}^{\e_j} (u_j, z_j, \bar z_j) \right\rangle . 
\end{align}
At this stage, it is important to note that the encoding of the massless $\mathcal{S}$-matrix in \eqref{def Carrollian amplitude} is redundant, and one has to fix the value of $\Delta_i$ to obtain a one-to-one correspondence between massless scattering amplitudes and Carrollian CFT correlators. A natural choice is setting $\Delta_i = 1$, which is consistent with the extrapolate dictionary, and for which \eqref{def Carrollian amplitude} reduces to the Fourier transform \cite{Donnay:2022aba,Donnay:2022wvx,Mason:2023mti}
\begin{align}
    \mathcal{C}_n^{{1, \dots , 1}} \Big(\{ u_j, z_j, \bar{z}_j \}^{\epsilon_j} \Big)  = \int_0^{+\infty} \prod_{j=1}^n \frac{d\om_j}{2\pi}  e^{-i\e_j \om_j u_j}  \mathcal{A}_n \left(\{ \omega_j, z_j, \bar{z}_j \}^{\epsilon_j} \right).
\label{eq:deltaonecarramp}
\end{align}
As we shall see in Section \ref{sec:Bulk perspective in flat space}, in the context of ABJM, the value of $\D_i$ will be dictated by the $R$-symmetry properties of the primary. 

Carrollian amplitudes are the flat space analogues of holographic correlators in AdS. To see this, we briefly review the correspondence between flat space limit in the bulk theory and Carrollian limit in the boundary theory \cite{Alday:2024yyj}. The AdS$_4$ line element can be written in Bondi coordinates as
\begin{equation}
    ds^2_{AdS_4} =-  \frac{r^2}{\ell^2} du^2 - 2 du dr + 2 r^2 dz d\bar z,
\label{Bondi coordinates}
\end{equation} where the dimensions of length are $\ell \sim L$, $u \sim L$, $r \sim L$, $z\sim L^0$ and $\bar z \sim L^0$. The flat limit is obtained by taking $ \frac{\ell}{r} \gg 1$
which is distinct from the large $N$ limit $\frac{\ell}{\ell_{11}} \gg 1$ discussed in Section \ref{sec:M-Theory/ABJM correspondence}. Hence one could in principle consider the flat space limit term by term in the $1/N$ expansion (cf. Appendix \ref{sec:higherdercamp}). An advantage of the Bondi coordinates \eqref{Bondi coordinates} is that the flat limit can simply be obtained by formally taking $\ell \to \infty$, so that \eqref{Bondi coordinates} reduces to
\begin{equation}
    ds^2_{\mathbb{R}^{3,1}} = - 2 du dr + 2 r^2 dz d\bar z,
\end{equation} 
which is the line element of $\mathbb{R}^{3,1}$. Furthermore, the boundary metric of AdS$_4$ is the flat space Lorentzian metric
\begin{equation}
    ds^2_{\partial AdS_4} =  -\frac{du^2}{\ell^2}+ 2 dz d\zb.
\label{eq:minkmetric}
\end{equation} Implementing the flat limit in the bulk yields the degenerate metric 
\begin{equation}
    ds^2_{\mathscr{I}} = 0\,du^2 + 2 dz d\bar{z},
\end{equation} 
which is part of the Carrollian structure at null infinity \cite{1977asst.conf....1G,Henneaux:1979vn,Duval:2014uva,Ashtekar:2014zsa}, the boundary of 4D flat space. Notice that the $1/\ell^2$ in \eqref{eq:minkmetric} appears at the same place and plays the same role as if we were to restore the speed of light $c$ in a 3D Minkowski line element and take the Carrollian limit $c\to 0$ \cite{Levy1965}. Therefore, we have a correspondence between flat space limit in the bulk of AdS and Carrollian limit at the boundary, which is formally implemented in Bondi coordinates by the following identification:
\begin{equation}
    c_{\text{boundary}} \equiv \frac{1}{\ell_{\text{bulk}}}
\label{identification bulk boundary}
\end{equation} 

This correspondence was solidified in \cite{Alday:2024yyj}, where it was shown that the Carrollian limit of holographic CFT correlators yields Carrollian amplitudes. For the convenience of the reader, we now briefly review the relevant results of that paper. The procedure for obtaining the Carrollian amplitude for massless scalars $\an{\Phi_{\D_1} \dots \Phi_{\D_n}}$ from its Euclidean CFT counterpart $\an{\mo_{\D_1} \dots \mo_{\D_n}}$ is:
\begin{itemize}
    \item Analytically continue the correlator to Lorentzian/Kleinian signature.
    \item Compute $\lim_{c \to 0} \, c^{\sum_i \D_i-1} \an{\mo_{\D_1} \dots \mo_{D_n}}$ by keeping track of distributional terms and identify the rescaled operator $c^{\D-1}\mo_{\D}$ with $\Phi_{\D}$ up to a normalization.
\end{itemize}
The resulting object is the Carrollian amplitude. We will outline how this works for $2$, $3$ and $4$ point correlators below. 
\paragraph{2 points: } The two point function is completely fixed by conformal symmetry. After analytic continuation to Lorentzian signature it is given by
\begin{align}
    \an{\mo_{\D} \left(x_1\right) \mo_{\D} \left(x_2\right)}  = \frac{\mathcal{N}_2}{\left(x_{12}^2+ i \e\right)^{\D}},
\end{align}
where $\mathcal{N}_2$ is a normalization and $x_{ij}^2 = -c^2 u_{ij}^2 + 2 |z_{ij}|^2$. Following the procedure above, we compute
\begin{equation}
   \lim_{c \to 0} c^{2\D-2}  \an{\mo_{\D} \left(x_1\right) \mo_{\D} \left(x_2\right)} =  \frac{\mathcal{N}_2\, \delta^{2}(z_{12})}{2(\Delta -1) (-u_{12}+i \varepsilon)^{2\Delta - 2}} \propto \an{\Phi_{\D}^{\e_1}\Phi_{\D}^{\e_2} },
\end{equation} 
where we have suppressed the coordinate dependence of the Carrollian amplitude. After appropriate normalization and setting $\e_1 = -\e_2 = -1$, the above proportionality can be turned into an equality.
 
\paragraph{3 points:} Here it is convenient to work in Klein signature in the bulk since it allows for non-trivial 3-point amplitudes.\footnote{In Minkowski space, they are non-zero only when all momenta are collinear.} This amounts to treating $z_i, \zb_i$ as real and independent. The time-ordered correlator with $z_i, \zb_i$ real and independent is:
 \begin{align}
\an{\mo_{\D_1}\left(x_1\right)\mo_{\D_2}\left(x_2\right)\mo_{\D_3}\left(x_3\right)}_K = \frac{\mathcal{N}_3}{c}\frac{1}{\left(x_{12}^2 + i \varepsilon\right)^{\D_{12}} \left(x_{23}^2 + i \varepsilon\right)^{\D_{23}} \left(x_{13}^2+ i \varepsilon\right)^{\D_{13}}} \ .
\end{align} where $\mathcal{N}_3$ is once again a normalization and $\D_{ij} = \D_i + \D_j - \frac{1}{2}\sum_{k=1}^3 \D_k$. Applying the procedure outlined above, we get 
\begin{align}
      & \lim_{c\to 0} c^{3- \sum_{j=1}^3 \Delta_j}\langle \mathcal{O}_{\Delta_1}(u_1, z_1, \bar z_1) \mathcal{O}_{\Delta_2}(u_2, z_2, \bar z_2) \mathcal{O}_{\Delta_3}(u_3, z_3, \bar z_3)   \rangle \\
      & \nonumber \qquad= \frac{ \tilde{\mathcal{N}_3}\, \delta (\bar z_{12}) \delta (\bar{z}_{23})\Theta\left(z_{12}z_{31}\right)\Theta\left(z_{13}z_{23}\right) z_{12}^{\Delta_3-2}z_{23}^{\Delta_1}z_{13}^{\Delta_2-2}}{\left(u_1 z_{23}+u_2 z_{31}+u_3 z_{12}+i \text{sign }z_{23} \varepsilon \right)^{\sum_{j=1}^3 \Delta_j-4}} \propto  \langle \Phi_{\Delta_1} \Phi_{\Delta_2} \Phi_{\Delta_3} \rangle.
  \end{align}
We refer the reader to \cite{Alday:2024yyj} for the normalization factor $\tilde{\mathcal{N}}_3$. This coincides with a $3$-point Carrollian scalar amplitude with $\e_1 = -\e_2 = -\e_3 = 1$ obtained from the $3$-point amplitude in $(2,2)$ signature momentum space by using \eqref{def Carrollian amplitude}. 

\paragraph{4 points:} 
We will only need to consider the Carrollian limit of scalar contact diagrams in the bulk, which take the form
\begin{equation}
    \langle \mathcal{O}_{\Delta_1}(x_1) \mathcal{O}_{\Delta_2}(x_2) \mathcal{O}_{\Delta_3}(x_3) \mathcal{O}_{\Delta_4}(x_4) \rangle   \propto \bar{D}_{\D_1, \D_2, \D_3, \D_4}\left(U,V\right) \,
\end{equation} 
where we have dropped numerical factors and a coordinate-dependant one which encodes the conformal weights, and 
\begin{equation}
    U = \frac{x_{12}^2x_{34}^2}{x_{13}^2 x_{24}^2} = Z \Zb \ \ , \qquad V = \frac{x_{23}^2x_{14}^2}{x_{13}^2x_{24}^2} =(1-Z)(1-\Zb)
\end{equation} 
are the conformal cross ratios. The definition of $\bar{D}$ functions and various useful properties can be found in appendix D of \cite{Arutyunov:2002fh}. The $\bar{D}$ function becomes singular as $Z \to \Zb$ upon analytic continuation to Lorentzian signature  \cite{Gary:2009ae,Maldacena:2015iua} and its leading singularity is \cite{Alday:2024yyj}
\begin{align}
\label{eq:lsgen}
     \bar{D}_{\D_1, \D_2, \D_3, \D_4}\left(U,V\right) \xrightarrow[]{Z \to \Zb} \hat{\Phi}^{l.s}_{\D_1, \D_2, \D_3, \D_4} \equiv \mathcal{K}_\D \frac{Z^{\D_{3}+\D_{4}-2}(1-Z)^{\D_{1}+\D_{4}-2}}{(Z-\bar{Z})^{\sum_{i=1}^4 \Delta_i-3}}.
\end{align}
 The Carrollian limit is non-trivial only on the support of this leading singularity and
 \begin{align}
     \lim_{c \to 0} \hat{\Phi}^{l.s}_{\D_1, \D_2, \D_3, \D_4} = \mathcal{R}\left(u_i, z_i\right) \delta\left(z-\zb\right),
 \end{align}
 where $z = \frac{z_{12} z_{34}}{z_{13}z_{24}}$ is the 2D cross ratio and $\mathcal{R}\left(u_i, z_i\right)$ is a complicated function of the coordinates whose expression can be found in \cite{Alday:2024yyj}. Using these results, we can show that applying the procedure outlined at the beginning of this section to the correlator corresponding to the four point contact diagram results in
\begin{equation}
    \begin{split}
        &\lim_{c\to 0} c^{4 - \sum_\Delta}  \langle \mathcal{O}_{\Delta_1}(x_1) \mathcal{O}_{\Delta_2}(x_2) \mathcal{O}_{\Delta_3}(x_3) \mathcal{O}_{\Delta_4}(x_4) \rangle  \\
        &=  \mathcal{N}\,\left(\frac{\left|z_{23}\right|^2}{\left|z_{34}\right|^2\left|z_{24}\right|^2}\right)^{\frac{4-\Sigma_{\D}}{2}} \frac{z^{2-\D_1-\D_2}\left(1-z\right)^{\D_1+\D_4-2} \delta\left(z-\zb\right)}{\mathcal{U}^{\sum_{i=1}^4 \D_i -4}} \propto \langle \Phi_{\Delta_1}^{\e_1} \Phi_{\Delta_2}^{\e_2} \Phi_{\Delta_3}^{\e_3} \Phi_{\Delta_4}^{\e_4} \rangle,
    \end{split}
\end{equation}
where
\begin{align}
\label{eq:Carrollden}
    &\mathcal{U} =  u_4 - u_1 z \left|\frac{z_{24}}{z_{12}}\right|^2+u_2 \frac{1-z}{z}\left|\frac{z_{34}}{z_{23}}\right|^2 - u_3\frac{1}{1-z}\left|\frac{z_{14}}{z_{13}}\right|^2 
\end{align} 
is the translation-invariant denominator appearing in the four-point Carrollian amplitude. Depending on the details of the analytic continuation, we can have $z<0, 0<z<1$ or $z>1$. This coincides with the allowed values of $z$ for which the Carrollian amplitude is non-zero. Focussing on $0<z<1$, the proportionality can once again be turned into an equality after an appropriate choice of normalization and setting $\e_1 = \e_2 = -\e_3 = -\e_4 = 1$. Let us emphasize that the Carrollian limit discussed above is taken intrinsically in the CFT, without referring to the bulk spacetime. It is valid for any scalar subsector of holographic CFTs in three dimensions. We will apply this to ABJM correlators in section \ref{sec:Carrollian limit of ABJM correlators} to obtain correlators of a Carrollian ABJM theory living at null infinity.

\section{ABJM correlators in position space}
\label{sec:ABJM correlators in position space}
In this section, we present 2, 3 and 4-point correlation functions of $\frac{1}{2}$-BPS operators in the ABJM theory. The 2 and 3-point functions are computed directly in position space from the dual supergravity action. The 4-point function has been computed using bootstrap methods in Mellin space, the results of which we review in appendix \eqref{sec:ABJMmellinamp}. Here we rewrite these results in position space in a way that makes the computation of their Carrollian limit feasible. At 4 points, We will restrict our attention to correlators in the supergravity approximation, i.e the leading terms in the $\frac{1}{N}$ expansion, relegating a discussion of higher derivative corrections to appendix \eqref{sec:higherdercamp}. 

\subsection{Two and three-point functions}

On the supergravity side, we can identify the operator \eqref{eq:contractedop} as the source for one of the scalar fluctuations around the AdS${}_4$ $\times $ $S^7$ background. Denoting this bulk scalar by $s$, we can expand it in Kaluza Klein (KK) modes on the 7-sphere as  \cite{Bastianelli:1999vm} 
\begin{align}
\label{eq:scalarharexp}
    s = \sum_{k\ge 0} Y_k^{(7)} \, s_k \, = \sum_{k \ge 0} \frac{s_k}{\ell^k} \, \mathcal{C}^{(k)}_{I_1 \dots I_k} Z^{I_1} \dots Z^{I_k}
\end{align}
where $Y_k^{(7)}$ are spherical harmonics on S$^7$ and $Z^I$ are embedding coordinates for the S$^7$, notably coordinates in $\mathbb{R}^8$ such that $Z^I Z_I =1$. In the second equality, we have represented the spherical harmonics as homogeneous polynomials encoded by the traceless, symmetric tensor $\mathcal{C}_{I_1 \dots I_k}$ \cite{Lee:1998bxa}. The action for the scalar fields $s_k$ on AdS$_4$ can be derived from the 11D $\mathcal{N}=1$ supergravity action after integrating over the $S^7$ \cite{Bastianelli:1999en} and is given by \footnote{The modes $k=0, 1$ decouple from the action.}
\begin{align}
    \label{eq:scalaraction}
    S = \frac{243}{\kappa^2}\int_{\text{AdS}_4} d^4 y \, \sqrt{-\bar{g}_4} \Big\lbrace  &\sum_{k\geq 2} \frac{\left(2\ell\right)^7}{2}\,  A_k s_k \left(\Box_{AdS} - m_k^2\right)  s_k\an{\mathcal{C}^{(k)}\mathcal{C}^{(k)}} \nonumber \\
    &\qquad\qquad +\sum_{k_i\geq 2}\frac{\left(2\ell\right)^5}{3}\an{\mathcal{C}^{(k_1)}\mathcal{C}^{(k_2)}\mathcal{C}^{(k_3)}} g_{123} \,s_{k_1}\, s_{k_2}\, s_{k_3}\Big\rbrace,
\end{align}
where $\kappa$ is the 11D gravitational coupling and is related to the 11D Planck length by $4\kappa^2 = \left(2\pi\right)^5 \ell_{11}^9$. The other constants appearing in the action are 
\begin{align}
    \label{eq:constants}
    &A_k = \frac{4\pi^4 k (k-1)}{3\times 2^k (k+1)(k+2)^2} , \qquad  m_k^2 = \frac{k\left(k-6\right)}{4\ell^2},\\
    &\nonumber g_{123} = \frac{192 \pi^4  \left(\alpha^2-9\right)\, \left(\alpha^2-1\right)\, \left(\alpha+2\right)}{\left(2\alpha+6\right)!!} \prod_{i=1}^3 \frac{k_i!}{\left(k_i+2\right)\Gamma\left(\alpha_i\right)}.
\end{align} 
Here  $\an{\mathcal{C}^{(k_1)} \dots \mathcal{C}^{(k_n)} }$ is the unique SO$(7)$ invariant contraction of the tensors representing the spherical harmonics and
\begin{align}
\label{eq:alphadef}
    \alpha_i = \frac{1}{2}\sum_{j=1}^3 k_j -k_i, \qquad \alpha = \frac{1}{2}\sum_{i=1}^3 k_i
\end{align}The scalar fields $s_k$ couple to the operators $\mo_k$ in the dual ABJM theory via 
\begin{align}
    S_{int} = \int_{\partial\text{AdS}_4} d^3y  \, w_k \, s_k^{(0)} \mo_k,
\end{align}
where $w_k$ are proportionality factors, $s^{(0)}_k$ is the boundary value of $s_k$ and $\mo_k = \mo_{I_1 \dots I_k} \mathcal{C}^{I_1 \dots I_k}$. From \eqref{eq:constants} and the standard relation $\Delta(\Delta-d)=m^2 \ell^2$ (where $d$ is the boundary dimension), we see that the spectrum of scaling dimensions of the operators dual to the scalars $s_k$ is indeed $k/2$. 
We can make contact with the operators in \eqref{eq:contractedop} if we set 
\begin{align}
    \label{eq:connection}
    \mathcal{C}^{k}_{I_1 \dots I_k} = t_{I_1} \dots t_{I_k}.
\end{align}
Note that this choice implies the normalization
\begin{align}
    \label{eq:cnormalization}
  \an{\mathcal{C}^{(k_1)}\mathcal{C}^{(k_2)}} \equiv  \mathcal{C}^{(k_1)}_{I_1 \dots I_{k_1}}\mathcal{C}^{(k_2)I_1 \dots I_{k_2}} = t_{12}^{k_1} \, \delta_{k_{1}, k_{2}},
\end{align}
where $t_{12} \equiv t_1 \cdot t_2$. In the rest of this paper, we will tacitly assume that this choice has been made. For more details, we refer the reader to \cite{Bastianelli:1999en, Bastianelli:1999vm}.

\paragraph{Two-point functions} The two-point function of the operators $\mo_k$ can be computed by evaluating the supergravity action \eqref{eq:scalaraction} on-shell and differentiating it with respect to the scalars. This yields
\begin{align}
    \label{eq:2ptfuncgen}
    \an{\mo_{k_1}\left(x_1, t_1\right) \mo_{k_2} \left(x_2, t_2\right) } = 243 A_k\frac{\left(2\ell\right)^7}{\kappa^2} \frac{\left(k-3\right)}{\pi^{\frac{3}{2}}} \frac{\Gamma\left(\frac{k}{2}\right)}{\Gamma\left(\frac{k-3}{2}\right)}\frac{w_{k_1}^2 t_{12}^{k_1} \delta_{k_1, k_2}}{\left(x_{12}^2\right)^{k_1}}
\end{align}
We choose the constants $w_k$ such that the two-point function has the normalization
\begin{equation}
    \label{eq:ABJM2ptfunc} 
    \an{\mo_{k_1}\left(x_1, t_1\right) \mo_{k_2} \left(x_2, t_2\right) } = \frac{\delta_{k_1, k_2}\, t_{12}^{k_1}}{\left(x_{12}^2\right)^{\frac{k_1}{2}}}. 
\end{equation}
Plugging in the value of $A_k$ from \eqref{eq:constants}, replacing $\kappa$ by the 11D Planck length and simplifying the resulting expression, we get
\begin{align}
    \label{eq:wk}
    w_k = \frac{\left(2\pi\right)^{\frac{3}{2}}\sqrt{2}\ell}{9} \left(\frac{\ell_{11}}{2\ell}\right)^{\frac{9}{2}} \frac{(k+2)}{(k-3)(k-1)}\frac{\sqrt{\Gamma(2+k)}}{\Gamma\left(\frac{k}{2}+1\right)}
\end{align}

\paragraph{Three-point functions} The three point function of $\mo_k$ derived from the supergravity action is \cite{Bastianelli:1999en}
\begin{align}
   \label{eq:ABJM3ptfunc}
\an{\mo_{k_1}\,\mo_{k_2}\,\mo_{k_3}} =  \left(\frac{\ell_{11}}{\ell}\right)^{\frac{9}{2}}R_{k_1, k_2, k_3} \frac{t_{12}^{\alpha_3}t_{23}^{\alpha_1}t_{13}^{\alpha_2}}{x_{12}^{\alpha_3}x_{23}^{\alpha_1}x_{13}^{\alpha_2}},
\end{align}
with 
\begin{align}
    \left(\frac{\ell_{11}}{\ell}\right)^{\frac{9}{2}}R_{k_1, k_2, k_3} = \frac{7776 (2\ell)^6}{\ell_{11}^9 \left(2\pi\right)^8}\Gamma\left(\frac{\alpha -3}{2}\right) \prod_{i=1}^3\frac{ \Gamma\left(\frac{\alpha_i}{2}\right)w_{k_i}}{\Gamma\left(\frac{k_i-3}{2}\right)}  g_{123}.
\end{align}
Plugging in the value of $w_{k_i}$ from \eqref{eq:wk} and simplifying, we get 
\begin{align}
   \label{eq:ABJM3ptcoefficient}
    \left(\frac{\ell_{11}}{\ell}\right)^{\frac{9}{2}} R_{k_1, k_2, k_3} = \frac{\pi}{2^{\frac{5}{2}}} \left(\frac{\ell_{11}}{\ell}\right)^{\frac{9}{2}}\frac{2^{-\alpha}}{\Gamma\left(1+\frac{\alpha}{2}\right)} \prod_{i=1}^3 \frac{\sqrt{\Gamma\left(k_i+2\right)}}{\Gamma\left(\frac{\alpha_i+1}{2}\right)} = \frac{\pi }{N^{\frac{3}{4}}}\frac{2^{-\alpha-\frac{1}{4}}}{\Gamma\left(1+\frac{\alpha}{2}\right)} \prod_{i=1}^3 \frac{\sqrt{\Gamma\left(k_i+2\right)}}{\Gamma\left(\frac{\alpha_i+1}{2}\right)}.
\end{align}
In arriving at the last equality, we have used the relationship between $\ell$ an $N$ from \eqref{eq:Nlrel} and set $k_{CS}=1$ for simplicity. The numerator was obtained by evaluating the contraction 
\begin{align}
    \label{eq:3ptnum}
    \an{\mathcal{C}^{k_1}\mathcal{C}^{k_2}\mathcal{C}^{k_3}} = t_{12}^{\alpha_3}t_{23}^{\alpha_1}t_{13}^{\alpha_2}
\end{align}
In particular, note that the 3 point function is finite when $k_i = 2$ or $\D_i = \frac{k_i}{2} = 1$. This is in contrast with the three point couplings considered in \cite{Alday:2024yyj}.

\subsection{Four-point functions}
\label{sec:Four point correlators in ABJM}

Four-point functions of the superconformal primaries in ABJM can be written as \cite{Dolan:2004mu,Alday:2020dtb}
\begin{align}
\label{eq:4ptcorr}
    \an{\mo_{k_1}\dots \mo_{k_4}} = \prod_{i<j} \left(\frac{t_{ij}^2}{x_{ij}^{2}}\right)^{\frac{\gamma_{ij}^0}{2}} \left(\frac{t_{12}^2 t_{34}^2}{x_{12}^{2}x_{34}^{2}}\right)^{\frac{\mathcal{E}}{2}} \mathcal{G}_{k_1, \dots, k_4}\left(U, V,\sigma, \tau\right).
\end{align}
Here $t_{ij} = t_i \cdot t_j, x^2_{ij} = -c^2 u_{ij}^2 + 2 z_{ij} \zb_{ij}$ and 
\begin{align}
\label{eq:crossratios}
    &U = \frac{x_{12}^2 x_{34}^2}{x_{13}^2x_{24}^2} = Z \Zb, \qquad V = \frac{x_{14}^2 x_{23}^2}{x_{13}^2 x_{24}^2} = \left(1-Z\right) \left(1-\Zb\right), \qquad \sigma = \frac{t_{13} t_{24}}{t_{12} t_{34}}, \qquad \tau = \frac{t_{14} t_{23}}{t_{12} t_{34}}.
\end{align}
The extremality $\mathcal{E}$ is
\begin{align}
\label{eq:extremality}
     \mathcal{E} = \begin{cases}
        \frac{k_1+k_2+k_3-k_4}{2} \qquad\qquad &\text{Case I}: k_1+k_4 \geq k_2+k_3,\\
        k_1  \qquad\qquad  &\text{Case II}: k_1+k_4 < k_2+k_3,
    \end{cases}
\end{align}
and the exponents $\gamma_{ij}^0$ are  given by
\begin{align}
    \label{eq:gammas}
    &\gamma_{12}^0 = \gamma_{13}^0 = 0, \qquad \gamma_{34}^0 = \frac{\kappa_s}{2}, \, \gamma_{24}^0 = \frac{\kappa_u}{2},\\
    &\nonumber \text{Case I: } \gamma_{14}^0 = \frac{\kappa_t}{2}, \qquad \gamma_{23}^0 = 0, \qquad \text{Case II: } \gamma_{14}^0 = 0, \qquad \gamma_{23}^0 = \frac{\kappa_t}{2} ,
\end{align}
where 
\begin{align}
\label{eq:kappas}
    \kappa_s = \left|k_1 + k_2 - k_3 - k_4\right|, \qquad   \kappa_t = \left|k_1 + k_4 - k_2 - k_3\right|, \qquad  \kappa_u = \left|k_2 + k_4 - k_1 - k_3\right|.
\end{align}

These correlators admit a large $c_T$ expansion of the form
 \begin{align}
     \label{eq:genGlargeN}
     \mathcal{G}_{k_1, \dots , k_4} =   \mathcal{G}^0_{k_1, \dots , k_4} +   \frac{1}{c_T}\mathcal{G}^R_{k_1, \dots ,  k_4} + \dots,  
 \end{align}
where $ \mathcal{G}^0_{k_1, \dots ,k_4}$ is the disconnected part of the correlator which is described by generalized free fields. We will ignore this contribution for the rest of this paper and focus only on the connected part. The leading contribution in the large $c_T$  limit comes from tree-level supergravity in the bulk and has been computed in \cite{Alday:2020dtb}. The stress tensor belongs to the $k=2$ multiplet and these correlators are of particular interest. Corrections in $\frac{1}{c_T}$ arise from higher derivative and loop corrections to supergravity in the bulk. These have been computed in \cite{Alday:2022rly}. 
In \cite{Alday:2020dtb}, the authors exploit the ambiguity inherent in the definition of exchange diagrams to absorb all contact terms into them and write
\begin{align}
    \mathcal{G}_{k_1, k_2, k_3, k_4}^R =  \mathcal{G}_{k_1, k_2, k_3, k_4, s}^R+ \mathcal{G}_{k_1, k_2, k_3, k_4, t}^R+ \mathcal{G}_{k_1, k_2, k_3, k_4, u}^R, 
\end{align} where the subscripts stand for $s-$, $t-$ and $u-$channels. 
All of these correlators have been computed in Mellin space. The connected 4-point correlator 
\begin{align}
    \label{eq:conncorr}
\mathcal{G}^c_{k_1, \dots k_4}\left(U, V,\sigma, \tau\right) \equiv \mathcal{G}_{k_1, \dots k_4}\left(U, V,\sigma, \tau\right) - \mathcal{G}^0_{k_1, \dots k_4}\left(U, V,\sigma, \tau\right)
\end{align}
admits the following Mellin representation:
\begin{align}
\label{eq:Mellinrep}
     &\mathcal{G}^c_{k_1, \dots k_4}\left(U, V,\sigma, \tau\right) = \int_{-i\infty}^{i \infty} \frac{ds \, dt}{\left(4\pi i\right)^2} \, U^{\frac{s}{2}-a_s} \, V^{\frac{t}{2}-a_t} \, \mathcal{M}_{k_1, \dots k_4}\left(s, t; \sigma, \tau\right) \, \Gamma_{\left\lbrace k_i \right\rbrace},
\end{align}
where
\begin{align}
\label{eq:gengammas}
     & \Gamma_{\left\lbrace k_i \right\rbrace} = \Gamma\left(\frac{k_1+k_2}{4}-\frac{s}{2}\right)\Gamma\left(\frac{k_3+k_4}{4}-\frac{s}{2} \right)\Gamma\left(\frac{k_1+k_4}{2}-\frac{t}{2} \right) \\
     &\qquad\qquad \nonumber \Gamma\left(\frac{k_2+k_3}{4}-\frac{t}{2} \right)\Gamma\left(\frac{k_1+k_3}{4}-\frac{u}{2} \right)\Gamma\left(\frac{k_2+k_4}{4}-\frac{u}{2} \right)\nonumber, \\
    &\nonumber a_s = \frac{1}{4}\left(k_1+k_2\right) - \frac{1}{2} \mathcal{E}, \qquad a_t = \frac{1}{4} \text{Min} \left\lbrace k_1+k_4, k_2+k_3\right\rbrace, \qquad s+t+u = \frac{1}{2} \sum_{i=1}^4 k_i.
\end{align}  We have relegated the discussion of the details of the ABJM Mellin amplitudes to Appendix \ref{sec:ABJMmellinamp}. We will convert these expressions to position space by recognizing certain pieces in them as $\bar{D}$ functions by comparing with their Mellin representation \cite{Symanzik:1972wj,Mack:2009gy,Penedones:2010ue}
\begin{align}
\label{eq:dbarnewrep}
\bar{D}_{\Delta_{1} \Delta_{2} \Delta_{3} \Delta_{4}}(U, V)=\int_{-i \infty}^{i \infty} \frac{d j_{1} d j_{2}}{(2 \pi i)^{2}} U^{j_{1}} V^{j_{2}} \Gamma\left(j_{1}+j_{2}+\Delta_{2}\right) \Gamma\left(j_{1}+j_{2}+\Delta-\Delta_{4}\right)& \nonumber\\
\times \Gamma\left(-j_{1}\right) \Gamma\left(-j_{2}\right) \Gamma\left(-j_{1}-\Delta+\Delta_{3}+\Delta_{4}\right) \Gamma\left(-j_{2}+\Delta-\Delta_{2}-\Delta_{3}\right),
\end{align}
where $2\D = \sum_{i=1}^4 \D_i$. This definition holds for zero, half-integer and negative integer weights. 

In the rest of the paper, we will focus on connected correlators \eqref{eq:conncorr} and drop the superscript $c$. Furthermore, we will focus on the $\frac{1}{c_T}$ contributions 
and drop the superscript $R$. Hence, we denote $\mathcal{G}_{k_1,k_2,k_3,k_4}^{c,R}\equiv \mathcal{G}_{k_1,k_2,k_3,k_4}$.

\subsubsection{$\mathcal{G}_{2,2,2,2}$}
The four-point function involving operators with weight $\D_i = \frac{k_i}{2} = 1$ is of particular interest since they are at the bottom of the stress tensor multiplet. The definition of the Mellin amplitude \eqref{eq:Mellinrep} adapted to the case $k_1 = k_2 = k_3 = k_4 =2$ gives 
\begin{align}
    \mathcal{G}_{2,2,2,2} \left(U, V, \sigma, \tau\right) &\equiv \int_{-i\infty}^{i \infty}\, \frac{ds \, dt}{\left(4\pi i\right)^2} \, U^{\frac{s}{2}} \, V^{\frac{t}{2}-1} \, \mathcal{M}_{2,2,2,2}  \,\Gamma^2\left(\frac{2-s}{2}\right) \Gamma^2\left(\frac{2-t}{2}\right)\Gamma^2\left(\frac{s+t-2}{2}\right).
\label{mellin2222}    
\end{align}
The contact contributions are polynomials in $s,t$ and can be directly written as $\bar{D}$ functions in position space. The $s$-channel contribution to the position space correlator can be evaluated by starting from \eqref{eq:Mellin2222sclosed}, plugging it into \eqref{mellin2222}, cancelling the $s(s+2)$ poles by shifting the arguments of various $\Gamma$ functions, comparing with \eqref{eq:dbarnewrep} and writing it as a sum of $\bar{D}$ functions. In doing so, we arrive at the following expression :
\begin{align}
\label{eq:2222sposspaces}
      \mathcal{G}_{2,2,2,2,s}  = -\frac{6}{\sqrt{8N^3 \pi^3}}&\left[\left(3 \sqrt{\pi} U \bar{D}_{3,1,0,0}-\sqrt{\pi} \, U^2 \bar{D}_{4,2,0,0} -2 \sqrt{U} \bar{D}_{\frac{5}{2}, \frac{1}{2},0,0}\right)\right. \\
      &\nonumber \left. \,\,+ \sigma \left(3 \sqrt{\pi} U \bar{D}_{2,1,0,1}-\sqrt{\pi}U^2 \bar{D}_{3,2,0,1}-2\sqrt{U} \bar{D}_{\frac{3}{2},\frac{1}{2},0,1} \right) \right.\\
      &\nonumber \left.\,\,+\tau \left(3\sqrt{\pi} U \bar{D}_{2,1,1,0} -\sqrt{\pi} U^2 \bar{D}_{3,2,1,0} -2\sqrt{U} \bar{D}_{\frac{3}{2},\frac{1}{2},1,0} \right)\right].
\end{align}
The $t, u$ channel contributions can be expressed in terms of the $s$ channel one by using (this follows from \eqref{eq:Mellintu}) 
\begin{align}
    \label{eq:posspacetu}
    &\mathcal{G}^R_{2,2,2,2,t} \left(U, V, \sigma, \tau\right) = \tau^2 \frac{U}{V} \mathcal{G}^R_{2,2,2,2,s}\left(V, U, \frac{\sigma}{\tau}, \frac{1}{\tau} \right), \\
    &\nonumber \mathcal{G}^R_{2,2,2,2,u} \left(U, V, \sigma, \tau\right) = \sigma^2 U  \mathcal{G}_{2,2,2,2,s}\left(\frac{1}{U}, \frac{V}{U}, \frac{1}{\sigma}, \frac{\tau}{\sigma} \right),
\end{align}
along with the $\bar{D}$ function identities
\begin{align}
    \label{eq:dbaridentities}
    &\bar{D}_{\D_1, \D_2, \D_3, \D_4}\left(V, U \right) =   \bar{D}_{\D_3, \D_2, \D_1, \D_4}\left(U, V \right),\\  
    &\nonumber \bar{D}_{\D_1, \D_2, \D_3, \D_4}\left(\frac{1}{U}, \frac{V}{U} \right) =  U^{\D_2} \bar{D}_{\D_4, \D_2, \D_3, \D_1}\left(U, V \right).
\end{align}
We can now write down the position space correlator from \eqref{eq:4ptcorr}. Since $k_1 = k_2 = k_3 = k_4 = 2$, equations \eqref{eq:extremality}, \eqref{eq:gammas} and \eqref{eq:kappas} give
\begin{align}
    \label{eq:2222parameters}
    \mathcal{E} = 2, \quad \kappa_s = \kappa_t = \kappa_u = 0, \qquad \gamma_{ij}^0 = 0,
\end{align}
and we have 
\begin{align}
\label{eq:11114pt}
    \an{\mo_2 \left(x_1, t_1\right) \dots \mo_2 \left(x_4, t_4\right)} = \frac{t_{12}^2 t_{34}^2}{x_{12}^2 x_{34}^2} \mathcal{G}_{2,2,2,2} \left(U, V, \sigma, \tau\right) 
\end{align}

\subsubsection{$\mathcal{G}_{2,2,k,k}$}
We will follow a similar procedure to evaluate the position space correlator $\mathcal{G}_{2,2,k,k}$ whose Mellin representation is 
\begin{align}
    \label{eq:G22kkdef}
    \mathcal{G}_{2,2,k,k}\left(U,V,\sigma, \tau\right) = \int_{-i \infty}^{i \infty}  \frac{ds \, dt}{\left(4\pi i\right)^2} \, &U^{\frac{s}{2}} V^{\frac{t}{2}-\frac{k}{4}-\frac{1}{2}} \mathcal{M}_{2,2,k,k} \\
    \nonumber &\times\Gamma\left(1-\frac{s}{2}\right) \Gamma\left(\frac{k-s}{2}\right) \Gamma^2\left(\frac{1}{2}+\frac{k}{4}-\frac{t}{2}\right) \Gamma^2\left(\frac{s+t-1}{2}-\frac{k}{4}\right)
\end{align}
The $s-$channel contribution from \eqref{eq:M22kks} is 
\begin{align}
\label{eq:G22kkk}
    \mathcal{G}_{2,2,k,k,s} = &\frac{6}{\sqrt{8\pi^3 N^3}}\left[\frac{\sqrt{\pi}}{\Gamma\left(\frac{k}{2}\right)} \left(\left(1-k\right)U \partial_U -k\right)\left(\frac{V}{U}\bar{D}_{-1,1,\frac{k}{2}+1,\frac{k}{2}+1} + \sigma V \bar{D}_{0,1,\frac{k}{2}+1,\frac{k}{2}} +\tau \bar{D}_{0,1,\frac{k}{2},\frac{k}{2}+1}\right)   \right. \nonumber \\
    &\left.  \qquad \qquad\qquad +k \left(\frac{V}{U}\bar{D}_{-1,1,\frac{3}{2},\frac{3}{2}} + V\sigma  \bar{D}_{0,1,\frac{3}{2},\frac{1}{2}} + \tau\bar{D}_{0,1,\frac{1}{2},\frac{3}{2}} \right) \right]
\end{align}
While this can be simplified and expressed fully in terms of $\bar{D}$ functions, we will not do so here as the above form is more suitable for taking the Carrollian limit. The $t-$channel contribution is more complicated even in Mellin space and is given in \eqref{eq:M22kkt}. Upon converting to position space, we get
\begin{align}
     &\mathcal{G}_{2,2,k,k,t} =(-1)^{\frac{k}{2}}\frac{12k\tau U}{\sqrt{2N^3}}\frac{\Gamma\left(\frac{k}{2}+1\right)}{\Gamma\left(\frac{k}{2}-\frac{1}{2}\right)} \left[\frac{1}{4\pi}\left(\bar{D}_{\frac{1}{2},2-\frac{k}{2},0,\frac{1+k}{2}}+\sigma \bar{D}_{\frac{1}{2},1-\frac{k}{2},1,\frac{1+k}{2}} + \tau \bar{D}_{\frac{1}{2},1-\frac{k}{2},0,\frac{3+k}{2}}\right) \right.\nonumber \\
     &\qquad \left.-\sum_{i=0}^{\lceil \frac{k-1}{2}\rceil} 2^i x_i \left(k\right) \left(V \partial_V + \frac{k}{4}+\frac{1}{2}\right)^i \left(\bar{D}_{1,2-\frac{k}{2},0,1+\frac{k}{2}}+ \sigma  \bar{D}_{1,1-\frac{k}{2},1,1+\frac{k}{2}}+ \tau\bar{D}_{1,1-\frac{k}{2},0,2+\frac{k}{2}}\right) \right],
\end{align}
where $x_i\left(k\right)$ is defined in \eqref{eq:xkformula}. Finally, the $u-$channel contribution can be obtained from the $t-$channel one by 
\begin{align}
    \label{eq:u22kk}
\mathcal{G}_{2,2,k,k,u}\left(U,V,\sigma,\tau\right) = \mathcal{G}_{2,2,k,k,t}\left(\frac{U}{V},\frac{1}{V},\tau, \sigma\right).
\end{align}
We can now write down the position space correlator from \eqref{eq:4ptcorr}. Since $k_1 = k_2 = 2$ and $k_3 = k_4 = k$, Equations \eqref{eq:extremality}, \eqref{eq:gammas}, \eqref{eq:kappas} give
\begin{align}
    \label{eq:22kkparameters}
    \mathcal{E} = 2, \quad \kappa_s = 2k-4,\, \kappa_t = \kappa_u = 0, \qquad \gamma_{12}^0 = \gamma_{13}^0 =\gamma_{14}^0 =\gamma_{23}^0 =\gamma_{24}^0 = 0,\,\gamma_{34}^0 = k-2.
\end{align}
and \eqref{eq:4ptcorr} reduces to
\begin{align}
\label{eq:22kk}
    \an{\mo_2\mo_2 \mo_k \mo_k} = \left(\frac{t_{34}^{2} }{x_{34}^{2}}\right)^{\frac{k-2}{2}} \left(\frac{t^2_{12}t^2_{34}}{x_{12}^{2}x_{34}^{2}}\right)\mathcal{G}_{2,2,k,k} \left(U, V, \sigma, \tau\right).
\end{align} 

\subsubsection{$\mathcal{G}_{k_1, k_2, k_3, k_4}$}

The full correlator for generic $k_i$ cannot be expressed as a finite sum of $\bar{D}$ functions in position space and we will restrict our attention to the leading high energy term of the Mellin amplitude (cf. \eqref{eq:MellinHEgenk}). As shown in Appendix \ref{sec:MCequivalence}, this is sufficient to compute the Carrollian limit.\footnote{We have presented a working proof of this in Appendix[\ref{sec:MCequivalence}]. We will address the connection more completely in a future publication.} We begin by assuming that $k_1+k_2$, $k_1+k_3$, $k_1+k_4 \in 2\mathbb{Z}^+$. The final result will be valid in all cases. We  can use the formula
\begin{align}
    \frac{1}{s}\Gamma\left(\frac{k_1+k_2}{4}-\frac{s}{2}\right) =-\frac{1}{2}\Gamma\left(-\frac{s}{2}\right) \prod_{n=1}^{\frac{k_1+k_2}{4}-1} \left[\frac{k_1+k_2}{4}-\frac{s}{2}-n\right],
\end{align}
and analogous ones for $t, u$ to express \eqref{eq:MellinHEgenk} in position space as
\begin{align}
\label{eq:HEgenk}
     \mathcal{G}^{HE}_{k_1, k_2, k_3, k_4} = -&\frac{1}{2}\mathcal{N}_{k_i} P_{k_i} \left(\sigma, \tau\right) (-1)^{\frac{k_1+k_2}{2}+\frac{k_1+k_4}{2}+\frac{k_1+k_3}{4}} \left((1-\alpha)U \partial_U +V\partial_V\right)^2 \left((1-\bar{\alpha})U\partial_U + V\partial_V\right)^2\nonumber\\
     & \sum_{r} \binom{\frac{k_1+k_3}{4}-1}{r} U^{\frac{k_1+k_2}{4}-1-a_s+r} \, V^{\frac{2k_1+k_3+k_4}{4}-2-a_t-r} \bar{D}_{a_1+r-1,a_2-3,a_3-r-2, a_4},
\end{align}
where
\begin{align}
    &a_1 = \frac{k_2-k_4}{4}, \quad a_2 = \frac{2k_1+k_2+k_4}{2}, \quad a_3 = \frac{k_1-k_2+k_3+k_4}{4}, \quad a_4 = \frac{-k_1+k_2+k_3+k_4}{4}.
\end{align}
The superscript serves as a reminder that this is not the full position space correlator but merely the one corresponding to the leading high energy (HE) behaviour of the Mellin amplitude.  With this, we have
\begin{align}
    \label{eq:4ptfuncHEgenk}
    \an{\mo_{k_1}\dots \mo_{k_4}}^{HE} = \prod_{i<j} \left(\frac{t_{ij}^2}{x_{ij}^{2}}\right)^{\frac{\gamma_{ij}^0}{2}} \left(\frac{t_{12}^2 t_{34}^2}{x_{12}^{2}x_{34}^{2}}\right)^{\frac{\mathcal{E}}{2}} \mathcal{G}^{HE}_{k_1, \dots, k_4}\left(U, V,\sigma, \tau\right).
\end{align}

\section{Carrollian limit of ABJM correlators}
\label{sec:Carrollian limit of ABJM correlators}

In this section, we implement the Carrollian limit of the ABJM correlators derived in the previous sections. 
We will follow the procedure presented in \cite{Alday:2024yyj} and reviewed in Section \ref{sec:Carrollian amplitudes and Carrollian/flat space limit correspondence}. 
The flat space limit of the full AdS$_4 \times S^7$ line element, 
\begin{equation}
    ds^2_{AdS_4 \times S^7} = ds^2_{AdS_4} + 4 \ell^2 ds^2_{S^7}
\end{equation} is more subtle than the flat space limit of the AdS$_4$ factor alone described around \eqref{Bondi coordinates}. Indeed, the $S^7$ factor decompactifies, so that $\lim_{\ell \to \infty}ds^2_{AdS_4 \times S^7} = ds^2_{\mathbb{R}^{10,1}}$, yielding an infinite tower of massless KK modes. One of the objectives of our analysis is to understand how the decompactification of $S^7$ is seen from the 3D boundary perspective when taking the Carrollian limit. 

Throughout this section, we will work with the coordinates $(u,z,\bar{z})$ for which the metric on 3D Minkowski space, where the ABJM theory is living, is given by \eqref{eq:minkmetric}, and we will implement the Carrollian limit $c \equiv \frac{1}{\ell} \to 0$ on the CFT correlators. 
We define the electric Carrollian operators $\Phi_k$ by 
\begin{align}
    \label{eq:operatorexp}
    \mo_k =\sigma_k \, \ell^{\frac{k}{2} - 1}  \, \Phi_k, \qquad \sigma_k =  \frac{2\pi}{\sqrt{\Gamma\left(k-1\right)}}. 
\end{align}
The normalization $\sigma_k$ has been chosen so that the Carrollian limit of the two point function agrees with the 2 point Carrollian amplitude \eqref{eq:2ptcarrollian}. In particular, the scaling with $\ell$ in \eqref{eq:operatorexp} is consistent with the one used in Section \ref{sec:Carrollian amplitudes and Carrollian/flat space limit correspondence}, upon identification \eqref{identification bulk boundary}.

\subsection{Two and three-point functions}

We start off by computing the Carrollian limit of the two and three-point functions. The electric limit of \eqref{eq:ABJM2ptfunc}, after analytic continuation to Lorentzian signature, is 
\begin{align}
\label{eq:2ptcarrolllim}
   \lim_{\ell \to \infty} \frac{\an{\mo_{k_1}\left(x_1\right) \mo_{k_2}\left(x_2\right)}}{\ell^{\frac{k_1+k_2}{2}-2}\sigma_{k_1} \sigma_{k_2}} = \an{\Phi_{k_1} \Phi_{k_2}} = \frac{\delta_{k_1, k_2}}{\left(2\pi\right)^2}\frac{(-1)^{\frac{k_1}{2}-1} \Gamma(k_1-2)}{\left(u_{12}-i \e\right)^{k_1-2} } \,t_{12}^{k_1}\, \delta^2\left(z_{12}\right),
\end{align}
which is the electric two-point Carrollian amplitude \eqref{eq:2ptcarrollian} whose precise relation with a two-point flat space amplitude will be discussed in section \ref{sec:Two and Three point amplitudes from 11D SUGRA}. The dependence of three-point correlators in ABJM \eqref{eq:ABJM3ptfunc} on $\ell$ is different from the scalar correlators considered in \cite{Alday:2024yyj}. This reflects the fact that these arise from a theory on AdS${}_4 \times S^7$ rather than AdS${}_4$. Applying the analysis of section 4.3 of \cite{Alday:2024yyj} and reviewed in section \ref{sec:Carrollian amplitudes and Carrollian/flat space limit correspondence}, which involves analytic continuation to $(2,2)$ Kleinian signature in the bulk of AdS$_4$, we see that 
\begin{align}
      \frac{\an{\mo_{k_1}\mo_{k_2}\mo_{k_3}}}{\ell^{\frac{k_1+k_2+k_3}{2}-3}} \xrightarrow[]{\ell \to \infty} \mathcal{O}\left(\frac{1}{\ell^{\frac{11}{2}}}\right).
\end{align}
While this might seem troubling, this behaviour must be compared with Carrollian amplitudes of appropriately normalized scalars. We will show in Section \ref{sec:Two and Three point amplitudes from 11D SUGRA} that such amplitudes also vanish at an identical rate with respect to an IR cut-off. It is thus useful to compute the leading order term in the limit: 
\begin{align}
    \label{eq:carroll3ptabjm}
    \lim_{\ell\to \infty} \ell^{\frac{11}{2}} \frac{\an{\mo_{k_1}\mo_{k_2}\mo_{k_3}}}{\ell^{\frac{k_1+k_2+k_3}{2}-3}\prod_{i=1}^3 \sigma_{k_i}} =& \frac{\Gamma\left(\frac{\alpha}{2}-2\right)\Theta\left(z_{12}z_{31}\right)\Theta\left(z_{13}z_{23}\right)}{\Gamma\left(\frac{\alpha_1}{2}\right)\Gamma\left(\frac{\alpha_2}{2}\right)\Gamma\left(\frac{\alpha_3}{2}\right)}\pi^2 R_{k_1, k_2, k_3}\prod_{i=1}^3 \frac{1}{\sigma_{k_i}} \\
    &\nonumber \qquad \times \delta (\bar z_{12}) \delta (\bar{z}_{23})\frac{t_{12}^{\alpha_3}\,t_{23}^{\alpha_1}\,t_{13}^{\alpha_2}z_{12}^{\frac{k_3}{2}-2}z_{23}^{\frac{k_1}{2}-2}z_{13}^{\frac{k_2}{2}-2}}{\left(u_1 z_{23}+u_2 z_{31}+u_3 z_{12}+i \varepsilon \right)^{\frac{k_1+k_2+k_3}{2}-4}} .
\end{align}

\subsection{Four-point functions}

In Section \ref{sec:Four point correlators in ABJM}, we expressed all four-point functions\footnote{For the case of generic $k_i$, we mean the part relevant for the flat space limit $\mathcal{G}_{k_1, k_2, k_3, k_4}^{HE}$.} in terms of $\bar{D}$ functions. We will start by analyzing the Carrollian limits of $\mathcal{G}_{2,2,2,2}$ and $\mathcal{G}_{2,2,k,k}$ before moving on to the generic case. As in the previous section, we will restrict out attention to the leading term in the $\frac{1}{N}$ expansion, treating higher derivative corrections in appendix \eqref{sec:higherdercamp}.

\subsubsection{Carrollian limit of $\mathcal{G}_{2,2,2,2}$}

In order to compute the Carrollian limit, we will use the following formula \cite{Alday:2024yyj}\footnote{This formula was previously derived assuming integer scaling dimensions. We extend this result to non-integer values by analytic continuation.}: 
\begin{align}
    \label{eq:carrlimitmap}
   U^a \, V^b \,  \bar{D}_{\D_1, \D_2, \D_3, \D_4}  \xrightarrow[]{\ell \to \infty}\frac{\ell^{-4+\Sigma_{\D}} \mathcal{K}}{\mathcal{U}^{\Sigma_{\D}-4}} \left(\frac{\left|z_{23}\right|^2}{\left|z_{34}\right|^2 \left|z_{24}\right|^2}\right)^{\frac{4-\Sigma_{\D}}{2}} \frac{\left(1-z\right)^{\D_1+\D_4-2+2b}}{z^{\D_1+\D_2-2-2a}} \delta\left(z-\zb\right)\Theta\left(z\right)\Theta\left(1-z\right) ,
\end{align}
where $z=\frac{z_{12} z_{34}}{z_{13} z_{24}}$ is the is the $2d$ cross-ratio, $\mathcal{U}$ was defined in \eqref{eq:Carrollden} and
\begin{align}
\label{eq:Kdef}
\mathcal{K} =(-1)^{\D_1+\D_3} 2^{\frac{\Sigma_{\D}}{2}} \pi^2 \Gamma\left(\frac{\Sigma_{\D}-4}{2}\right).
\end{align} 
As we explained in Section \eqref{sec:Carrollian amplitudes and Carrollian/flat space limit correspondence}, this formula involves a choice of analytic continuation. In writing \eqref{eq:carrlimitmap}, we have chosen a particular one such that $0<z<1$. The leading terms in the Carrollian limit are those which have $\bar{D}$ functions with the highest weight. Applying the above formula to \eqref{eq:2222sposspaces}, the leading terms are
\begin{align}
    \mathcal{G}_{2,2,2,2,s} &\xrightarrow[]{\ell \to \infty}  \frac{48\pi \,\ell^2}{\sqrt{2N^3}\mathcal{U}^2} \, \left(\frac{\left|z_{34}\right|^2 \left|z_{24}\right|^2}{\left|z_{23}\right|^2}\right)\left(1-z\right)\left(1-\alpha z\right)^2\left(1-\bar{\alpha} z\right)^2 \delta\left(z-\zb\right) \Theta\left(z\right)\Theta\left(1-z\right),
\end{align}
where we have set $\sigma = \alpha \bar{\alpha}, \tau = \left(1-\alpha\right)\left(1-\bar{\alpha}\right)$. Combining the results of the other two channels and and accounting for the pre-factor in \eqref{eq:11114pt} we get
\begin{align}
\label{eq:delta1carrlimit}
    \an{\mo_2 \left(x_1, t_1\right) \dots \mo_2 \left(x_4, t_4\right)} \xrightarrow[]{\ell \to \infty}
    & \frac{3 \ell_{11}^9 \pi}{8\ell^7\mathcal{U}^2} \, \left(\frac{\left|z_{24}\right|^2}{\left|z_{12}\right|^2\left|z_{23}\right|^2}\right)t_{12}^2 \, t_{34}^2\left(1-\alpha z\right)^2 \left(1-\bar{\alpha} z\right)^2  \delta\left(z-\zb\right)\Theta\left(z\right)\Theta\left(1-z\right),
\end{align}
where we have used the relation \eqref{eq:Nlrel} with $k_{CS}=1$. As we will see in Section \ref{sec:Bulk perspective in flat space}, the flat space counterpart of this correlator also vanishes at an identical rate. In order to get a non-zero result as $\ell \to \infty$, we multiply by the volume of $S^7$, $V_7 = \frac{\left(2\ell\right)^7 \pi^4}{3}$:
\begin{align}
\label{eq:clabjm2222}
   \lim_{\ell \to \infty} \frac{V_7}{\left(2\pi\right)^4}& \an{\mo_1 \left(x_1, t_1\right) \dots \mo_1 \left(x_4, t_4\right)} = \an{\Phi_2\left(u_1, z_1, \zb_1\right)\,\dots \Phi_2\left(u_4, z_4, \zb_4\right)}\\
   &=  \frac{\pi \ell_{11}^9}{2\mathcal{U}^2} \, \left(\frac{\left|z_{24}\right|^2}{\left|z_{12}\right|^2\left|z_{23}\right|^2}\right)  t_{12}^2 \, t_{34}^2\left(1-\alpha z\right)^2 \left(1-\bar{\alpha} z\right)^2  \delta\left(z-\zb\right) \Theta\left(z\right)\Theta\left(1-z\right)\nonumber.
\end{align}

\subsubsection{Carrollian limit of $\mathcal{G}_{2,2,k,k}$}
We will compute the Carrollian limit of the correlators $\mathcal{G}_{2,2,k,k}$ by once again applying the formula \eqref{eq:carrlimitmap}. The terms from the $s-$ and $t-$channel contributions that will dominate in the Carrollian limit are
\begin{align}
\label{eq:G22kkls}
    &\mathcal{G}_{2,2,k,k,s} \xrightarrow[]{\ell \to \infty} \frac{-3U}{\sqrt{2N^3}\pi}\frac{\left(1-k\right)}{\Gamma\left(\frac{k}{2}\right)}\left(V\,\bar{D}_{0,2,\frac{k}{2}+1,\frac{k}{2}+1} + \sigma\, U \, V\, \bar{D}_{1,2,\frac{k}{2}+1,\frac{k}{2}} +\tau\, U\, \bar{D}_{1,2,\frac{k}{2},\frac{k}{2}+1}\right) \\
     &\mathcal{G}_{2,2,k,k,t}\xrightarrow[]{\ell \to \infty}\frac{12k\,\tau\, U}{\sqrt{2N^3}}\frac{\Gamma\left(\frac{k}{2}+1\right)}{\Gamma\left(\frac{k}{2}-\frac{1}{2}\right)} 2^{\frac{k}{2}} \, x_{\frac{k}{2}} \left(\bar{D}_{1,2,\frac{k}{2},1+\frac{k}{2}}+ \sigma  \bar{D}_{1,1,1+\frac{k}{2},1+\frac{k}{2}}+ \tau\bar{D}_{1,1,\frac{k}{2},2+\frac{k}{2}}\right), \nonumber
\end{align}
with $x_{\frac{k}{2}}$ being given by \eqref{eq:xkformula}. The dominant term in the $u-$ channel can be simply obtained from the $t-$channel one by using the relation \eqref{eq:u22kk}. Combining all of this, accounting for the prefactor in \eqref{eq:22kk}, the relations in \eqref{eq:operatorexp} and \eqref{eq:Nlrel} and multiplying by the volume of $S^7$ we find, 
\begin{align}
    \label{eq:G22kkcarrlimit}
     \lim_{\ell \to \infty}V_7 \left[\ell^{2-k}\frac{\an{\mo_2\mo_2 \mo_{k} \mo_{k}}}{\sigma_k^2} \right]=& \frac{\pi \ell_{11}^9}{2} (1-\alpha  z)^2 (1-\bar{\alpha} z)^2 (-1)^{\frac{k}{2}-1} \, t_{34}^k t_{12}^2 \\
     & \nonumber \qquad \times\frac{\left|z_{24}\right|^k}{\left|z_{12}\right|^2\left|z_{23}\right|^k} \frac{ \Gamma\left(k\right)(1-z)^{\frac{k}{2}-1} \delta\left(z-\zb\right)}{\mathcal{U}^k}.
\end{align}

\subsubsection{Carrollian limit of $\mathcal{G}_{k_1, k_2, k_3, k_4}$}

Following a similar procedure starting from \eqref{eq:HEgenk}, we arrive at 
 \begin{align}
      \mathcal{G}_{k_1, k_2, k_3, k_4}^{HE} \xrightarrow[]{\ell \to \infty} &\frac{\mathcal{N}_{k_i} \mathcal{P}_{k_i}}{N^{\frac{3}{2}}\mathcal{U}^{\frac{\sum_i k_i}{2}-2}} (-1)^{\frac{k_1+k_3}{2}}z^{-2a_s} \left(1-z\right)^{\frac{\sum_i k_i}{2}-2-2a_t}\pi^{\frac{5}{2}}2^{4-\frac{\sum_i k_i}{4}}\frac{\Gamma\left(\frac{\sum_i k_i}{2}-5\right)\Gamma\left(\frac{\sum_i k_i}{4}-1\right)}{\Gamma\left(\frac{\sum_i k_i}{4}-4\right)\Gamma\left(\frac{\sum_i k_i}{4}-\frac{1}{2}\right)}\nonumber\\
      & \times \left(1-\alpha z\right)^2 \left(1-\bar{\alpha}z\right)^2 \left(\frac{\left|z_{23}\right|^2}{\left|z_{34}\right|^2 \left|z_{24}\right|^2}\right)^{1-\frac{\sum_i k_i}{4}}\delta\left(z-\zb \right)\left(\frac{1}{\ell}\right)^{2-\frac{\sum_i k_i}{2}}.
 \end{align}
 We can use this to compute the Carrollian limit of the correlator \eqref{eq:4ptcorr}. Accounting for all the prefactors, using \eqref{eq:operatorexp} and \eqref{eq:Nlrel}, we get
 \begin{align}
 \label{G2kkkk Carrollian}
     \lim_{\ell \to \infty} V_7\frac{\an{\mo_{k_1} \dots \mo_{k_4}}}{\prod_{i=1}^4 \ell^{\frac{k_i}{2}-1}\sigma_k}= & \tilde{\mathcal{N}}\, \left[\prod_{i<j} \left(\frac{t_{ij}}{\left|z_{ij}\right|}\right)^{\gamma_{ij}^0}\left(\frac{t_{12}t_{34}}{\left|z_{12}\right|\left|z_{34}\right|}\right)^{\mathcal{E}} \left(\frac{\left|z_{23}\right|^2}{\left|z_{34}\right|^2 \left|z_{24}\right|^2}\right)^{1-\frac{\sum_i k_i}{4}}\right] \\
     &\qquad \times \left[z^{-2a_s} \left(1-z\right)^{\frac{\sum_i k_i}{2}-2-2a_t} \frac{\delta\left(z-\zb \right)}{\mathcal{U}^{\frac{\sum_i k_i}{2}-2}}\right] \times \left[\left(1-\alpha z\right)^2 \left(1-\bar{\alpha}z\right)^2 \mathcal{P}_{k_i}\left(\sigma,\tau\right)\right]\nonumber,
 \end{align}
where
\begin{align}
  \tilde{\mathcal{N}} = \frac{V_7\ell_{11}^9}{\left(2\pi\right)^4}\mathcal{N}_{k_i} \pi^{\frac{5}{2}}2^{-\frac{1+\sum_i k_i}{2}}\frac{\Gamma\left(\frac{\sum_i k_i}{2}-5\right)\Gamma\left(\frac{\sum_i k_i}{4}-1\right)}{\Gamma\left(\frac{\sum_i k_i}{4}-4\right)\Gamma\left(\frac{\sum_i k_i}{4}-\frac{1}{2}\right)}(-1)^{\frac{k_1+k_3}{2}} .
\end{align}

\subsubsection{Carrollian limit of superconformal Ward identities}
In this section, we will compute the Carrollian limit of the superconformal Ward identities satisfied by correlators of $\frac{1}{2}$-BPS operators  \cite{Dolan:2004mu}, which for ABJM take the form
\begin{align}
    \label{eq:WIABJM}
    &\left(Z \partial_Z - \frac{\alpha}{2}\partial_{\alpha}\right) \mathcal{G}_{k_1, k_2, k_3, k_4}\left(Z,\Zb, \alpha, \bar{\alpha}\right) \Big|_{\alpha = \frac{1}{Z}} =  \left(\Zb \partial_{\Zb} - \frac{\alpha}{2}\partial_{\alpha}\right) \mathcal{G}_{k_1, k_2, k_3, k_4}\left(Z,\Zb, \alpha, \bar{\alpha}\right)\Big|_{\alpha = \frac{1}{\Zb}} = 0 ,\\
    &\left(Z \partial_Z - \frac{\bar{\alpha}}{2}\partial_{\bar{\alpha}}\right) \mathcal{G}_{k_1, k_2, k_3, k_4}\left(Z,\Zb, \alpha, \bar{\alpha}\right)\Big|_{\bar{\alpha} = \frac{1}{Z}} = \left(\Zb \partial_{\Zb} - \frac{\bar{\alpha}}{2}\partial_{\bar{\alpha}}\right) \mathcal{G}_{k_1, k_2, k_3, k_4}\left(Z,\Zb, \alpha, \bar{\alpha}\right)\Big|_{\bar{\alpha} = \frac{1}{\Zb}} = 0. \nonumber
\end{align}
Since the Carrollian limit is obtained from the leading singularity of the four point function as $Z \to \Zb$, we first expand
\begin{align}
    \mathcal{G}_{k_1, k_2, k_3, k_4}\left(Z, \Zb, \alpha, \bar{\alpha}\right) = \frac{\mathcal{G}^0\left(Z, \alpha, \bar{\alpha}\right)}{\left(Z-\Zb\right)^p} + \mo\left(\frac{1}{\left(Z-\Zb\right)^{p-1}}\right).
\end{align}
Here, $\mathcal{G}^0$ is the expression that eventually turns into the numerator of the Carrollian amplitude. Plugging this into \eqref{eq:WIABJM} and retaining only the leading piece leads to 
\begin{align}
    \label{eq:WIflat}
    \mathcal{G}^0\left(z, \alpha =\frac{1}{z}, \bar{\alpha}\right) = \mathcal{G}^0\left(z, \alpha, \bar{\alpha} =\frac{1}{z}\right) = 0.
\end{align}
This is easily seen to be satisfied by the Carrollian limit of $\mathcal{G}_{2,2,2,2}$ in \eqref{eq:delta1carrlimit}, of $\mathcal{G}_{2,2,k,k}$ in \eqref{eq:G22kkcarrlimit} and of $\mathcal{G}_{k,k,k,k}$ in \eqref{G2kkkk Carrollian}. It would be interesting to find an intrinsically Carrollian derivation of these identities. 

\section{Bulk perspective in flat space}
\label{sec:Bulk perspective in flat space}

In this section we will explain how to obtain the Carrollian ABJM correlators derived in the previous section from a bulk point of view. At two and three points, we will follow the strategy of expanding the supergravity action in AdS$_4 \times$S$^7$ in modes on 7-sphere, truncating the sum over modes, integrating out the 7-sphere to obtain a four-dimensional effective action in AdS$_4$, and taking the flat space limit. The resulting Lagrangian can then be used to derive scattering amplitudes which reproduce the results of the previous section after performing a modified Mellin transform. At four points, we follow a different strategy: starting from the 11d supergravity amplitude in flat space we will take the external kinematics to be four dimensional and the polarisation vectors to point along the other seven directions. After performing a modified Mellin transform, we obtain the lowest-charge ($k=2$) 4-point correlator. We can then obtain higher-charge correlators by conformally compactifying the internal space to a seven-sphere and making an appropriate choice of external states. This approach can also be used to obtain 2 and 3-point Carrollian ABJM correlators, as we explain in Appendix \ref{sec:3ptamps}.   
 
\subsection{Two and three-point amplitudes} 
\label{sec:Two and Three point amplitudes from 11D SUGRA}

In this section, we will provide an interpretation of the Carrollian limit of ABJM correlators in terms of flat space physics. 
First, we will compute the two and three-point amplitudes from the flat limit of the SUGRA action in \eqref{eq:scalaraction} and compare them to those obtained from the Carrollian limit of ABJM. Kaluza-Klein modes $s_k$ whose conformal dimensions scale with the AdS radius become massive in the flat space limit, consistently with $\D \left(\D-3\right) = m^2 \ell^2$. Since the scaling dimension of $s_k$ is $k/2$, where $k$ is the R-charge of the dual CFT operator, we will truncate the sum over KK modes in \eqref{eq:scalaraction} to a finite maximum value $k_{\text{max}}$ in order to ensure that the scalars become massless when we take $\ell \to \infty$. 
On taking the flat limit $\ell \to \infty$ of \eqref{eq:scalaraction} and rescaling the fields such that the kinetic terms are canonical we then get
\begin{align}
    \label{eq:scalaractionflat}
    S = \int_{\mathbb{R}^{3,1}} d^4 x \left\lbrace \sum_{k=2}^{k_{\text{max}}} \frac{t_{12}^{k}}{2}\, s_k \Box  s_k +\sum_{k_1, k_2, k_3=2}^{k_{\text{max}}} \frac{t_{12}^{\alpha_3}t_{23}^{\alpha_1}t_{13}^{\alpha_2}}{2\ell}\left(\frac{\ell_{11}}{2\ell}\right)^{\frac{9}{2}} \frac{\tilde{g}_{123}}{3}\, s_{k_1}\, s_{k_2}\, s_{k_3} 
    \right\rbrace,  
\end{align}
where 
\begin{align}
    \tilde{g}_{123} = \frac{144\sqrt{3} \,2^{\alpha}\left(\alpha^2-9\right)\, \left(\alpha^2-1\right)\, \left(\alpha+2\right)}{\left(2\alpha+6\right)!!\pi^2} \prod_{i=1}^3\frac{\Gamma\left(k_i-1\right)}{\Gamma\left(\alpha_i\right)} \sqrt{(k_i+1)k_i(k_i-1)}.
\end{align} Parametrizing the null momenta as in \eqref{eq:mompar}, the two point amplitude of two massless scalars computed from the action \eqref{eq:scalaractionflat} is
\begin{align}
    \label{eq:2ptfromZ}
    \mathcal{A}_{k_1, k_2} = \frac{t_{12}^{k_1}\delta_{k_1, k_2}}{\om_1} \, \delta_{\e_1, -\e_2}\delta\left(\om_1 - \om_2\right) \delta^2\left(z_{12}\right).
\end{align}
The corresponding Carrollian amplitude is obtained by computing the following modified Mellin transform \eqref{def Carrollian amplitude}:
\begin{align}
    \mathcal{C}_{k_1, k_2}^{\D_1, \D_2} = \int \prod_{j=1}^2 \frac{d\om_j}{2\pi} \om_j^{\D_j-1} e^{i \, u_j \om_j \e_j } \mathcal{A}_{k_1, k_2}\Big|_{\e_1 = -\e_2 = -1} = \frac{\delta_{k_1, k_2}}{\left(2\pi\right)^2}\frac{(-1)^{\D_1-1} \Gamma(\D_1+\D_2-2)}{\left(u_{12}-i \e\right)^{\D_1+\D_2-2} } \,t_{12}^{k_1}\, \delta^2\left(z_{12}\right).
\end{align}
This is in agreement with \eqref{eq:2ptcarrolllim} \emph{only} if we set $\D_k = \frac{k}{2}$, in which case we get
\begin{align}
    \label{eq:2ptcarrollian}
    \mathcal{C}_{k_1, k_2}^{\frac{k_1}{2}, \frac{k_2}{2}} = \frac{\delta_{k_1, k_2}}{\left(2\pi\right)^2}\frac{(-1)^{\frac{k_1}{2}-1} \Gamma(k_1-2)}{\left(u_{12}-i \e\right)^{k_1-2} } \,t_{12}^{k_1}\, \delta^2\left(z_{12}\right).
\end{align}
 It is interesting to note that we have to use the modified Mellin transform with a fixed value of $\Delta_k$ which differs for each operator field $s_k$. The three point amplitude can be read off from the cubic term to be
\begin{align}
    \label{eq:3ptamp}
    \mathcal{A}_{k_1, k_2, k_3} = \frac{\tilde{g}_{123}}{2\ell} \left(\frac{\ell_{11}}{2\ell}\right)^{\frac{9}{2}} t_{12}^{\alpha_3}t_{23}^{\alpha_1}t_{13}^{\alpha_2} \,\delta^{(4)}\left(p_1+p_2+p_3\right).
\end{align}
Note that this amplitude vanishes in the strict $\ell \to \infty$ limit. This is consistent with the behaviour of the 11D three-point graviton amplitude after dimensional reduction to 4D, 
(see Appendix \ref{sec:3ptamps} for more details). Three-point amplitudes are non-trivial only in $(2,2)$ signature. We can obtain the Carrollian amplitude from \eqref{eq:3ptamp} by parametrizing the momentum in $(2,2)$ signature, found by Wick rotating \eqref{eq:mompar}, and applying the modified Mellin transform \eqref{def Carrollian amplitude}:
\begin{align}
    \mathcal{C}_{k_1, k_2, k_3}^{\D_1, \D_2, \D_3} = \int \prod_{j=1}^3 \frac{d\om_j}{2\pi} \om_j^{\D_j-1} e^{i \, u_j \om_j \e_j } \mathcal{A}_{k_1, k_2, k_3}.
\end{align}
This gives 
\begin{multline}
        C_{k_1, k_2, k_3}^{\D_1, \D_2, \D_3} = \frac{-i\e_1\e_2\e_3}{(2\pi)^3} \frac{\tilde{g}_{123}}{2\ell} \left(\frac{\ell_{11}}{2\ell}\right)^{\frac{9}{2}}t_{12}^{\alpha_3}t_{23}^{\alpha_1}t_{13}^{\alpha_2}\left(z_{12}\right)^{\D_1-2}\left(z_{13}\right)^{\D_2-2}\left(z_{23}\right)^{\D_3-2} \delta\left(\zb_{12}\right)\delta\left(\zb_{23}\right)\\
    \times  \Theta\left(-\frac{z_{13}}{z_{23}}\e_1\e_2\right)\Theta\left(\frac{z_{12}}{z_{23}}\e_1\e_3\right) \frac{\Gamma\left(\sum_{i=1}^3 \D_i -4\right)}{\left(z_{23}u_1 + z_{31}u_2+z_{12}u_3 - i \varepsilon \e_1\, \text{sign} (z_{23}) \right)^{\sum_{i=1}^3 \D_i-4}}.
\end{multline} which can be matched with \eqref{eq:carroll3ptabjm} with the exact factor by taking again $\Delta_k = \frac{k}{2}$ and setting $\e_1 = -\e_2 = -\e_3 = 1$. 

\subsection{Four-point amplitudes} 
\label{sec:4ptfrom11DSUGRA}
In the previous section, we first computed the flat limit of the effective action \eqref{eq:scalaraction} for scalar fluctuations around AdS${}_4 \times S^7$ to obtain \eqref{eq:scalaractionflat}. The 2 and 3 point amplitudes in flat space followed directly from the quadratic and cubic terms in it. However, the generalization of \eqref{eq:scalaraction} to the quartic level is not known. We will instead start from the tree-level, 4 point graviton amplitude in 11D $\mathcal{N}=1$ supergravity and make contact with the Carrollian limits \eqref{eq:clabjm2222}, \eqref{eq:G22kkcarrlimit}, \eqref{G2kkkk Carrollian} by evaluating the amplitude in certain special configurations. The amplitude is \cite{Sannan:1986tz, Chowdhury:2019kaq} 
\begin{align}
    A_{4} =  \frac{-a_4\ell_{11}^9 }{stu}&\left( \frac{1}{2} e_2 \cdot e_3 \left( s\, e_1 \cdot P_3 \, e_4 \cdot P_2 + t\, e_1 \cdot P_2 \, e_4 \cdot P_3 \right) + \frac{1}{2} e_1 \cdot e_4\, \left( s \,e_2 \cdot P_4\, e_3 \cdot P_1 + t\, e_2 \cdot P_1 \,e_3 \cdot P_4 \right) \right. \nonumber\\
&\left. + \frac{1}{2} e_2 \cdot e_4 \left( s \,e_1 \cdot P_4\, e_3 \cdot P_2 + u \,e_1 \cdot P_2 \,e_3 \cdot P_4 \right) + \frac{1}{2} e_1 \cdot e_3\left( s\, e_2 \cdot P_3 \,e_4 \cdot P_1 + u \,e_2 \cdot P_1 \,e_4 \cdot P_3 \right)\right.\nonumber \\
&\left. + \frac{1}{2} e_3 \cdot e_4 \left( t\, e_1 \cdot P_4 \,e_2 \cdot P_3 + u\, e_1 \cdot P_3\, e_2 \cdot P_4 \right) + \frac{1}{2} e_1 \cdot e_2 \left( t\, e_3 \cdot P_2 \,e_4 \cdot P_1 + u \,e_3 \cdot P_1 \, e_4 \cdot P_2 \right) \right.\nonumber\\
&\left. -\frac{1}{4} s \, t\, e_1 \cdot e_4 \,e_2 \cdot e_3 - \frac{1}{4} s \, u\, e_1 \cdot e_3\, e_2 \cdot e_4 - \frac{1}{4} t \, u \, e_1 \cdot e_2\, e_3 \cdot e_4 \right)^2\delta^{(11)}\left(\sum_{i=1}^4 P_i\right),
\label{11D graviton}
\end{align}
where $s,t, u$ are the Mandelstam variables ($s+t+u=0$), $e_{\mu \nu, i} = e_{\mu, i} e_{\nu, i}$ are the polarization vectors for the gravitons and $a_4$ is a normalization constant. 

We need to choose the momenta and polarizations in a specific way to make contact with the Carrollian limits of ABJM correaltors. As we will see, this choice leads to divergences which need to be regulated. A natural way of doing this is by introducing a sphere of radius $2\ell$ with $\ell \to \infty$. We explain the various choices involved below. A similar procedure has been utilized in Mellin space\cite{Alday:2021odx, Chester:2018aca}. 

\paragraph{Momenta: } We first pick four directions which will later be identified as arising from the flat limit of AdS${}_4$. The momentum of the particle $i$ decomposes as 
\begin{align}
   P_i^{\alpha} = \left(p_i^{\mu}, \tilde{p}_i^{I}\right), \qquad p_i \in \mathbb{R}^{1,3}, \tilde{p}_i \in \mathbb{R}^7, \qquad p_i^{\mu} \sim \mo\left(1\right),  \tilde{p}_i^I \sim \mo\left(\frac{1}{\ell}\right) \approx 0.
\end{align}
Here $\ell$ is a large parameter with dimensions of length. We expect to land on such a configuration on taking the flat limit of AdS${}_4 \times S^7$ with  $\ell$ being the AdS radius. $\alpha = 0, 1, \dots , 10$,  $\mu = 0, \dots , 3$ and $I = 4, \dots , 10$.
\paragraph{Polarizations: }We are interested in massless scalars in $\mathbb{R}^{1,3}$ arising from dimensional reduction of the 11D graviton. We will set
\begin{align}
    e_i^{\alpha} = \left(0,0,0,0, \xi_i^I\right),
\end{align}
where $\xi_i^2 = 0$ since $e_i^2=0$. Later on we will conformally compactify $\mathbb{R}^{7}$ to S$^7$ and express the latter in terms of 8D embedding coordinates $Z\cdot Z=1$. We can then embed the 7D null vector $\xi_i$ into an 8D vector and identify
\begin{align}
    t_i = \left(0, \xi_i\right).
\end{align}
where $t_i^A$ is the $R$-symmetry null vector. 
This has the property 
\begin{align}
\label{eq:flatspaceconstraints}
e_i \cdot e_j = t_i \cdot t_j \equiv t_{ij}, \qquad e_i \cdot p_j \sim \mo\left(\frac{1}{\ell}\right) \approx 0,
\end{align}

\paragraph{Wavefunctions: }Since the fields on AdS${}_4 \times S^7$ are expanded in spherical harmonics, a natural choice for the wavefunctions of the level $k$ KK mode of the 11D graviton is
\begin{align}
    \label{eq:naivewvfns}
    h_j^{\alpha \beta} \left(X\right)= \mathcal{N}_j \,e_j^\alpha e_j^\beta \, e^{i p_j \cdot x}  \left(\xi_j \cdot \tilde{x}\right)^{k_j-2},
\end{align}
with $X = (x,\tilde{x})\in \mathbb{R}^{1,10}$, $x \in\mathbb{R}^{1,3}$ and $\tilde{x} \in \mathbb{R}^7$. It is easy to see that this wavefunction solves the equations of motion for a free, massless spin-2 field in 11D flat space in de Donder gauge ($\partial^{\alpha} \bar{h}_{\alpha \beta} = 0$), 
\begin{align}
\Box \bar{h}_{\alpha \beta} =0,    \qquad \bar{h}_{\alpha \beta} = h_{\alpha \beta} - \frac{1}{2}\eta_{\alpha \beta} h_\gamma^\gamma,  
\end{align}
since $p_i^2 = \xi_i^2 =0$. However, it is not normalizable and leads to divergences when computing amplitudes as seen simply by computing the inner product
\begin{align}
    \int_{\mathbb{R}^{1,3}} d^4x\int_{\mathbb{R}^7}d^7\tilde{x} \, h_1^{\alpha\beta} h_{2,\alpha\beta} &= \mathcal{N}_1 \mathcal{N}_2 t_{12}^2 \left(2\pi\right)^4 \delta^{(4)} \left(P_1+P_2\right) \int_{\mathbb{R}^7} d^7\tilde{x} \left(\xi_1 \cdot \tilde{x}\right)^{k_1-2}  \left(\xi_2 \cdot \tilde{x}\right)^{k_2-2} \\ 
    &= \mathcal{N}_1 \mathcal{N}_2 t_{12}^{2}\left(2\pi\right)^4 \delta^{(4)} \left(P_1+P_2\right) \times \frac{2\pi^{\frac{7}{2}}t_{12}^{2k_1-2}\delta_{k_1, k_2} }{\Gamma\left(\frac{7}{2}\right)}\int_{0}^{\infty} d\left|\tilde{x}\right| \left|\tilde{x}\right|^{k_1+k_2+2}. \nonumber
\end{align}
We evaluated the integral using the methods in \cite{Lee:1998bxa, Bastianelli:1999vm, Chen:2020ipe} and replaced $\xi_1 \cdot \xi_2$ by $t_{12}$. It is easy to see that such divergences will also occur in the four point function. We will regulate these divergences by replacing the integral over $\mathbb{R}^7$ by an integral over $S^7$ of radius $2\ell$.\footnote{The choice of $2\ell$ for the radius is arbitrary. The exact numerical factor is irrelevant since we only match with the Carrollian limit up to a numerical factor.} Choosing an appropriate normalization and replacing $\xi_i \cdot \tilde{x} \rightarrow t_i \cdot Z$, where $Z \in \mathbb{R}^8$ are embedding coordinates for the sphere and $Z \cdot Z =1$, the wavefunction is
\begin{align}
    \label{eq:k>2wavefn}
      h_j^{\alpha \beta}\left(X\right) = \frac{1}{\sqrt{V_7}}e_i^\alpha e_i^\beta e^{i p_j \cdot x} \frac{\left(t_j \cdot Z\right)^{k_j-2}}{\left(2\ell\right)^{k_j-2}}.
\end{align}  
This wavefunction now solves the free equations of motion on $\mathbb{R}^{1,3} \times$ S$^7$ and is a solution of the free field equations on $\mathbb{R}^{1,3} \times \mathbb{R}^7$ in the limit $\ell \to \infty$. To see this, note that the scalar Laplacian in S$^7$ takes the same form in embedding coordinates as a Laplacian in flat space. With this, we are now in a position to connect the 11D supergravity amplitude with the Carrollian limit of ABJM correlators. It is instructive to understand this connection separately for correlators involving operators with $k=2$ and $k>2$. 
\subsubsection{Amplitudes of $k=2$ KK modes}
The Carrollian limit of $\an{\mo_2 \dots \mo_2}$ in \eqref{eq:clabjm2222} corresponds to the amplitude for in/out states with the wavefunctions
\begin{align}
\label{eq:k=2wavefn}
   h_j^{\alpha \beta}\left(X\right) = \frac{1}{\sqrt{V_7}}e_i^\alpha e_i^\beta e^{i p_j \cdot x},
\end{align} 
with $p_j$ being a null momentum parametrized by \eqref{eq:mompar}. The dimensional reduction can be carried out by plugging in \eqref{eq:flatspaceconstraints}. In addition to this, since the wavefunction in \eqref{eq:k=2wavefn} does not involve plane waves in the $\tilde{x}^I$ directions, the amplitude for these states does not produce $\delta^{(11)} \left(\sum_{i=1}^4 P_i\right)$. We should replace the $\delta$ function by
\begin{align}
    \delta^{(11)} \left(\sum_{i=1}^4 P_i\right) \to \frac{1}{\left(2\pi\right)^7 V_7^2}\int_{\mathbb{R}^{3,1}} \frac{d^4x}{\left(2\pi\right)^4} \,e^{i \sum_{j=1}^4 p_j \cdot x} \int_{S^7} d^8 Z \, \delta\left(Z \cdot Z -4\ell^2\right) = \frac{\delta^{(4)}\left(\sum_{i=1}^4  p_i \right)}{\left(2\pi\right)^7 V_7}
\end{align}
Putting all of this together, we get
 \begin{align}
       \mA^{2,2,2,2}_{4} = \frac{-a_4\ell_{11}^9 t^2_{12}t^2_{34}}{\left(2\pi\right)^7\, V_7  stu}\left(s\, t\,\tau + s\, u \sigma + t \,u \right)^2 \times \delta^{(4)}\left(\sum_{i=1}^4 p_i\right).
 \end{align}
This amplitude can also be derived from the 4D $\mathcal{N}=8$ supergravity amplitude. We refer the reader to Appendix C of \cite{Chester:2018aca} for this connection. Note that this amplitude vanishes as $\ell \to \infty$ as mentioned in Section \ref{sec:Carrollian limit of ABJM correlators}. The Carrollian amplitude corresponding to this can be obtained simply via a Fourier transform \eqref{eq:deltaonecarramp}.  Setting $\e_1 = -\e_2 = \e_3 = -\e_4 = 1$, we get
\begin{align}
   V_7\, \mathcal{C}_4^{{1, \dots , 1}} \Big(\{ u_j, z_j, \bar{z}_j \}^{\epsilon_j} \Big)  = \frac{a_4 \ell_{11}^9}{2\left(2\pi\right)^{11}\mathcal{U}^2}
 \left(\frac{\left|z_{24}\right|^2}{\left|z_{12}\right|^2\left|z_{23}\right|^2}\right)  t^2_{12}t^2_{34}
\left(1-\alpha z\right)^2\left(1-\bar{\alpha}z\right)^2\delta\left(z-\zb\right)  \Theta\left(z\right)\Theta\left(1-z\right), 
\end{align}
where $\mathcal{U}$ was defined in \eqref{eq:Carrollden}. This agrees with \eqref{eq:clabjm2222} up to a normalization. We only find this agreement if we choose to perform the Fourier transform or equivalently, the modified Mellin transform \eqref{def Carrollian amplitude} with $\D_i=1$.   

\subsubsection{Amplitudes of $k>2$ KK modes}
We can also make contact with the Carrollian limit of ABJM correlators involving operators with $k>2$ by dimensionally reducing 11D supergravity amplitude using the wavefunction \eqref{eq:k>2wavefn} which  implies that $\delta^{(11)}\left(\sum_i P_i\right)$ is replaced by  
 \begin{align}
        \delta^{(11)} \left(P_1+P_2+P_3 + P_4\right) \longrightarrow\, & \delta^{(4)} \left(\sum_{k=1}^4 p_k\right) \,\frac{1}{(2\pi)^7 V_7^2}\int d^8Z \delta\left(Z \cdot Z-4\ell^2\right) \prod_{j=1}^4 \frac{\left(t_j \cdot Z\right)^{k_j-2}}{\left(2\ell\right)^{k_j-2}} \\
\nonumber &= \frac{\tilde{N}_{k_i}}{V_7}\prod_{i<j} t_{ij}^{\gamma_{ij}^0} \left(t_{12}t_{34}\right)^{\mathcal{E}-2}\, \mathcal{P}_{k_i}\left(\sigma, \tau\right) \delta^{(4)} \left(\sum_{j=1}^4 p_j\right),
\label{conserving delta}
\end{align} 
where $\mathcal{P}_{k_i}\left(\sigma, \tau\right) $ is a polynomial defined in \eqref{eq:polyk} which depends on  ratios of polarization vectors which are equal to the $R$-symmetry cross ratios $\sigma, \tau$ due to \eqref{eq:flatspaceconstraints}, 
 \begin{align}
     \frac{\e_1\cdot \e_3 \, \e_2 \cdot \e_4}{ \e_1 \cdot \e_2 \, \e_3 \cdot \e_4} = \frac{t_{13}t_{24}}{t_{12}t_{34}} \equiv \sigma, \qquad \frac{\e_2\cdot \e_3 \, \e_1 \cdot \e_4}{ \e_1 \cdot \e_2 \, \e_3 \cdot \e_4} = \frac{t_{23}t_{14}}{t_{12}t_{34}} \equiv \tau.
 \end{align}
 Implementing these changes in the 11D graviton amplitude \eqref{11D graviton}, we get for the 4D amplitude of higher KK modes ($k_i >2$),
\begin{align}
    \mathcal{A}_{4}^{k_1,k_2,k_3,k_4} = -\frac{\tilde{N}_{k_i}}{V_7}\prod_{i<j} t_{ij}^{-\gamma_{ij}^0}\frac{\left(t_{12}t_{34}\right)^{2-\mathcal{E}}}{4\ell_{11}^9 s\, t\, u}\left(t \, u + s \, u\, \sigma + s \, t \, \tau\right)^2 \mathcal{P}_{k_i}\left(\sigma, \tau\right)
\end{align}
From this, we can compute the Carrollian amplitude using the modified Mellin transform \eqref{def Carrollian amplitude}. Setting $\e_1 = -\e_2 = \e_3 = -\e_4 = 1$ and $\D_i = \frac{k_i}{2}$ gives
\begin{align}
\label{eq:Carrampdef}
     &\mathcal{C}_{k_1,k_2,k_3,k_4}^{\frac{k_1}{2},\frac{k_2}{2},\frac{k_3}{2},\frac{k_4}{2}} = \int_0^{+\infty} \prod_{j=1}^4 \frac{d\om_j}{2\pi} \, (-i \epsilon_j \omega_j)^{\frac{k_j}{2}-1} e^{-i\e_j \om_j u_j}  \mA_4^{k_1, k_2, k_3, k_4}\\
\nonumber =& -\frac{\tilde{N}_{k_i} (-1)^{\frac{k_1+k_3}{2}} i^{\sum_{i=1}^4 \frac{k_i}{2}}\Gamma\left(-2+\sum_{i=1}^4 \frac{k_i}{2}\right)  }{V_74\ell_{11}^9 \left(2\pi\right)^4}\left[\left(t_{12}t_{34}\right)^{2-\mathcal{E}}\prod_{i<j} t_{ij}^{-\gamma_{ij}^0}\left(1-\alpha z\right)^2\left(1-\bar{\alpha}z\right)^2 \right] \nonumber \\
  &\times  \left[
  \frac{\left|z_{14}\right|^{k_3-2}\left|z_{24}\right|^{k_1+2}\left|z_{34}\right|^{k_2-4}}{\left|z_{12}\right|^{k_1+2}\left|z_{13}\right|^{k_3-4}\left|z_{23}\right|^{k_2}}\right] \times \left[\frac{\delta\left(z-\zb\right)  \Theta\left(z\right)\Theta\left(1-z\right)z^{\frac{k_1-k_2+4}{2}}  \left(1-z\right)^{\frac{k_2-k_3}{2}}}{\mathcal{U}^{-2+\sum_{i=1}^4 \frac{k_i}{2}}}\right]\nonumber,
\end{align} 
which agrees with \eqref{G2kkkk Carrollian} up to an overall $k$ dependant normalization factor.

\section{Super conformal Carrollian correlators}
\label{sec:Super conformal Carrollian correlators}

In the previous sections, we obtained position space correlators at null infinity, which are interpreted as scalar correlators in a Carrollian ABJM theory. In this section, we discuss some basic kinematic properties of this theory by $(i)$ deriving the superconformal Carrollian algebra, $(ii)$ defining super conformal Carrollian primaries, $(iii)$ relating the correlators of these operators with the above position space correlators at $\mathscr{I}$.

\subsection{Superconformal Carrollian algebra}
\label{sec:Superconformal Carrollian algebra}

The Carrollian limit of the superconformal algebra has been studied in \cite{Bagchi:2022owq,Zheng:2025cuw}. In this section, we revisit this discussion by keeping $\mathcal{N}$ arbitrary and carefully treating the Majorana reality conditions for $d=3$. We start from the superconformal algebra and follow the conventions of \cite{Park:1999cw}. 

The bosonic generators are given by the Lorentz transformations $J_{\mu\nu} = J_{[\mu\nu]}$ ($J_{ij}$ are the spatial rotations and $B_i = J_{0i}$ the boosts), the translations $P_\mu = (-H, P_i)$, the dilation $D$, and the special conformal transformations $K_\mu = (-K, K_i)$. They form the standard conformal algebra $\mathfrak{so}(3,2)$. Furthermore, the fermionic generators $Q^I_\alpha$ and $S^I_{\alpha}$ ($I= 1, \ldots , \mathcal{N}$, $\alpha = 1,2$) satisfy the anticommutation relations
\begin{equation}
\begin{split} \label{fermionic anticomm}
    &\{Q^{I\alpha}, \bar{Q}_{J\beta} \} = 2 {\delta^I}_J {\gamma^{\mu \alpha}}_\beta P_\mu , \quad \{S^{I\alpha}, \bar{S}_{J\beta} \} = 2 {\delta^I}_J {\gamma^{\mu \alpha}}_\beta K_\mu , \\
    &\{ Q^{\alpha I} , \bar{S}_{\beta J}  \} = -i {\delta^{I}}_J ( 2 {\delta^\alpha}_\beta  D + {(\gamma^{[\mu} \gamma^{\nu] })^\alpha}_{\beta} M_{\mu\nu} ) + 2 i  {\delta^\alpha}_{\beta} {R^I}_J 
\end{split}
\end{equation} where we defined the Majorana conjugation 
\begin{equation} \label{Majo}
    \bar{Q}_{J} = ({Q}^J)^\dagger \gamma^0 = - (Q^J)^T \epsilon , \quad \bar{S}_{J} = ({S}^J)^\dagger \gamma^0 =- (S^J)^T \epsilon .
\end{equation} Here $\epsilon = (\epsilon_{\alpha\beta}) = (\epsilon_{[\alpha\beta]})$ with $\epsilon_{01} =1$ is the charge conjugation matrix. The $2 \times 2$ matrices $\gamma^\mu$, $\mu = 0,1,2$, are given by 
\begin{equation}
    \gamma^0 = \sigma^3, \quad \gamma^1 = i \sigma^1, \quad \gamma^2 = i \sigma^2
\end{equation} with $\sigma^1,\sigma^2,\sigma^3$ the Pauli matrices, and satisfy the Clifford algebra $\{ \gamma^\mu, \gamma^\nu  \} = 2 \eta^{\mu\nu}$. Thus, there are $2\mathcal{N}$ independent fermionic generators. Finally, the $R$-symmetry generators $R_{IJ} = R_{[IJ]}$ form an $\mathfrak{so}(\mathcal{N})$ algebra,
\begin{equation} \label{R symmetry comm}
    [R_{IJ} , R_{KL}] = i (\delta^{IK} R^{JL} + \delta^{JL} R^{IK} - \delta^{IL} R^{JK} - \delta^{JK} R^{IL} ) .
\end{equation} They commute with the bosonic generators, and rotate the fermionic generators
\begin{equation} \label{R Q symmetry}
    [R_{IJ}, Q^{K}   ] = i ({\delta_I}^K \delta_{JD} - {\delta_J}^K \delta_{ID} )  Q^{D} , \qquad [R_{IJ}, S^{K}   ] = i ({\delta_I}^K \delta_{JD} - {\delta_J}^K \delta_{ID} )  S^{D} .
\end{equation} All the above generators constitute the superconformal algebra, $\mathfrak{osp}(\mathcal{N} | 4 ,\mathbb{R})$.

We now implement the Carrollian limit of this algebra, corresponding to an İnönü-Wigner contraction. We start with the bosonic sector. We rescale the generators
\begin{align}
    H \to  \frac{1}{c} H , \quad B_i \to \frac{1}{c} B_i, \quad  K \to \frac{1}{c} K
\end{align} and keep the other bosonic generators untouched. Taking $c \to 0$, the $\mathfrak{so}(3,2)$ algebra contracts into the global conformal Carrollian algebra $\mathfrak{CCarr}^{\text{glob}}_3$. This algebra admits an infinite-dimensional enhancement with supertranslations (and possibly superrotations), leading to the conformal Carrollian algebra, $\mathfrak{CCarr}_3 \simeq \mathfrak{bms}_4$. In this work, we focus on the finite-dimensional global subalgebra. Possible extensions of the above contractions to the fermionic sector have been discussed in \cite{Bagchi:2022owq}. Here, we consider the symmetric (or ``democratic") rescaling, 
\begin{equation}
    Q^{I\alpha} \to \frac{1}{\sqrt{c}}  Q^{I\alpha}, \quad S^{I\alpha} \to \frac{1}{\sqrt{c}} S^{I\alpha} .
\label{contraction fermionic}
\end{equation} Furthermore, we do not rescale the $R$-symmetry generators, $R\to R$, to keep a non-trivial $R$-symmetry algebra in the limit. Taking the $c\to 0$ limit on \eqref{fermionic anticomm}, \eqref{R symmetry comm} and \eqref{R Q symmetry}, we get \begin{align}
        &\{ Q^{I\alpha} , \bar Q_{J\beta} \} = - 2 {\delta^I}_J {\gamma^{0\alpha}}_\beta H , \qquad \{ S^{I\alpha} , \bar S_{J\beta} \} = - 2 {\delta^I}_J {\gamma^{0\alpha}}_\beta K , \label{eq75}\\
        &\{ Q^{\alpha I} , \bar{S}_{\beta J}  \} = - 2 i {\delta^{I}}_J    {(\gamma^{[0} \gamma^{i] })^\alpha}_{\beta} B_i  , \label{eq76}\\
         &[R_{IJ} , R_{KL}] = i (\delta^{IK} R^{JL} + \delta^{JL} R^{IK} - \delta^{IL} R^{JK} - \delta^{JK} R^{IL} ), \label{soN} \\
          &[R_{IJ}, Q^{K}   ] = i ({\delta_I}^K \delta_{JD} - {\delta_J}^K \delta_{ID} )  Q^{D} , \qquad [R_{IJ}, S^{K}   ] = i ({\delta_I}^K \delta_{JD} - {\delta_J}^K \delta_{ID} )  S^{D} . \label{rotationsR}
\end{align}  
This defines the global superconformal Carrollian algebra in $d=3$, $\mathfrak{sCCarr}_3^{\text{glob},\mathcal{N}}$.  Analogously to the bosonic case, this algebra admits an infinite-dimensional enhancement with both bosonic and fermionic supertranslations, $\mathfrak{sCCarr}_3^{\mathcal{N}}$ \cite{Bagchi:2022owq}, and also with (bosonic) superrotations \cite{Barnich:2009se,Barnich:2011ct,Barnich:2010eb}. Here we focus on the finite-dimensional global subalgebra.  

We now show that $\mathfrak{sCCarr}_3^{\text{glob},\mathcal{N}}$ is isomorphic to the $\mathcal{N}$-extended super-Poincar\'e algebra in four dimensions, $\mathfrak{spoin}(\mathcal{N},4)$. To show that, let us recall the isomorphism between the the global conformal Carrollian algebra in three dimensions and the Poincar\'e algebra in four dimensions, 
\begin{equation}
\mathfrak{CCarr}_3^{\text{glob}} \simeq \mathfrak{iso}(3,1)
\label{bosonic isomorphism}
\end{equation} (see e.g. Appendix B of \cite{Donnay:2022wvx} or Section 3 of \cite{Nguyen:2023vfz}). Hence, the bosonic sector of $\mathfrak{sCCarr}_3^{\text{glob},\mathcal{N}}$ is already taken care of. For the fermionic sector, the $2\mathcal{N}$ bulk supersymmetry generators satisfy the algebra
\begin{equation}
    \{ \boldsymbol{\mathcal{Q}}^I_\alpha, \bar{\boldsymbol{\mathcal{Q}}}^J_{\dot\alpha}  \} = 2 \delta^{IJ} \sigma^{\mu}_{\alpha\dot{\alpha}} \boldsymbol{\mathcal{P}}_{\mu}
\label{susy 4d}
\end{equation} where $\sigma^\mu = ( \mathbb{I} , \sigma^i)$, $\bar{\boldsymbol{\mathcal{Q}}}^I = (\boldsymbol{\mathcal{Q}}^I)^\dagger$ and $\boldsymbol{\mathcal{P}}_{\mu}$ the four-dimensional translation generator. Matching the $R$-symmetry structure between the two algebras is non-trivial an deserves further comments. The $R$-symmetry of $\mathfrak{spoin}(\mathcal{N},4)$ is typically $\mathfrak{u}(\mathcal{N})$ or $\mathfrak{su}(\mathcal{N})$ with the supercharges transforming in the fundamental representation. There appears to be a mismatch with the $\mathfrak{so}(\mathcal{N})$ $R$-symmetry of $\mathfrak{sCCarr}_3^{\text{glob},\mathcal{N}}$ induced from the $c \to 0$ limit, and we do not have an obvious isomorphism. It would be interesting to investigate whether the $R$-symmetry at the boundary is enhanced beyond the naive $\mathfrak{so}(\mathcal{N})$ to $\mathfrak{su}(\mathcal{N})$ in holographic theories. Here, in order to make contact with the algebra at the boundary, we simply project onto $\mathfrak{so}(\mathcal{N})$. This projection is similar to the one done at the amplitude level in Appendix C of \cite{Chester:2018aca}. One can then show that \eqref{susy 4d} reproduces \eqref{eq75} and \eqref{eq76} by performing the identifications 
\begin{equation}
    \boldsymbol{\mathcal{P}}_0 = - \frac{1}{2}( H+K), \quad  ,  \boldsymbol{\mathcal{P}}_1 = - B_1, \quad \boldsymbol{\mathcal{P}}_2 = - B_2, \quad \boldsymbol{\mathcal{P}}_3 = \frac{1}{2}(H- K)
\end{equation} together with
\begin{equation}
\boldsymbol{\mathcal{Q}}^I_1 = S^{I1} = -\bar{S}_{I2} , \quad  \bar{\boldsymbol{\mathcal{Q}}}^I_{\dot{1}} = \bar{S}^{I1} = {S}^{I2}, \quad \boldsymbol{\mathcal{Q}}^I_2 =  \bar{Q}_{I1} = Q^{I2}, \quad 
\bar{\boldsymbol{\mathcal{Q}}}_{\dot{2}}= Q^{I1} = - \bar{Q}_{I2} \label{fermionic identif}
\end{equation} where in the second equalities, we used the Majorana reality conditions \eqref{Majo}. Upon the above mentioned projection of the $R$-symmetry representation from $\mathfrak{su}(\mathcal{N})$ to $\mathfrak{so}(\mathcal{N})$, the $R$-symmetry generators of the two algebras can simply be identified as $\boldsymbol{\mathcal{R}}_{IJ} \equiv R_{IJ}$, ensuring that they satisfy the $\mathfrak{so}(\mathcal{N})$ algebra \eqref{soN}. It is then straightforward to show that 
\begin{equation}  [\boldsymbol{\mathcal{R}}_{IJ}, \boldsymbol{\mathcal{Q}}^{K}   ] = i ({\delta_I}^K \delta_{JD} - {\delta_J}^K \delta_{ID} )  \boldsymbol{\mathcal{Q}}^{D} , \qquad [\boldsymbol{\mathcal{R}}_{IJ}, \bar{\boldsymbol{\mathcal{Q}}}^{K}   ] = i ({\delta_I}^K \delta_{JD} - {\delta_J}^K \delta_{ID} )  \bar{\boldsymbol{\mathcal{Q}}}^{D} , 
\end{equation} together with \eqref{fermionic identif} reproduce correctly \eqref{rotationsR}. Therefore, upon the $R$-symmetry projection $\mathfrak{su}(\mathcal{N}) \to \mathfrak{so}(\mathcal{N})$ in the right-hand side, we have established the important isomorphism
\begin{equation}
\mathfrak{sCCarr}_3^{\text{glob},\mathcal{N}} \simeq \mathfrak{spoin}(\mathcal{N},4)
\end{equation} generalizing the isomorphism \eqref{bosonic isomorphism} to the supersymmetric case. This matching of supersymmetries between the four-dimensional bulk and  the three-dimensional boundary constitutes a strong hint towards Carrollian holography. Again, this isomorphism can be lifted to the infinite-dimensional algebras, where $\mathfrak{sCarr}_3^{\mathcal{N}}$ is isomorphic to the super BMS algebra discussed in \cite{Fotopoulos:2020bqj,Henneaux:2020ekh,Fuentealba:2021xhn,Bagchi:2022owq} for $\mathcal{N}=1$.

\subsection{Superconformal Carrollian primaries and correlators}

Massless flat space amplitudes have been shown to be encoded in terms of boundary correlators of Carrollian CFT primaries at null infinity  (we refer to \cite{Donnay:2022aba,Bagchi:2022emh,Donnay:2022wvx,Salzer:2023jqv,Saha:2023abr,Nguyen:2023vfz,Nguyen:2023miw,Mason:2023mti,Bagchi:2023cen,Liu:2024nfc,Stieberger:2024shv,Adamo:2024mqn,Alday:2024yyj,Ruzziconi:2024zkr,Jorstad:2024yzm,Ruzziconi:2024kzo} for recent developments). In this section and the next, we extend this statement to supersymmetric correlators. Conformal Carrollian primaries have been defined in \cite{Bagchi:2016bcd,Donnay:2022aba,Nguyen:2023vfz,Saha:2023hsl}. This definition is naturally found by taking the Carrollian limit of the definition of a conformal primary in CFT and rescaling the operators consistently (see \eqref{eq:operatorexp}). Here we focus on singlets of scalar primaries, which are relevant to encode correlators of bulk scalar fields considered in the previous section. They are defined through the action of the subalgebra of operators preserving the origin: 
\begin{equation}
\begin{split} \label{bosonic def conf}
&[J_{ij}, \phi_\Delta (0)] = 0, \quad [B_i, \phi_\Delta (0)] = 0 , \\
&[D, \phi_\Delta (0)]  = - i \Delta \phi_{\Delta} (0), \quad  [K, \phi_{\Delta} (0)] = 0 , \quad [K_i, \phi_{\Delta} (0)] = 0
\end{split}
\end{equation} where $\Delta$ is the conformal dimension. We can extend this definition to superconformal Carrollian primaries by replacing the two last conditions in \eqref{bosonic def conf} by
\begin{equation}
    [S^{I \alpha} , \phi_\Delta (0) ] = 0 =[\bar{S}_{I \alpha} , \phi_\Delta (0) ] 
\end{equation} where second equality automatically follows from the first one via the Marjorana condition \eqref{Majo}. They transform in spin-$s$ $\mathfrak{so}(\mathcal{N})$ representations of the $R$-symmetry algebra:
\begin{equation}
    [R_{IJ}, \phi_{\Delta}(0)] = \mathcal{R}^{(s)} \cdot \phi_{\Delta} (0) .
\end{equation} Analogously to \eqref{eq:contractedop}, we can contract the $R$-symmetry indices with null vectors $t_I$ to obtain $R$-symmetry scalars 
\begin{equation}
    \phi_k   = \phi^{I_1 \ldots I_k} t_{I_1} \ldots t_{I_k}
\end{equation}
and, for the case of interest arising from the Carrollian limit of ABJM, we will have $\Delta_k = \frac{k}{2}$.

It is convenient to introduce fermionic coordinates $\theta^{I\alpha}$ and $\bar{\theta}_{I\alpha}$ satisfying the Majorana reality condition \eqref{Majo}, i.e. $\bar{\theta}_{I\alpha} = -\epsilon (\theta^{I})^T$. Superconformal Carrollian primaries can be seen as fields on the superspace, $\phi_\Delta (u,z,\bar z, \theta^{I\alpha})$. Using the translation operator on the superspace, 
\begin{equation}
     \phi_\Delta (x, {\theta}) = U  \phi_\Delta (0,0) U^{-1} , \qquad U = e^{-i ( - H u + P_i x^i + \bar{Q}_{I \alpha} {\theta}^{I\alpha}) }
\end{equation}
and following similar steps than those presented in \cite{Bagchi:2022owq} (but keeping $\mathcal{N}$ general), one can deduce the action of all the super conformal Carrollian operators on the fields at any point of the superspace $(x,\theta) = (u,z,\bar z, \theta^{I\alpha})$. Denoting the infinitesimal variation of the field as the commutator
\begin{align}
    &\delta {\phi}_\Delta (x, \theta)\\
    &=    i \Big[a H +  b^j B_j  +  k K  +  a^j P_j + \frac{1}{2}  r^{jk} J_{jk} + \lambda D +  k^j K_j + \epsilon^I \bar{Q}_I + \kappa^I \bar{S}_I + 
 \frac{1}{2}\omega^{IJ} R_{IJ},\phi_\Delta (x, \theta)  \Big] \nonumber
\end{align} where $a$, $b^j$, $k$, $a^j$, $r^{jk} = r^{[jk]}$, $\lambda$, $k^j$, $\epsilon^I$, $\kappa^I$, and $\omega^{IJ} = \omega^{[IJ]}$ are the transformation parameters, the superconformal Carrollian Ward identities read as
\begin{equation}
    \sum_{j = 1}^n \langle \phi_{\Delta_1} (x_1, \theta^I_1)  \ldots 
 \delta \phi_{\Delta_j} (x_j, \theta^I_j)   \ldots   \phi_{\Delta_n}  (x_n, \theta^I_n)   \rangle = 0 . 
 \label{SCCWI}
\end{equation}
These Ward identities are associated with the global superconformal Carrollian algebra $\mathfrak{sCCarr}_{3}^{\text{glob},\mathcal{N}}$. One could also write Ward identities for the infinite-dimensional algebra $\mathfrak{sCCarr}_3^\mathcal{N}$. The additional constraints on the correlators would be related to soft physics in the bulk (the fermionic supertranslation Ward identities have been shown to be equivalent to the leading soft gravitino theorem in the bulk \cite{Fotopoulos:2020bqj}). 

The superconformal Carrollian primary encode the information of a superconformal multiplet. One could expand it in terms of the fermionic coordinates as follows: 
\begin{equation}
    \phi_{\Delta}(x^a,\theta^{I\alpha}) = \Phi_\Delta (x^a) + \theta^{I\alpha} \bar{\Psi}_{I\alpha}(x^a) + \ldots
\end{equation} Each component $\Phi_\Delta$, $\bar{\Psi}_{I\alpha}$, $\ldots$ is a standard conformal Carrollian primary with a certain spin. In this paper, we were interested by the scalar components $ \phi_{\Delta}(x^a,0) = \Phi_\Delta (x^a)$ and their associated correlators $\langle \Phi_{\Delta_1} (x_1) \ldots  \Phi_{\Delta_n} (x_n)  \rangle$. These are the type of correlators found in the Carrollian limit of holographic correlators in Section \ref{sec:Carrollian limit of ABJM correlators}, or by modified Mellin transform in Section \ref{sec:Bulk perspective in flat space}. Indeed, this discussion provides an intrinsic Carrollian CFT definition for what the operators appearing in the left-hand side of the following integral transform are: 
\begin{align}
     &\langle \Phi_{k_1} (x_1) \ldots  \Phi_{k_n} (x_n) \rangle  = \int_0^{+\infty} \prod_{j=1}^n \frac{d\om_j}{2\pi} \, (-i \epsilon_j \omega_k)^{\frac{k_j}{2}-1} e^{-i\e_j \om_j u_j}  \mA_n^{k_1, \ldots, k_n}. 
\end{align} In the right-hand side, $k_1, \ldots , k_n$ label the scalar KK modes. As discussed in the previous sections, the conformal dimension is completely fixed for each operator: $\Delta_i = \frac{k_i}{2}$.

\section{Conclusion}
\label{sec:Conclusion}

In this paper we have taken the first step towards the ambitious goal of deriving a flat space Carrollian hologram from a canonical example of AdS/CFT, which relates the ABJM theory to M-theory in AdS$_4 \times$ S$^7$. Our strategy was to take the $c\rightarrow 0$ limit of ABJM correlators of protected operators and match them against the flat space limit of bulk supergravity calculations after integrating out the 7-sphere. Crucially, when doing so we worked at fixed KK mode number (corresponding to operators of fixed $R$-charge in the dual CFT), yielding four-dimensional bulk scattering amplitudes  of $\mathcal{N}=8$ supergravity, dual to three dimensional Carrollian correlators. We also showed that the superconformal algebra of ABJM, $\mathfrak{osp}\left(4|8\right)$ contracted to a subalgebra of super Poincar{\'e} algebra of $\mathcal{N}=8$ supergravity.

This paper opens up a number of new directions worth pursuing. Perhaps the most pressing of all is that it is still unclear how to obtain the scattering amplitudes of this paper from a 3D Carrollian boundary theory from first principles. As a first step, we may take the $c\rightarrow 0$ limit of the ABJM theory, but 
various conceptual difficulties arise. As shown in \cite{Bagchi:2024efs}, the Carrollian limit of 3D Chern-Simons matter theories contain kinetic terms of the form $(\partial_u\phi)^2$, where we restrict to scalar fields for simplicity. The resulting propagators are therefore of the form $u\, \delta^2(z)$, leading to a proliferation of delta functions which are incompatible with the structure of Carrollian correlators obtained by performing a modified Mellin transform of 4D supergravity amplitudes. A related problem is that the Carrollian correlators in \eqref{eq:carroll3ptabjm} and \eqref{G2kkkk Carrollian} have non-local poles and branch cuts in the $u$-variables whose origin from Carrollian Feynman rules is completely unclear. We can simultaneously resolve these two issues by uplifting the Carrollian propagator to a 3D Lorentz-invariant propagator $1/(-c^2 u^2+2 z \bar{z})$ and carefully taking $c\rightarrow 0$, effectively treating $c$ as a regulator. On the other hand if we restore the $c$-dependence of the propagators, nothing prevents us from doing so for the interaction vertices. Another motivation for restoring $c$-dependence is to note that in the $c\rightarrow 0$ limit, Chern-Simons matter theories have infinite dimensional Carrollian conformal symmetry (or BMS$_4$ symmetry), but this symmetry must be broken to the global Carroll group in order to probe bulk physics beyond universal soft limits \cite{Bagchi:2024gnn}. The question would then be how to do this in a minimal way such that the resulting theory encodes 4D scattering amplitudes in flat space without the additional baggage of an infinite series of curvature corrections. This naturally raises the question of whether Carrollian theories can be defined without resorting to a limiting procedure. Investigating the Carrollian analogues of conformal blocks and crossing symmetry would be an obvious place to start. We hope to address these question more systematically in the future.

Another important question is how to think about the flat space limit. In this paper we have treated AdS and the 7-sphere asymmetrically by holding charges of the dual operators fixed and integrating out the 7-sphere when taking the flat space limit, yielding 4D scattering amplitudes. This was motivated by the desire to understand how holography might work in 4d flat space. On the other hand, the radius of AdS and the 7-sphere cannot be taken to infinity independently, so when we take the flat space limit the bulk theory really becomes 11D flat space, which may be dual to a 10D Carrollian CFT. In principle, this should be visible from the CFT side if we take the charges to infinity at the appropriate rate. Alternatively, we could describe the resulting 11D amplitudes in terms of 4D amplitudes with massive kinematics. Perhaps the nicest context to explore such questions would be in $\mathcal{N}=4$ SYM, whose correlators exhibit a hidden 10D symmetry (at least in the supergravity limit \cite{Caron-Huot:2018kta,
Abl:2020dbx,Aprile:2020mus,Caron-Huot:2021usw}) which allows them to be repackaged into 10D master correlators whose Carrollian limit can in principle then be matched with 10D supergravity amplitudes in the flat space limit.

We hope that this paper sharpens the questions that need to be answered in order to realise the ambitious goal of deriving a concrete example of Carrollian holography from AdS/CFT and provides crucial data that any such proposal must satisfy.

\section*{Acknowledgements}

We thank Luis Fernando Alday, Arjun Bagchi, Paul Heslop, Lionel Mason, Ana-Maria Raclariu, Atul Sharma, David Skinner, and Andrew Strominger for helpful discussions. AL is supported by an STFC Consolidated Grant ST/X000591/1. RR is supported by the Titchmarsh Research Fellowship at the Mathematical Institute and by the Walker Early Career Fellowship at Balliol College.  AYS is supported by the STFC grant ST/X000761/1.

\appendix
\section{Mellin amplitudes in ABJM}
\label{sec:ABJMmellinamp}
In this Appendix, we will summarize the known tree-level Mellin amplitudes in ABJM while providing some of the explicit details needed for the computations in the main body of the paper. 

\subsection{$\mathcal{M}_{2,2,2,2}$}
The best understood case is the one with $k_1 = \dots k_4 = 2$ corresponding to $\D_1 = \dots \D_4 = 1$. The Mellin amplitude admits an expansion in $c_T$ 
\begin{align}
    \label{eq:M2222largect}
    \mathcal{M}_{2,2,2,2} &= \frac{1}{c_T} \mathcal{M}_{2,2,2,2}^{R} + \frac{1}{c_T^{\frac{5}{3}}} B^{R^4} \mathcal{M}_{2,2,2,2}^{\left(4\right)} \\
    &\qquad\nonumber + \frac{1}{c_T^{\frac{7}{3}}}\left( B_4^{D^6 R^4} \mathcal{M}_{2,2,2,2}^{\left(4\right)}  +  B_6^{D^6 R^4} \mathcal{M}_{2,2,2,2}^{\left(6\right)} + B_7^{D^6 R^4}\mathcal{M}_{2,2,2,2}^{\left(7\right)} \right) + \dots .
\end{align}
$\mathcal{M}_{2,2,2,2}^{R} $ is the tree-level supergravity term and is given by \cite{Zhou:2017zaw, Alday:2020dtb}
\begin{align}
    &\mathcal{M}^R_{2,2,2,2}\left(s,t; \sigma, \tau \right) =  \mathcal{M}^R_{2,2,2,2, s} + \mathcal{M}^R_{2,2,2,2, t} + \mathcal{M}^R_{2,2,2,2, u} \nonumber\\
    &\label{eq:Mellin2222s}\mathcal{M}^R_{2,2,2,2, s}\left(s,t; \sigma, \tau \right)  = \sum_{m=0}^{\infty} - \frac{3\left((t-2)(u-2)+(s+2)((t-2)\sigma+(u-2)\tau)\right)}{\sqrt{2\pi} N^{\frac{3}{2}}\Gamma\left(\frac{1}{2}-m\right)^2 \, m! \, \Gamma\left(m+\frac{5}{2}\right)(s-1-2m)}.
\end{align}
The sum over $m$ in \eqref{eq:Mellin2222s} can be performed to get
\begin{align}
    \label{eq:Mellin2222sclosed}
     \mathcal{M}^R_{2,2,2,2, s}\left(s,t; \sigma, \tau \right)  = \frac{3}{\sqrt{8 \pi^3 N^3}} \frac{1}{s(s+2)}&\left[ \left(t-2\right) \left(s+t-2\right)-\sigma \, \left(s+2\right)\left(t-2\right) + \tau \, \left(s+2\right) \left(s+t-2\right) \right] \nonumber\\
     & \times \left[\sqrt{\pi} \left(s+4\right) - 4\frac{\Gamma\left(\frac{1-s}{2}\right)}{\Gamma\left(1-\frac{s}{2}\right)}\right]
\end{align}
The $t,u$ channel Mellin amplitudes can be obtained from the $s$ channel one via
\begin{align}
    \label{eq:Mellintu}
    \mathcal{M}^R_{2,2,2,2, t}\left(s,t; \sigma, \tau \right) = \tau^2 \mathcal{M}^R_{2,2,2,2, s}\left(t, s; \frac{\sigma}{\tau}, \frac{1}{\tau}\right), \quad \mathcal{M}^R_{2,2,2,2, u}\left(s,t; \sigma, \tau \right) = \sigma^2 \mathcal{M}^R_{2,2,2,2, s}\left(u,t;\frac{1}{\sigma}, \frac{\tau}{\sigma} \right).
\end{align}
$\mathcal{M}_{2,2,2,2}^{4}, \mathcal{M}_{2,2,2,2}^{6}, \mathcal{M}_{2,2,2,2}^{7}$ are  correction $\frac{1}{N}$ to the Supergravity approximation. These are discussed in more detail in Appendix \eqref{sec:higherdercamp}.
 large polynomials in $s,t$. 
\subsection{$\mathcal{M}_{2,2,k,k}$}
We will also be interested in the amplitude with $k_1 = k_2 = 2, k_3 = k_4 = k$. This also admits a large $c_T$ expansion but we will only focus on the leading term, 
\begin{align}
    \label{eq:M22kklargect}
    \mathcal{M}_{2,2,k,k} = \frac{1}{c_T}\mathcal{M}_{2,2,k,k}^R.
\end{align}
The leading contributions are presented explicitly in \cite{Alday:2020dtb}.  Performing the sum over $m$, the $s-$ channel Mellin amplitude is
\begin{align}
\label{eq:M22kks}
    \mathcal{M}_{2,2,k,k,s} = \frac{3}{8 \sqrt{2} \pi ^{3/2} N^{3/2} s(s+2)}&\left[(k-2 t+2) (k-2 u+2)-2 (s+2) \sigma  (k-2 t+2)-2 (s+2) \tau  (k-2 u+2)\right]\nonumber\\
    & \times \left(\frac{\sqrt{\pi } (s-k (s+2))}{\Gamma \left(\frac{k}{2}\right)}+\frac{2 k \Gamma
   \left(\frac{1}{2}-\frac{s}{2}\right)}{\Gamma \left(\frac{k-s}{2}\right)}\right)
\end{align}
The $t-$ channel Mellin amplitude is more complicated and is given by
\begin{align}
     \mathcal{M}_{2,2,k,k,t} =& \frac{-3 k \tau  \Gamma \left(\frac{k}{2}+1\right) \left[(k+2 t+2) (k-2 u+2)+2 \sigma  (k-s) (k+2 t+2)-2 \tau  (k-s) (k-2 u+2)\right]}{4 \sqrt{2} \pi  \sqrt{N^3} (k-2 t) \Gamma
   \left(\frac{k-1}{2}\right) \Gamma
   \left(\frac{3+k}{2}\right)}\nonumber \\
   &\qquad \times   _3F_2\left(\frac{1}{2},\frac{1}{2},\frac{k}{4}-\frac{t}{2};\frac{k}{2}+\frac{3}{2},\frac{k}{4}-\frac{t}{2}+1;1\right)
\end{align}
For even $k$, we can replace the hypergeometric function  by (see the Mathematica file supplied with \cite{Alday:2021ymb})
\begin{align}
    _3F_2\left(\frac{1}{2},\frac{1}{2},\frac{k}{4}-\frac{t}{2};\frac{k}{2}+\frac{3}{2},\frac{k}{4}-\frac{t}{2}+1;1\right) =  \frac{\pi \left(k-2t\right) \Gamma\left(\frac{3+k}{2}\right)}{\left(\frac{2-k+2t}{4}\right)_{\frac{2+k}{2}}} \left[\frac{\Gamma \left(\frac{k}{4}-\frac{t}{2}\right)}{4\pi \Gamma \left(\frac{k}{4}+\frac{1}{2}-\frac{t}{2}\right)}- \sum_{i=0}^{\lceil \frac{k-1}{2}\rceil} x_it^i \right],
\end{align}
with the coefficients $x_i$ being determined in the Mathematica file. For our purposes, we will only need the coefficient of the highest power of $t$. For even $k$, since $\lceil\frac{k-1}{2}\rceil = \frac{k}{2}$, this is
\begin{align}
    \label{eq:xkformula}
    x_{\frac{k}{2}} = \frac{\Gamma\left(\frac{1+k}{2}\right)}{4\pi\, 2^{\frac{k}{2}}\Gamma^2\left(1+\frac{k}{2}\right)}
\end{align}
Plugging this into $\mathcal{M}_{2,2,k,k,t}$, we get
\begin{align}
\label{eq:M22kkt}
     \mathcal{M}_{2,2,k,k,t} =& \frac{-3 k \tau  \Gamma \left(\frac{k}{2}+1\right) \left[-(k+2 t+2) (k-2 u+2)+2 \sigma  (k-s) (k+2 t+2)-2 \tau  (k-s) (k-2 u+2)\right]}{4 \sqrt{2} \sqrt{N^3} \Gamma \left(\frac{k-1}{2}\right)  \left[\prod_{n=0}^{\frac{k}{2}} \left(\frac{t}{2}+\frac{k}{4}+\frac{1}{2}-n \right) \right]}\nonumber \\
   &\qquad \times  \left[\frac{\Gamma \left(\frac{k}{4}-\frac{t}{2}\right)}{4\pi \Gamma \left(\frac{k}{4}+\frac{1}{2}-\frac{t}{2}\right)}- \sum_{i=0}^{\lceil \frac{k-1}{2}\rceil} x_it^i \right]
\end{align}
Finally, $\mathcal{M}_{2,2,k,k,u} \left(s,t, \sigma,\tau\right)= \mathcal{M}_{2,2,k,k,t} \left(s,u ,\tau, \sigma\right)$.

\subsection{$\mathcal{M}_{k_1, k_2, k_3, k_4}$}
It is not possible to easily express correlator for general $k_i$ as a sum over a finite number of $\bar{D}$ functions. In this case, we appeal to the equivalence of the Carrollian and high energy limits shown in Appendix[\ref{sec:MCequivalence}] and just present the leading high energy behavior in Mellin space which can be easily converted to position space. From \cite{Alday:2020dtb}, we have
\begin{align}
    \label{eq:MellinHEgenk}
    \lim_{s, t \to \infty} \mathcal{M}_{k_1, k_2, k_3, k_4} = \frac{\mathcal{N}_{k_i}}{N^{\frac{3}{2}}} \frac{\left(t  u + s t \sigma + \tau s u \right)^2}{s t u} \mathcal{P}_{k_i} \left(\sigma, \tau\right) = \frac{\mathcal{N}_{k_i}}{N^{\frac{3}{2}}} \frac{\left(s+t - s\alpha\right)^2\left(s+t - s\bar{\alpha}\right)^2}{s t u} \mathcal{P}_{k_i} \left(\sigma, \tau\right), 
\end{align}
where
\begin{align}
\label{eq:polyk}
    \mathcal{P}_{k_i} \left(\sigma, \tau\right) = \underset{\underset{0\leq i,j,k\leq \mathcal{E}-2}{i+j+k = \mathcal{E}-2}}{\sum} \frac{\left(\mathcal{E}-2\right)! \sigma^i \tau^j}{i!  j! \left(i + \frac{\kappa_u}{2}\right)! \left(j + \frac{\kappa_t}{2}\right)! \left(\frac{\kappa_u}{2}\right)! },
\end{align}
and in the final equality in \eqref{eq:MellinHEgenk}, we have defined new variables $\alpha, \bar{\alpha}$ by 
\begin{align}
    \sigma = \alpha \bar{\alpha}, \qquad \tau= \left(1-\alpha\right) \left(1-\bar{\alpha}\right),
\end{align}
and used the fact that $s+t+u = 0$.

\section{High energy limit versus Carrollian limit}
\label{sec:MCequivalence}
An efficient way of computing the flat space limit starting from the Mellin amplitude is to take its high energy limit ($s, t \to \infty$)\cite{Fitzpatrick:2011hu, Fitzpatrick:2011ia, Penedones:2010ue}\footnote{This procedure results in a flat space amplitude with massless external particles. We will restrict to this case in this paper.}. In this section, we will demonstrate that this is equivalent to taking the Carrollian limit in position space. We will do this by showing that the following procedures yield identical results:
\begin{itemize}
    \item First compute the leading term in the high-energy limit of a generic Mellin amplitude. Convert this to position space using \eqref{eq:Mellingen}.
    \item First compute the position space correlator using \eqref{eq:Mellingen} and then take the Carrollian limit.
\end{itemize}
We will show this equivalence for tree-level contact and exchange diagrams involving external scalars. The internal operator in the case of exchange diagrams could have any spin. The following definitions will come in handy. 
\begin{align}
\label{eq:gen4pt}
    \langle O_1(x_1) \dots O_4(x_4) \rangle = \frac{1}{(x_{12}^2)^{\frac{\Delta_1 + \Delta_2}{2}} (x_{34}^2)^{\frac{\Delta_3 + \Delta_4}{2}}} \left( \frac{x_{14}^2}{x_{24}^2} \right)^a \left( \frac{x_{14}^2}{x_{13}^2} \right)^b \mathcal{G}(U, V),
\end{align}
where
\begin{align*}
    a = \frac{1}{2}(\Delta_2 - \Delta_1), \quad b = \frac{1}{2}(\Delta_3 - \Delta_4),
\end{align*}
and
\begin{align}
\label{eq:Mellingen}
    \mathcal{G}(U, V) = \int_{-i\infty}^{i\infty} \frac{ds \, dt}{(4 \pi i)^2} U^{\frac{s}{2}} V^{\frac{t}{2} - \frac{\Delta_2 + \Delta_3}{2}} \mathcal{M}(s, t) \,\, \Gamma_{\left\lbrace\D_i\right\rbrace},
\end{align}
with 
\begin{align}
\Gamma_{\left\lbrace\D_i\right\rbrace} =&\Gamma\left( \frac{\Delta_1 + \Delta_2 - s}{2} \right) \Gamma\left( \frac{\Delta_3 + \Delta_4 - s}{2} \right) \Gamma\left( \frac{\Delta_1 + \Delta_4 - t}{2} \right) \\
     &\nonumber \Gamma\left( \frac{\Delta_2 + \Delta_3 - t}{2} \right) \Gamma\left( \frac{\Delta_1 + \Delta_3 - u}{2} \right) \Gamma\left( \frac{\Delta_2 + \Delta_4 - u}{2} \right).
\end{align}
\subsection{Contact diagrams}
In Mellin space, contact diagrams are just polynomials in $s,t$. Let
\begin{align}
    \label{eq:genMellincontact}
    \mathcal{M}^c\left(s,t\right) = \sum_{a=a_0,b=b_0}^{a_1, b_1} \chi_{a,b} \, s^a \, t^b.
\end{align}
\paragraph{Position space correlator: }The position space correlator can be written as a sum of $\bar{D}$ functions in a straightforward manner. 
\begin{align}
    \mathcal{G}^c\left(U,V\right) = \sum_{a=a_0,b=b_0}^{a_1, b_1} \chi_{a,b} \left(2U \partial_U\right)^a \left(2V \partial_V + \D_2 + \D_3\right)^b \left(U^{\frac{\D_1+\D_2}{2}} \, \bar{D}_{\D_1, \D_2, \D_3, \D_4} \right)
\end{align}
For the Carrollian limit, we are only interested in the leading singular term as $Z \to \Zb$ which is
\begin{align}
    \label{eq:gencontactls}
     \mathcal{G}^{c, l.s}\left(U,V\right) = (-1)^{a_1+b_1} \, \chi_{a_1, b_1}\, \left(2U\right)^{a_1} \left(2V\right)^{b_1} \, U^{\frac{\D_1+\D_2}{2}} \, \phi^{l.s}_{\D_1+a_1, \D_2+a_1+b_1, \D_3+b_1, \D_4},
\end{align}
where $\phi^{l.s}_{\D_1+a_1, \D_2+a_1+b_1, \D_3+b_1, \D_4}$ is the leading singularity of the $\bar{D}$ function \eqref{eq:lsgen}.

\paragraph{High energy limit of ${\bf \mathcal{M}^c\left(s,t\right)}$:} In the limit $s,t \to \infty$ , $\mathcal{M}^c\left(s,t\right) \xrightarrow[]{s,t \to \infty} \chi_{a_1, b_1} s^{a_1} \, t^{b_1}$.
It is easy to see that the position space correlator corresponding to this agrees with \eqref{eq:gencontactls}, thus proving the equivalence of two limits for scalar contact diagrams.
\subsection{Exchange diagrams}
Exchange diagrams in Mellin space take the form 
\begin{align}
    \label{eq:genexchangeM}
    \mathcal{M}^{ex}_{\D_E, \ell_E} \left(s,t\right) = \sum_{m=0}^{\infty} \frac{f_{m, \ell_E} \, Q_{\ell_E} \left(t,u\right)}{s-\tau_E - 2m},
\end{align}
where $\tau_E = \D_E -\ell_E$ is the twist of the exchanged operator, with $\D_E, \ell_E$ being its conformal dimension and spin respectively. $f_{m, \ell_E}$ is a coefficient independent of $s, t, u$ and $Q_{\ell_E}$ is a polynomial of degree $\ell_E$ in $t,u$. Explicit expressions for these can be found in Appendix B of \cite{Alday:2020dtb}\footnote{Note that here these polynomials are called $Q_{m, \ell_E}$. However, they are independent of $m$ and we have chosen to drop the subscript $m$ here in order to avoid confusion.}. 
\paragraph{Position space correlator: } As shown in \cite{Alday:2024yyj}, the flat space limit of an exchange diagram is independent of the conformal dimension of the exchanged operator\footnote{We remind the reader that we are assuming that the conformal dimension doesn't scale with $\ell$ in this limit. }. We can therefore freely assume $\tau_E = \D_1 + \D_2 - 2m_0$ and write
\begin{align}
     \mathcal{M}^{ex}_{\D_E, \ell_E} \left(s,t\right) = \sum_{m=0}^{m_0} \frac{f_{m, \ell_E} \, Q_{\ell_E} \left(t,u\right)}{s-\left(\D_1+\D_2\right) +2k_m},
\end{align}
where $k_m = \left(m_0-m\right)$. The sum truncates since $f_{m, \ell_E}$ vanishes for $m > m_0$. We can write this as a finite sum of $\bar{D}$ functions in position space by using the identity (valid when $k_m \in \mathbb{Z}^+$).
\begin{align}
\label{eq:Gammaidentity}
    \frac{1}{s-\left(\D_1+\D_2\right)+2k_m} &\Gamma\left(\frac{\D_1+\D_2-s}{2}\right)= -\frac{1}{2}\Gamma\left(\frac{\D_1+\D_2-s}{2}-k_m\right) \prod_{n=1}^{k_m-1} \left[\frac{\D_1+\D_2-s}{2}-n\right],
\end{align}
in \eqref{eq:Mellingen} and arriving at
\begin{align}
    \label{eq:genexchangeP}
     \mathcal{G}^{ex}_{\D_E, \ell_E} \left(U, V\right)  = -\sum_{m=0}^{m_0} \frac{f_{m, \ell_E}}{2} \, \hat{Q}_{\ell_E} \prod_{n=1}^{k_m-1} \left[\frac{\D_1+\D_2}{2}-U \partial_U -n\right] \left(U^{\frac{\D_1+\D_2}{2}-k_m} \bar{D}_{\D_1-k_m,\D_2-k_m,\D_3,\D_4} \right),
\end{align}
where $\hat{Q}_{\ell_E}$ is a differential operator obtained from $Q_{\ell_E}$ by the replacements
\begin{align}
    t \rightarrow \mathcal{D}_t = 2V \partial_V +\D_2+\D_3, \qquad u \rightarrow \mathcal{D}_u = \D_1+\D_4 - 2U \partial_U - 2V \partial_V.
\end{align}
The most singular term in \eqref{eq:genexchangeP} arises from the terms of degree $\ell_E$ in $Q_{m, \ell_E}\left(t,u\right)$. To this end, let us write
\begin{align}
    \label{eq:Qls}
    \mathcal{Q}_{\ell_E} \left(s,t\right) =  \tilde{\mathcal{Q}} + \sum_{a+b = \ell_E} \chi_{a,b}\,  t^a\,  u^b 
\end{align}
where $\tilde{\mathcal{Q}}$ is a lower degree polynomial and $\chi_{a,b}$ are coefficients which are independent of $t,u$. Their explicit form can be easily extracted from Appendix B of \cite{Alday:2020dtb}. We now have
\begin{align}
    \label{eq:genexchangels}
     \mathcal{G}^{ex, l.s}_{\D_E, \ell_E} \left(U, V\right)  = -\frac{1}{2}\left[\sum_{m=0}^{m_0} f_{m, \ell_E} \,\right]  \sum_{a+b = \ell_E} (-1)^a \chi_{a,b}\,2^{a+b}  V^a U^{\frac{\D_1+\D_2}{2}-1} \Phi^{l.s}_{\D_1-1,\D_2-1+a+b,\D_3+a,\D_4+b} .
\end{align}
\paragraph{High energy limit of ${\bf \mathcal{M}^{ex}_{\D_E, \ell_E} \left(s,t\right)}$:} We will take the high energy limit by first writing $u = \Sigma_{\D} - s -t$ and then taking $s, t \to \infty$ which yields
\begin{align}
\label{eq:Mellinexhelimit}
    \mathcal{M}^{ex}_{\D_E, \ell_E} \left(s,t\right) \xrightarrow[]{s,t \to \infty} \frac{ \sum_{a+b = \ell_E} \chi_{a,b}\,  t^a\,  u^b }{s}\sum_{m=0}^{\infty} f_{m, \ell_E} .
\end{align}
We can use the identity \footnote{Even though this identity is valid only when $\frac{\D_1+\D_2}{2} \in \mathbb{Z}$, the final result in terms of $\bar{D}$ functions holds for all values of $\D_1, \D_2$.}
\begin{align}
     \frac{1}{s} &\Gamma\left(\frac{\D_1+\D_2-s}{2}\right)= -\frac{1}{2}\Gamma\left(-\frac{s}{2}\right) \prod_{n=1}^{k-1} \left[\frac{k-s}{2}-n\right].
\end{align}
and \eqref{eq:Mellingen} to convert the above expression to position space to get
\begin{align}
    \label{eq:highenergyls}
     \mathcal{G}^{ex, l.s}_{\D_E, \ell_E} \left(U, V\right)  =-\frac{1}{2} \left[\sum_{m=0}^{\infty} f_{m, \ell_E}\right] \, \sum_{a+b = \ell_E} (-1)^a  \, \chi_{a,b}\, 2^{a+b} V^a    U^{\frac{\D_1+\D_2}{2}-1} \Phi^{l.s}_{\D_1-1,\D_2-1+a+b,\D_3+a,\D_4+b}
\end{align}
which is in agreement with \eqref{eq:genexchangels} thus showing equivalence of the two limits for scalar exchange diagrams. 

\section{Lower-point Carrollian amplitudes}
\label{sec:3ptamps}
In section \eqref{sec:4ptfrom11DSUGRA}, we obtained the Carrollian limit of 4-point ABJM correlators from a bulk perspective by dimensionally reducing the 11d gravity amplitude and performing a modified Mellin transform. In this Appendix, we will describe an analogous calculation at two and three points. 

Let us start with the 2-point amplitude:
\begin{align}
\mathcal{A}_{2}= \left(e_{1}\cdot e_{2}\right)^{2} \, 2P^0_1\left(2\pi\right)^{10}\delta^{(10)}\left(P_1+P_2\right).
\end{align}
We will dimensionally reduce this to 4D by setting the momenta along all but four directions to zero. This results in $\delta^{(7)} \left(0\right)$ which we regulate by replacing it by $V_7$, the volume of the $S^7$ with radius $\ell$, with $\ell \to \infty$. Furthermore taking $\e_1 \cdot \e_2 = t_{12}$ gives
\begin{align}
\mathcal{A}_{2}=t_{12}^{2} \, 2P^0_1 V_7\left(2\pi\right)^3\delta^{(3)}\left(P_1+P_2\right),
\end{align}
which is in agreement with \eqref{eq:2ptfromZ}. We may then perform the Fourier transform \eqref{eq:deltaonecarramp} as described in section \eqref{sec:Carrollian amplitudes and Carrollian/flat space limit correspondence} to obtain the expression in \eqref{eq:2ptcarrolllim} with $k_1 = k_2 =2$. Moreover, to get the higher charge 2-point functions, we dress with external states given in \eqref{eq:k>2wavefn},  regulate the resulting divergent integral by placing it on $S^7$ and perform the modified Mellin transform \eqref{eq:Carrampdef} with $\D_i = \frac{k_i}{2}$. 

Now let us consider 3-point amplitude:
\begin{equation}
\begin{aligned}
\mathcal{A}_{3}=&\left(e_{1}\cdot e_{2}e_{3}\cdot P_{1}+{\rm cyclic}\right)^{2} \, \delta^{(11)}(\sum_{i} P_i)\\
=&\left(e_{1}\cdot e_{2}e_{3}\cdot P_{1}+e_{2}\cdot e_{3}e_{1}\cdot P_{2}-e_{3}\cdot e_{1}e_{2}\cdot P_{1}\right)^{2} \, \delta^{(11)}(\sum_i P_i),
\end{aligned}
\end{equation}
where we used momentum conservation and $e_{2}\cdot P_{2}=0$ to obtain third term in the second line. Now write out the terms in the product explicitly:
\begin{align}
\label{3pt}
\mathcal{A}_{3}=&\Big[ \left(e_{1}\cdot e_{2} \,e_{3}\cdot P_{1}\right)^{2}+\left(e_{2}\cdot e_{3} \, e_{1}\cdot P_{2}\right)^{2}+\left(e_{3}\cdot e_{1} \, e_{2}\cdot P_{1}\right)^{2}
+ 2e_{1}\cdot e_{2} \, e_{3}\cdot P_{1} \, e_{2}\cdot e_{3} \, e_{1}\cdot P_{2}
\nonumber\\
& \qquad -2e_{2}\cdot e_{3} \, e_{1}\cdot P_{2} \, e_{3}\cdot e_{1} \, e_{2}\cdot P_{1} -2e_{1}\cdot e_{2} \, e_{3}\cdot P_{1} \, e_{3}\cdot e_{1} \, e_{2}\cdot P_{1}\Big]\delta^{(11)}(\sum_i P_i).
\end{align}
Repeating the dimensional reduction procedure we performed for the two point-case and setting $e_i \cdot P_j = 0$, we find that amplitude scales as $\ell^5$ which is consistent with the cubic term in the action \eqref{eq:scalaraction}. Let us next fix the magnitude of the 7d momenta and integrate over their directions \cite{Polchinski:1999ry}:
\begin{equation}
P_{i}^{A}P_{j}^{B}\rightarrow\frac{1}{\ell^{2}}\int d^{6}\hat{p}_{i}^{A}d^{6}\hat{p}_{j}^{B}=\frac{1}{\ell^{2}}\delta_{ij}\delta^{AB}
\end{equation}
where $\hat{p}_{i}^{A}$ is a unit vector and we take the magnitude to be $1/\ell$ up to a numerical factor which we ignore. After performing
this integral, the terms in the first line of \eqref{3pt} vanish because they
are proportional to $e_{i}^{2}=0$, the terms in the second line
vanish because they are proportional to $\delta_{ij}=0$ for $i\neq j$,
and the third line yields 
\begin{equation}
\mathcal{A}_{3}\rightarrow -\frac{2V_7}{\ell^2} e_{1}\cdot e_{2} \, e_{2}\cdot e_{3} \, e_{3}\cdot e_{1}\delta^{(4)}(\sum_i P_i) = -\frac{2V_7}{\ell^2}t_{12} t_{23} t_{13}\delta^{(4)}(\sum_i P_i).
\end{equation}
Which reproduces the $R$-symmetry structure found in \eqref{eq:carroll3ptabjm} for $k_i =2$. To get higher charge correlators, we dress with external states in \eqref{eq:k>2wavefn} and regulate the divergence by integrating over $S^{7}$. We may then perform the modified Mellin transform with $\D_i = \frac{k_i}{2}$ to obtain \eqref{eq:carroll3ptabjm}.  

\section{Carrollian amplitudes of higher derivative corrections}
\label{sec:higherdercamp}

In this Appendix, we will compute the Carrollian amplitudes corresponding to $\frac{1}{N}$ corrections to $\mathcal{M}_{2,2,2,2}$. These correspond to higher derivative corrections to supergravity arising from M-theory. The Mellin amplitudes can be found in \cite{Chester:2018aca, Binder:2018yvd} and also in the Mathematica file accompanying \cite{Alday:2022rly}.
\begin{align}
    \mathcal{M}_{2,2,2,2} &= \frac{1}{c_T} \mathcal{M}_{2,2,2,2}^{R} + \frac{1}{c_T^{\frac{5}{3}}} B^{R^4} \mathcal{M}_{2,2,2,2}^{\left(4\right)} \\
    &\qquad\nonumber + \frac{1}{c_T^{\frac{7}{3}}}\left( B_4^{D^6 R^4} \mathcal{M}_{2,2,2,2}^{\left(4\right)}  +  B_6^{D^6 R^4} \mathcal{M}_{2,2,2,2}^{\left(6\right)} + B_7^{D^6 R^4}\mathcal{M}_{2,2,2,2}^{\left(7\right)} \right) + \dots ,
\end{align}
where
\begin{align}
    \label{eq:B2222s}
     &B_4^{R^4} = 1120\left(\frac{2}{9\pi^8 k_{CS}^2}\right)^{\frac{1}{3}} , \qquad  B_4^{D^6 R^4} = -\frac{1352960}{9}\left(\frac{36}{\pi^{10}k_{CS}^{4}}\right)^{\frac{1}{3}}, \nonumber\\
     &B_6^{D^6 R^4} = -220528 \left(\frac{36}{\pi^{10}k_{CS}^{4}}\right)^{\frac{1}{3}},\qquad  B_7^{D^6 R^4} = 16016 \left(\frac{36}{\pi^{10}k_{CS}^{4}}\right)^{\frac{1}{3}}.
\end{align}
Since we are only interested in the Carrollian limit, it suffices to consider the leading high energy terms in the Mellin amplitudes, following Appendix \eqref{sec:MCequivalence}: 
\begin{align}
    \label{eq:higherderHE}
   &\mathcal{M}_{2,2,2,2}^{(4), HE} = \left(s+t-s\alpha\right)^2  \left(s+t-s\bar{\alpha}\right)^2, \\
   &\nonumber \mathcal{M}_{2,2,2,2}^{(6), HE} = 2\left(s^2+t^2+st\right)\left(s+t-s\alpha\right)^2  \left(s+t-s\bar{\alpha}\right)^2, \\
   &\nonumber  \mathcal{M}_{2,2,2,2}^{(7), HE} = s t u \left(s+t-s\alpha\right)^2  \left(s+t-s\bar{\alpha}\right)^2. 
\end{align}
This $\frac{1}{N}$ expansion of the Mellin amplitude neatly organizes into a $u-$derivative expansion of the Carrollian amplitude as shown below:
\begin{align}
   \lim_{\ell \to \infty} \frac{V_7}{\left(2\pi\right)^4}\an{\mo_2 \dots \mo_2} = \left[1 + f_1  \left(\ell_{11}\partial_{u_4}\right)^6 + f_2 \left(\ell_{11}\partial_{u_4}\right)^{10} + f_3  \left(\ell_{11}\partial_{u_4}\right)^{11} \right]\tilde{\mathcal{C}}_4.
   \end{align}
The $f_i$ are functions depending on the coordinates on the celestial sphere and are given by  
\begin{align}
    f_3 = \mathcal{N}_3 f_1^2, \qquad   &f_2 = \mathcal{N}_2 \left|\frac{z_{24}}{z_{12}}\right|^4\left(\left(1-z\right)^2\left|\frac{z_{34}}{z_{23}}\right|^4 - z (1-z) \left|\frac{z_{34}}{z_{23}}\right|^2 +z^2\right)f_1,  \nonumber\\
    & \qquad\qquad f_1 = \mathcal{N}_1 z^2 \left|z_{24}\right|^2\left|z_{34}\right|^2\left|z_{14}\right|^4,
\end{align}
where $\mathcal{N}_i$ are numerical constants. This expansion is reminiscent of how the $\alpha'$ expansion of Carrollian string amplitudes is converted to a $u-$derivative expansion \cite{Stieberger:2024shv}. 
\section*{References}
\bibliographystyle{style}
\renewcommand\refname{\vskip -1.3cm}
\bibliography{Biblio}

\providecommand{\href}[2]{#2}\begingroup\raggedright\begin{thebibliography}{100}

\bibitem{Susskind:1998vk}
L.~Susskind, \emph{{Holography in the flat space limit}}, AIP Conf. Proc. {\bf
  493} (1999), no.~1, 98--112,
\href{http://www.arXiv.org/abs/hep-th/9901079}{{\tt hep-th/9901079}}

\bibitem{Polchinski:1999ry}
J.~Polchinski, \emph{{S-matrices from AdS space-time}},
\href{http://www.arXiv.org/abs/hep-th/9901076}{{\tt hep-th/9901076}}

\bibitem{Giddings:1999jq}
S.~B. Giddings, \emph{{Flat space scattering and bulk locality in the AdS / CFT
  correspondence}}, Phys. Rev. D {\bf 61} (2000) 106008,
\href{http://www.arXiv.org/abs/hep-th/9907129}{{\tt hep-th/9907129}}

\bibitem{deBoer:2003vf}
J.~de~Boer and S.~N. Solodukhin, \emph{{A Holographic reduction of Minkowski
  space-time}}, Nucl. Phys. {\bf B665} (2003) 545--593,
\href{http://www.arXiv.org/abs/hep-th/0303006}{{\tt hep-th/0303006}}

\bibitem{He:2015zea}
T.~He, P.~Mitra  and A.~Strominger, \emph{{2D Kac-Moody Symmetry of 4D
  Yang-Mills Theory}}, JHEP {\bf 10} (2016) 137,
\href{http://www.arXiv.org/abs/1503.02663}{{\tt 1503.02663}}

\bibitem{Pasterski:2016qvg}
S.~Pasterski, S.-H. Shao  and A.~Strominger, \emph{{Flat Space Amplitudes and
  Conformal Symmetry of the Celestial Sphere}}, Phys. Rev. {\bf D96} (2017),
  no.~6, 065026,
\href{http://www.arXiv.org/abs/1701.00049}{{\tt 1701.00049}}

\bibitem{Cheung:2016iub}
C.~Cheung, A.~de~la Fuente  and R.~Sundrum, \emph{{4D scattering amplitudes and
  asymptotic symmetries from 2D CFT}}, JHEP {\bf 01} (2017) 112,
\href{http://www.arXiv.org/abs/1609.00732}{{\tt 1609.00732}}

\bibitem{Pasterski:2017kqt}
S.~Pasterski and S.-H. Shao, \emph{{Conformal basis for flat space
  amplitudes}}, Phys. Rev. {\bf D96} (2017), no.~6, 065022,
\href{http://www.arXiv.org/abs/1705.01027}{{\tt 1705.01027}}

\bibitem{Strominger:2017zoo}
A.~Strominger, {\em {Lectures on the Infrared Structure of Gravity and Gauge
  Theory}}.
\newblock {Princeton University Press},
2018

\bibitem{Pasterski:2017ylz}
S.~Pasterski, S.-H. Shao  and A.~Strominger, \emph{{Gluon Amplitudes as 2d
  Conformal Correlators}}, Phys. Rev. {\bf D96} (2017), no.~8, 085006,
\href{http://www.arXiv.org/abs/1706.03917}{{\tt 1706.03917}}

\bibitem{Arcioni:2003xx}
G.~Arcioni and C.~Dappiaggi, \emph{{Exploring the holographic principle in
  asymptotically flat space-times via the BMS group}}, Nucl. Phys. B {\bf 674}
  (2003) 553--592,
\href{http://www.arXiv.org/abs/hep-th/0306142}{{\tt hep-th/0306142}}

\bibitem{Dappiaggi:2005ci}
C.~Dappiaggi, V.~Moretti  and N.~Pinamonti, \emph{{Rigorous steps towards
  holography in asymptotically flat spacetimes}}, Rev. Math. Phys. {\bf 18}
  (2006) 349--416,
\href{http://www.arXiv.org/abs/gr-qc/0506069}{{\tt gr-qc/0506069}}

\bibitem{Barnich:2006av}
G.~Barnich and G.~Compere, \emph{{Classical central extension for asymptotic
  symmetries at null infinity in three spacetime dimensions}}, Class. Quant.
  Grav. {\bf 24} (2007) F15--F23,
\href{http://www.arXiv.org/abs/gr-qc/0610130}{{\tt gr-qc/0610130}}

\bibitem{Barnich:2010eb}
G.~Barnich and C.~Troessaert, \emph{{Aspects of the BMS/CFT correspondence}},
  JHEP {\bf 05} (2010) 062,
\href{http://www.arXiv.org/abs/1001.1541}{{\tt 1001.1541}}

\bibitem{Bagchi:2010zz}
A.~Bagchi, \emph{{Correspondence between Asymptotically Flat Spacetimes and
  Nonrelativistic Conformal Field Theories}}, Phys. Rev. Lett. {\bf 105} (2010)
  171601,
\href{http://www.arXiv.org/abs/1006.3354}{{\tt 1006.3354}}

\bibitem{Barnich:2012xq}
G.~Barnich, \emph{{Entropy of three-dimensional asymptotically flat
  cosmological solutions}}, JHEP {\bf 10} (2012) 095,
\href{http://www.arXiv.org/abs/1208.4371}{{\tt 1208.4371}}

\bibitem{Barnich:2012rz}
G.~Barnich, A.~Gomberoff  and H.~A. Gonz\'alez, \emph{{Three-dimensional
  Bondi-Metzner-Sachs invariant two-dimensional field theories as the flat
  limit of Liouville theory}}, Phys. Rev. D {\bf 87} (2013), no.~12, 124032,
\href{http://www.arXiv.org/abs/1210.0731}{{\tt 1210.0731}}

\bibitem{Bagchi:2012xr}
A.~Bagchi, S.~Detournay, R.~Fareghbal  and J.~Sim\'on, \emph{{Holography of 3D
  Flat Cosmological Horizons}}, Phys. Rev. Lett. {\bf 110} (2013), no.~14,
  141302,
\href{http://www.arXiv.org/abs/1208.4372}{{\tt 1208.4372}}

\bibitem{Bagchi:2014iea}
A.~Bagchi, R.~Basu, D.~Grumiller  and M.~Riegler, \emph{{Entanglement entropy
  in Galilean conformal field theories and flat holography}}, Phys. Rev. Lett.
  {\bf 114} (2015), no.~11, 111602,
\href{http://www.arXiv.org/abs/1410.4089}{{\tt 1410.4089}}

\bibitem{Bagchi:2015wna}
A.~Bagchi, D.~Grumiller  and W.~Merbis, \emph{{Stress tensor correlators in
  three-dimensional gravity}}, Phys. Rev. D {\bf 93} (2016), no.~6, 061502,
\href{http://www.arXiv.org/abs/1507.05620}{{\tt 1507.05620}}

\bibitem{Bagchi:2016bcd}
A.~Bagchi, R.~Basu, A.~Kakkar  and A.~Mehra, \emph{{Flat Holography: Aspects of
  the dual field theory}}, JHEP {\bf 12} (2016) 147,
\href{http://www.arXiv.org/abs/1609.06203}{{\tt 1609.06203}}

\bibitem{Ciambelli:2018wre}
L.~Ciambelli, C.~Marteau, A.~C. Petkou, P.~M. Petropoulos  and K.~Siampos,
  \emph{{Flat holography and Carrollian fluids}}, JHEP {\bf 07} (2018) 165,
\href{http://www.arXiv.org/abs/1802.06809}{{\tt 1802.06809}}

\bibitem{Donnay:2022aba}
L.~Donnay, A.~Fiorucci, Y.~Herfray  and R.~Ruzziconi, \emph{{Carrollian
  Perspective on Celestial Holography}}, Phys. Rev. Lett. {\bf 129} (2022),
  no.~7, 071602,
\href{http://www.arXiv.org/abs/2202.04702}{{\tt 2202.04702}}

\bibitem{Bagchi:2022emh}
A.~Bagchi, S.~Banerjee, R.~Basu  and S.~Dutta, \emph{{Scattering Amplitudes:
  Celestial and Carrollian}}, Phys. Rev. Lett. {\bf 128} (2022), no.~24,
  241601,
\href{http://www.arXiv.org/abs/2202.08438}{{\tt 2202.08438}}

\bibitem{Donnay:2022wvx}
L.~Donnay, A.~Fiorucci, Y.~Herfray  and R.~Ruzziconi, \emph{{Bridging
  Carrollian and celestial holography}}, Phys. Rev. D {\bf 107} (2023), no.~12,
  126027,
\href{http://www.arXiv.org/abs/2212.12553}{{\tt 2212.12553}}

\bibitem{Saha:2023hsl}
A.~Saha, \emph{{Carrollian approach to 1 + 3D flat holography}}, JHEP {\bf 06}
  (2023) 051,
\href{http://www.arXiv.org/abs/2304.02696}{{\tt 2304.02696}}

\bibitem{Strominger:2013jfa}
A.~Strominger, \emph{{On BMS Invariance of Gravitational Scattering}}, JHEP
  {\bf 07} (2014) 152,
\href{http://www.arXiv.org/abs/1312.2229}{{\tt 1312.2229}}

\bibitem{Adamo:2014yya}
T.~Adamo, E.~Casali  and D.~Skinner, \emph{{Perturbative gravity at null
  infinity}}, Class. Quant. Grav. {\bf 31} (2014), no.~22, 225008,
\href{http://www.arXiv.org/abs/1405.5122}{{\tt 1405.5122}}

\bibitem{He:2014laa}
T.~He, V.~Lysov, P.~Mitra  and A.~Strominger, \emph{{BMS supertranslations and
  Weinberg's soft graviton theorem}}, JHEP {\bf 05} (2015) 151,
\href{http://www.arXiv.org/abs/1401.7026}{{\tt 1401.7026}}

\bibitem{Kapec:2014opa}
D.~Kapec, V.~Lysov, S.~Pasterski  and A.~Strominger, \emph{{Semiclassical
  Virasoro symmetry of the quantum gravity $ \mathcal{S}$-matrix}}, JHEP {\bf
  08} (2014) 058,
\href{http://www.arXiv.org/abs/1406.3312}{{\tt 1406.3312}}

\bibitem{Strominger:2014pwa}
A.~Strominger and A.~Zhiboedov, \emph{{Gravitational Memory, BMS
  Supertranslations and Soft Theorems}}, JHEP {\bf 01} (2016) 086,
\href{http://www.arXiv.org/abs/1411.5745}{{\tt 1411.5745}}

\bibitem{Stieberger:2018edy}
S.~Stieberger and T.~R. Taylor, \emph{{Strings on Celestial Sphere}}, Nucl.
  Phys. {\bf B935} (2018) 388--411,
\href{http://www.arXiv.org/abs/1806.05688}{{\tt 1806.05688}}

\bibitem{Stieberger:2018onx}
S.~Stieberger and T.~R. Taylor, \emph{{Symmetries of Celestial Amplitudes}},
  Phys. Lett. B {\bf 793} (2019) 141--143,
\href{http://www.arXiv.org/abs/1812.01080}{{\tt 1812.01080}}

\bibitem{Banerjee:2018gce}
S.~Banerjee, \emph{{Null Infinity and Unitary Representation of The Poincare
  Group}}, JHEP {\bf 01} (2019) 205,
\href{http://www.arXiv.org/abs/1801.10171}{{\tt 1801.10171}}

\bibitem{Adamo:2019ipt}
T.~Adamo, L.~Mason  and A.~Sharma, \emph{{Celestial amplitudes and conformal
  soft theorems}}, Class. Quant. Grav. {\bf 36} (2019), no.~20, 205018,
\href{http://www.arXiv.org/abs/1905.09224}{{\tt 1905.09224}}

\bibitem{Fotopoulos:2019vac}
A.~Fotopoulos, S.~Stieberger, T.~R. Taylor  and B.~Zhu, \emph{{Extended BMS
  Algebra of Celestial CFT}}, JHEP {\bf 03} (2020) 130,
\href{http://www.arXiv.org/abs/1912.10973}{{\tt 1912.10973}}

\bibitem{Banerjee:2020zlg}
S.~Banerjee, S.~Ghosh  and P.~Paul, \emph{{MHV graviton scattering amplitudes
  and current algebra on the celestial sphere}}, JHEP {\bf 02} (2021) 176,
\href{http://www.arXiv.org/abs/2008.04330}{{\tt 2008.04330}}

\bibitem{Pasterski:2020pdk}
S.~Pasterski and A.~Puhm, \emph{{Shifting spin on the celestial sphere}}, Phys.
  Rev. D {\bf 104} (2021), no.~8, 086020,
\href{http://www.arXiv.org/abs/2012.15694}{{\tt 2012.15694}}

\bibitem{Pasterski:2021fjn}
S.~Pasterski, A.~Puhm  and E.~Trevisani, \emph{{Celestial diamonds: conformal
  multiplets in celestial CFT}}, JHEP {\bf 11} (2021) 072,
\href{http://www.arXiv.org/abs/2105.03516}{{\tt 2105.03516}}

\bibitem{Freidel:2021dfs}
L.~Freidel, D.~Pranzetti  and A.-M. Raclariu, \emph{{Sub-subleading Soft
  Graviton Theorem from Asymptotic Einstein's Equations}},
\href{http://www.arXiv.org/abs/2111.15607}{{\tt 2111.15607}}

\bibitem{Freidel:2021ytz}
L.~Freidel, D.~Pranzetti  and A.-M. Raclariu, \emph{{Higher spin dynamics in
  gravity and $w_{1 + \infty}$ celestial symmetries}},
\href{http://www.arXiv.org/abs/2112.15573}{{\tt 2112.15573}}

\bibitem{Adamo:2021zpw}
T.~Adamo, W.~Bu, E.~Casali  and A.~Sharma, \emph{{Celestial operator products
  from the worldsheet}}, JHEP {\bf 06} (2022) 052,
\href{http://www.arXiv.org/abs/2111.02279}{{\tt 2111.02279}}

\bibitem{Strominger:2021mtt}
A.~Strominger, \emph{{$w_{1+\infty}$ Algebra and the Celestial Sphere: Infinite
  Towers of Soft Graviton, Photon, and Gluon Symmetries}}, Phys. Rev. Lett.
  {\bf 127} (2021), no.~22,
221601

\bibitem{Adamo:2021lrv}
T.~Adamo, L.~Mason  and A.~Sharma, \emph{{Celestial $w_{1+\infty}$ Symmetries
  from Twistor Space}}, SIGMA {\bf 18} (2022) 016,
\href{http://www.arXiv.org/abs/2110.06066}{{\tt 2110.06066}}

\bibitem{Donnay:2021wrk}
L.~Donnay and R.~Ruzziconi, \emph{{BMS flux algebra in celestial holography}},
  JHEP {\bf 11} (2021) 040,
\href{http://www.arXiv.org/abs/2108.11969}{{\tt 2108.11969}}

\bibitem{Donnay:2022hkf}
L.~Donnay, K.~Nguyen  and R.~Ruzziconi, \emph{{Loop-corrected subleading soft
  theorem and the celestial stress tensor}}, JHEP {\bf 09} (2022) 063,
\href{http://www.arXiv.org/abs/2205.11477}{{\tt 2205.11477}}

\bibitem{Hu:2021lrx}
Y.~Hu, L.~Ren, A.~Y. Srikant  and A.~Volovich, \emph{{Celestial dual
  superconformal symmetry, MHV amplitudes and differential equations}}, JHEP
  {\bf 12} (2021) 171,
\href{http://www.arXiv.org/abs/2106.16111}{{\tt 2106.16111}}

\bibitem{Mago:2021wje}
J.~Mago, L.~Ren, A.~Y. Srikant  and A.~Volovich, \emph{{Deformed $w_{1+\infty}$
  Algebras in the Celestial CFT}},
\href{http://www.arXiv.org/abs/2111.11356}{{\tt 2111.11356}}

\bibitem{Freidel:2022bai}
L.~Freidel and P.~Jai-akson, \emph{{Carrollian hydrodynamics from symmetries}},
\href{http://www.arXiv.org/abs/2209.03328}{{\tt 2209.03328}}

\bibitem{Freidel:2022skz}
L.~Freidel, D.~Pranzetti  and A.-M. Raclariu, \emph{{A discrete basis for
  celestial holography}}, JHEP {\bf 02} (2024) 176,
\href{http://www.arXiv.org/abs/2212.12469}{{\tt 2212.12469}}

\bibitem{Ren:2022sws}
L.~Ren, M.~Spradlin, A.~Yelleshpur~Srikant  and A.~Volovich, \emph{{On
  effective field theories with celestial duals}}, JHEP {\bf 08} (2022) 251,
\href{http://www.arXiv.org/abs/2206.08322}{{\tt 2206.08322}}

\bibitem{Bhardwaj:2022anh}
R.~Bhardwaj, L.~Lippstreu, L.~Ren, M.~Spradlin, A.~Yelleshpur~Srikant  and
  A.~Volovich, \emph{{Loop-level gluon OPEs in celestial holography}}, JHEP
  {\bf 11} (2022) 171,
\href{http://www.arXiv.org/abs/2208.14416}{{\tt 2208.14416}}

\bibitem{Bu:2022iak}
W.~Bu, S.~Heuveline  and D.~Skinner, \emph{{Moyal deformations, W$_{1+\infty}$
  and celestial holography}}, JHEP {\bf 12} (2022) 011,
\href{http://www.arXiv.org/abs/2208.13750}{{\tt 2208.13750}}

\bibitem{Mason:2022hly}
L.~Mason, \emph{{Gravity from holomorphic discs and celestial $Lw_{1+\infty }$
  symmetries}}, Lett. Math. Phys. {\bf 113} (2023), no.~6, 111,
\href{http://www.arXiv.org/abs/2212.10895}{{\tt 2212.10895}}

\bibitem{Melton:2023bjw}
W.~Melton, A.~Sharma  and A.~Strominger, \emph{{Celestial leaf amplitudes}},
  JHEP {\bf 07} (2024) 132,
\href{http://www.arXiv.org/abs/2312.07820}{{\tt 2312.07820}}

\bibitem{Pano:2023slc}
Y.~Pano, A.~Puhm  and E.~Trevisani, \emph{{Symmetries in Celestial CFT$_{d}$}},
  JHEP {\bf 07} (2023) 076,
\href{http://www.arXiv.org/abs/2302.10222}{{\tt 2302.10222}}

\bibitem{Sleight:2023ojm}
C.~Sleight and M.~Taronna, \emph{{Celestial Holography Revisited}}, Phys. Rev.
  Lett. {\bf 133} (2024), no.~24, 241601,
\href{http://www.arXiv.org/abs/2301.01810}{{\tt 2301.01810}}

\bibitem{Fiorucci:2023lpb}
A.~Fiorucci, D.~Grumiller  and R.~Ruzziconi, \emph{{Logarithmic Celestial
  Conformal Field Theory}},
\href{http://www.arXiv.org/abs/2305.08913}{{\tt 2305.08913}}

\bibitem{Agrawal:2023zea}
S.~Agrawal, L.~Donnay, K.~Nguyen  and R.~Ruzziconi, \emph{{Logarithmic soft
  graviton theorems from superrotation Ward identities}}, JHEP {\bf 02} (2024)
  120,
\href{http://www.arXiv.org/abs/2309.11220}{{\tt 2309.11220}}

\bibitem{Choi:2024ygx}
S.~Choi, A.~Laddha  and A.~Puhm, \emph{{Asymptotic Symmetries for Logarithmic
  Soft Theorems in Gauge Theory and Gravity}},
\href{http://www.arXiv.org/abs/2403.13053}{{\tt 2403.13053}}

\bibitem{Geiller:2024bgf}
M.~Geiller, \emph{{Celestial $w_{1+\infty}$ charges and the subleading
  structure of asymptotically-flat spacetimes}},
\href{http://www.arXiv.org/abs/2403.05195}{{\tt 2403.05195}}

\bibitem{Adamo:2024mqn}
T.~Adamo, W.~Bu, P.~Tourkine  and B.~Zhu, \emph{{Eikonal amplitudes on the
  celestial sphere}}, JHEP {\bf 10} (2024) 192,
\href{http://www.arXiv.org/abs/2405.15594}{{\tt 2405.15594}}

\bibitem{Cresto:2024fhd}
N.~Cresto and L.~Freidel, \emph{{Asymptotic Higher Spin Symmetries I: Covariant
  Wedge Algebra in Gravity}},
\href{http://www.arXiv.org/abs/2409.12178}{{\tt 2409.12178}}

\bibitem{Cresto:2024mne}
N.~Cresto and L.~Freidel, \emph{{Asymptotic Higher Spin Symmetries II: Noether
  Realization in Gravity}},
\href{http://www.arXiv.org/abs/2410.15219}{{\tt 2410.15219}}

\bibitem{Ball:2023qim}
A.~Ball, M.~Spradlin, A.~Yelleshpur~Srikant  and A.~Volovich,
  \emph{{Supersymmetry and the celestial Jacobi identity}}, JHEP {\bf 04}
  (2024) 099,
\href{http://www.arXiv.org/abs/2311.01364}{{\tt 2311.01364}}

\bibitem{Bhardwaj:2024wld}
R.~Bhardwaj and A.~Yelleshpur~Srikant, \emph{{Celestial soft currents at
  one-loop and their OPEs}}, JHEP {\bf 07} (2024) 034,
\href{http://www.arXiv.org/abs/2403.10443}{{\tt 2403.10443}}

\bibitem{Ball:2022bgg}
A.~Ball, \emph{{Celestial locality and the Jacobi identity}}, JHEP {\bf 01}
  (2023) 146,
\href{http://www.arXiv.org/abs/2211.09151}{{\tt 2211.09151}}

\bibitem{Ball:2023sdz}
A.~Ball, Y.~Hu  and S.~Pasterski, \emph{{Multicollinear singularities in
  celestial CFT}}, JHEP {\bf 02} (2024) 219,
\href{http://www.arXiv.org/abs/2309.16602}{{\tt 2309.16602}}

\bibitem{Guevara:2024ixn}
A.~Guevara, Y.~Hu  and S.~Pasterski, \emph{{Multiparticle Contributions to the
  Celestial OPE}},
\href{http://www.arXiv.org/abs/2402.18798}{{\tt 2402.18798}}

\bibitem{Kulp:2024scx}
J.~Kulp and S.~Pasterski, \emph{{Multiparticle States for the Flat Hologram}},
\href{http://www.arXiv.org/abs/2501.00462}{{\tt 2501.00462}}

\bibitem{Costello:2022wso}
K.~Costello and N.~M. Paquette, \emph{{Celestial holography meets twisted
  holography: 4d amplitudes from chiral correlators}}, JHEP {\bf 10} (2022)
  193,
\href{http://www.arXiv.org/abs/2201.02595}{{\tt 2201.02595}}

\bibitem{Costello:2023hmi}
K.~Costello, N.~M. Paquette  and A.~Sharma, \emph{{Burns space and
  holography}}, JHEP {\bf 10} (2023) 174,
\href{http://www.arXiv.org/abs/2306.00940}{{\tt 2306.00940}}

\bibitem{Costello:2022jpg}
K.~Costello, N.~M. Paquette  and A.~Sharma, \emph{{Top-down holography in an
  asymptotically flat spacetime}},
\href{http://www.arXiv.org/abs/2208.14233}{{\tt 2208.14233}}

\bibitem{Bittleston:2024efo}
R.~Bittleston, K.~Costello  and K.~Zeng, \emph{{Self-Dual Gauge Theory from the
  Top Down}},
\href{http://www.arXiv.org/abs/2412.02680}{{\tt 2412.02680}}

\bibitem{Bittleston:2023bzp}
R.~Bittleston, S.~Heuveline  and D.~Skinner, \emph{{The celestial chiral
  algebra of self-dual gravity on Eguchi-Hanson space}}, JHEP {\bf 09} (2023)
  008,
\href{http://www.arXiv.org/abs/2305.09451}{{\tt 2305.09451}}

\bibitem{Fernandez:2024tue}
V.~E. Fern\'andez, N.~M. Paquette  and B.~R. Williams, \emph{{Twisted
  holography on AdS$_3 \times S^3 \times$ K3 \& the planar chiral algebra}},
  SciPost Phys. {\bf 17} (2024), no.~4, 109,
\href{http://www.arXiv.org/abs/2404.14318}{{\tt 2404.14318}}

\bibitem{Fernandez:2024qnu}
V.~E. Fern\'andez and N.~M. Paquette, \emph{{Associativity is enough: an
  all-orders 2d chiral algebra for 4d form factors}},
\href{http://www.arXiv.org/abs/2412.17168}{{\tt 2412.17168}}

\bibitem{Costello:2023vyy}
K.~J. Costello, \emph{{Bootstrapping two-loop QCD amplitudes}},
\href{http://www.arXiv.org/abs/2302.00770}{{\tt 2302.00770}}

\bibitem{Dixon:2024tsb}
L.~J. Dixon and A.~Morales, \emph{{Rational QCD loop amplitudes and quantum
  theories on twistor space}}, JHEP {\bf 03} (2025) 188,
\href{http://www.arXiv.org/abs/2411.10967}{{\tt 2411.10967}}

\bibitem{Melton:2024akx}
W.~Melton, A.~Sharma, A.~Strominger  and T.~Wang, \emph{{Celestial Dual for
  Maximal Helicity Violating Amplitudes}}, Phys. Rev. Lett. {\bf 133} (2024),
  no.~9, 091603,
\href{http://www.arXiv.org/abs/2403.18896}{{\tt 2403.18896}}

\bibitem{Stieberger:2023fju}
S.~Stieberger, T.~R. Taylor  and B.~Zhu, \emph{{Yang-Mills as a Liouville
  theory}}, Phys. Lett. B {\bf 846} (2023) 138229,
\href{http://www.arXiv.org/abs/2308.09741}{{\tt 2308.09741}}

\bibitem{Stieberger:2022zyk}
S.~Stieberger, T.~R. Taylor  and B.~Zhu, \emph{{Celestial Liouville theory for
  Yang-Mills amplitudes}}, Phys. Lett. B {\bf 836} (2023) 137588,
\href{http://www.arXiv.org/abs/2209.02724}{{\tt 2209.02724}}

\bibitem{Aharony:2008ug}
O.~Aharony, O.~Bergman, D.~L. Jafferis  and J.~Maldacena, \emph{{N=6
  superconformal Chern-Simons-matter theories, M2-branes and their gravity
  duals}}, JHEP {\bf 10} (2008) 091,
\href{http://www.arXiv.org/abs/0806.1218}{{\tt 0806.1218}}

\bibitem{Levy1965}
J.-M. Lévy-Leblond, \emph{Une nouvelle limite non-relativiste du groupe de
  Poincaré}, Annales de l'I.H.P. Physique théorique {\bf 3} (1965), no.~1,
1--12

\bibitem{Barnich:2012aw}
G.~Barnich, A.~Gomberoff  and H.~A. Gonzalez, \emph{{The Flat limit of three
  dimensional asymptotically anti-de Sitter spacetimes}}, Phys. Rev. D {\bf 86}
  (2012) 024020,
\href{http://www.arXiv.org/abs/1204.3288}{{\tt 1204.3288}}

\bibitem{Barnich:2014cwa}
G.~Barnich, L.~Donnay, J.~Matulich  and R.~Troncoso, \emph{{Asymptotic
  symmetries and dynamics of three-dimensional flat supergravity}}, JHEP {\bf
  08} (2014) 071,
\href{http://www.arXiv.org/abs/1407.4275}{{\tt 1407.4275}}

\bibitem{Compere:2019bua}
G.~Comp\`ere, A.~Fiorucci  and R.~Ruzziconi, \emph{{The $\Lambda$-BMS$_4$ group
  of dS$_4$ and new boundary conditions for AdS$_4$}}, Class. Quant. Grav. {\bf
  36} (2019), no.~19, 195017, \href{http://www.arXiv.org/abs/1905.00971}{{\tt
  1905.00971}},
[Erratum: Class.Quant.Grav. 38, 229501 (2021)]

\bibitem{Compere:2020lrt}
G.~Comp\`ere, A.~Fiorucci  and R.~Ruzziconi, \emph{{The $\Lambda$-BMS$_4$
  charge algebra}}, JHEP {\bf 10} (2020) 205,
\href{http://www.arXiv.org/abs/2004.10769}{{\tt 2004.10769}}

\bibitem{Campoleoni:2023fug}
A.~Campoleoni, A.~Delfante, S.~Pekar, P.~M. Petropoulos, D.~Rivera-Betancour
  and M.~Vilatte, \emph{{Flat from anti-de Sitter}},
\href{http://www.arXiv.org/abs/2309.15182}{{\tt 2309.15182}}

\bibitem{Alday:2024yyj}
L.~F. Alday, M.~Nocchi, R.~Ruzziconi  and A.~Yelleshpur~Srikant,
  \emph{{Carrollian Amplitudes from Holographic Correlators}},
\href{http://www.arXiv.org/abs/2406.19343}{{\tt 2406.19343}}

\bibitem{deGioia:2022fcn}
L.~P. de~Gioia and A.-M. Raclariu, \emph{{Eikonal approximation in celestial
  CFT}}, JHEP {\bf 03} (2023) 030,
\href{http://www.arXiv.org/abs/2206.10547}{{\tt 2206.10547}}

\bibitem{deGioia:2023cbd}
L.~P. de~Gioia and A.-M. Raclariu, \emph{{Celestial Sector in CFT: Conformally
  Soft Symmetries}},
\href{http://www.arXiv.org/abs/2303.10037}{{\tt 2303.10037}}

\bibitem{deGioia:2024yne}
L.~P. de~Gioia and A.-M. Raclariu, \emph{{Celestial amplitudes from conformal
  correlators with bulk-point kinematics}},
\href{http://www.arXiv.org/abs/2405.07972}{{\tt 2405.07972}}

\bibitem{Bagchi:2023fbj}
A.~Bagchi, P.~Dhivakar  and S.~Dutta, \emph{{AdS Witten diagrams to Carrollian
  correlators}}, JHEP {\bf 04} (2023) 135,
\href{http://www.arXiv.org/abs/2303.07388}{{\tt 2303.07388}}

\bibitem{Bagchi:2023cen}
A.~Bagchi, P.~Dhivakar  and S.~Dutta, \emph{{Holography in Flat Spacetimes: the
  case for Carroll}},
\href{http://www.arXiv.org/abs/2311.11246}{{\tt 2311.11246}}

\bibitem{Marotta:2024sce}
R.~Marotta, K.~Skenderis  and M.~Verma, \emph{{Flat space spinning massive
  amplitudes from momentum space CFT}}, JHEP {\bf 08} (2024) 226,
\href{http://www.arXiv.org/abs/2406.06447}{{\tt 2406.06447}}

\bibitem{Rastelli:2017udc}
L.~Rastelli and X.~Zhou, \emph{{How to Succeed at Holographic Correlators
  Without Really Trying}}, JHEP {\bf 04} (2018) 014,
\href{http://www.arXiv.org/abs/1710.05923}{{\tt 1710.05923}}

\bibitem{Chester:2018aca}
S.~M. Chester, S.~S. Pufu  and X.~Yin, \emph{{The M-Theory S-Matrix From ABJM:
  Beyond 11D Supergravity}}, JHEP {\bf 08} (2018) 115,
\href{http://www.arXiv.org/abs/1804.00949}{{\tt 1804.00949}}

\bibitem{Alday:2020dtb}
L.~F. Alday and X.~Zhou, \emph{{All Holographic Four-Point Functions in All
  Maximally Supersymmetric CFTs}}, Phys. Rev. X {\bf 11} (2021), no.~1, 011056,
\href{http://www.arXiv.org/abs/2006.12505}{{\tt 2006.12505}}

\bibitem{Alday:2021ymb}
L.~F. Alday, S.~M. Chester  and H.~Raj, \emph{{ABJM at strong coupling from
  M-theory, localization, and Lorentzian inversion}}, JHEP {\bf 02} (2022) 005,
\href{http://www.arXiv.org/abs/2107.10274}{{\tt 2107.10274}}

\bibitem{Alday:2022rly}
L.~F. Alday, S.~M. Chester  and H.~Raj, \emph{{M-theory on
  AdS$_{4}$\texttimes{} S$^{7}$ at 1-loop and beyond}}, JHEP {\bf 11} (2022)
  091,
\href{http://www.arXiv.org/abs/2207.11138}{{\tt 2207.11138}}

\bibitem{Chester:2024esn}
S.~M. Chester, T.~Hansen  and D.-l. Zhong, \emph{{The type IIA Virasoro-Shapiro
  amplitude in AdS$_4$$\times$ CP$^3$ from ABJM theory}},
\href{http://www.arXiv.org/abs/2412.08689}{{\tt 2412.08689}}

\bibitem{Chester:2024bij}
S.~M. Chester, R.~Dempsey  and S.~S. Pufu, \emph{{Higher-derivative corrections
  in M-theory from precision numerical bootstrap}},
\href{http://www.arXiv.org/abs/2412.14094}{{\tt 2412.14094}}

\bibitem{Binder:2021cif}
D.~J. Binder, S.~M. Chester  and M.~Jerdee, \emph{{ABJ Correlators with Weakly
  Broken Higher Spin Symmetry}}, JHEP {\bf 04} (2021) 242,
\href{http://www.arXiv.org/abs/2103.01969}{{\tt 2103.01969}}

\bibitem{Benna:2008zy}
M.~Benna, I.~Klebanov, T.~Klose  and M.~Smedback, \emph{{Superconformal
  Chern-Simons Theories and AdS(4)/CFT(3) Correspondence}}, JHEP {\bf 09}
  (2008) 072,
\href{http://www.arXiv.org/abs/0806.1519}{{\tt 0806.1519}}

\bibitem{Bandres:2008ry}
M.~A. Bandres, A.~E. Lipstein  and J.~H. Schwarz, \emph{{Studies of the ABJM
  Theory in a Formulation with Manifest SU(4) R-Symmetry}}, JHEP {\bf 09}
  (2008) 027,
\href{http://www.arXiv.org/abs/0807.0880}{{\tt 0807.0880}}

\bibitem{Kapustin:2010xq}
A.~Kapustin, B.~Willett  and I.~Yaakov, \emph{{Nonperturbative Tests of
  Three-Dimensional Dualities}}, JHEP {\bf 10} (2010) 013,
\href{http://www.arXiv.org/abs/1003.5694}{{\tt 1003.5694}}

\bibitem{Klebanov:1996un}
I.~R. Klebanov and A.~A. Tseytlin, \emph{{Entropy of near extremal black
  p-branes}}, Nucl. Phys. B {\bf 475} (1996) 164--178,
\href{http://www.arXiv.org/abs/hep-th/9604089}{{\tt hep-th/9604089}}

\bibitem{Klose:2010ki}
T.~Klose, \emph{{Review of AdS/CFT Integrability, Chapter IV.3: N=6
  Chern-Simons and Strings on AdS4xCP3}}, Lett. Math. Phys. {\bf 99} (2012)
  401--423,
\href{http://www.arXiv.org/abs/1012.3999}{{\tt 1012.3999}}

\bibitem{Klebanov:2009sg}
I.~R. Klebanov and G.~Torri, \emph{{M2-branes and AdS/CFT}}, Int. J. Mod. Phys.
  A {\bf 25} (2010) 332--350,
\href{http://www.arXiv.org/abs/0909.1580}{{\tt 0909.1580}}

\bibitem{Lambert:2019khh}
N.~Lambert, \emph{{M-Branes: Lessons from M2's and Hopes for M5's}}, Fortsch.
  Phys. {\bf 67} (2019), no.~8-9, 1910011,
\href{http://www.arXiv.org/abs/1903.02825}{{\tt 1903.02825}}

\bibitem{Mason:2023mti}
L.~Mason, R.~Ruzziconi  and A.~Yelleshpur~Srikant, \emph{{Carrollian amplitudes
  and celestial symmetries}}, JHEP {\bf 05} (2024) 012,
\href{http://www.arXiv.org/abs/2312.10138}{{\tt 2312.10138}}

\bibitem{Salzer:2023jqv}
J.~Salzer, \emph{{An Embedding Space Approach to Carrollian CFT Correlators for
  Flat Space Holography}},
\href{http://www.arXiv.org/abs/2304.08292}{{\tt 2304.08292}}

\bibitem{Saha:2023abr}
A.~Saha, \emph{{w$_{1+\infty}$ and Carrollian holography}}, JHEP {\bf 05}
  (2024) 145,
\href{http://www.arXiv.org/abs/2308.03673}{{\tt 2308.03673}}

\bibitem{Nguyen:2023vfz}
K.~Nguyen and P.~West, \emph{{Carrollian conformal fields and flat
  holography}},
\href{http://www.arXiv.org/abs/2305.02884}{{\tt 2305.02884}}

\bibitem{Nguyen:2023miw}
K.~Nguyen, \emph{{Carrollian conformal correlators and massless scattering
  amplitudes}},
\href{http://www.arXiv.org/abs/2311.09869}{{\tt 2311.09869}}

\bibitem{Liu:2024nfc}
W.-B. Liu, J.~Long  and X.-Q. Ye, \emph{{Feynman rules and loop structure of
  Carrollian amplitudes}}, JHEP {\bf 05} (2024) 213,
\href{http://www.arXiv.org/abs/2402.04120}{{\tt 2402.04120}}

\bibitem{Stieberger:2024shv}
S.~Stieberger, T.~R. Taylor  and B.~Zhu, \emph{{Carrollian Amplitudes from
  Strings}}, JHEP {\bf 04} (2024) 127,
\href{http://www.arXiv.org/abs/2402.14062}{{\tt 2402.14062}}

\bibitem{Ruzziconi:2024zkr}
R.~Ruzziconi, S.~Stieberger, T.~R. Taylor  and B.~Zhu, \emph{{Differential
  Equations for Carrollian Amplitudes}},
\href{http://www.arXiv.org/abs/2407.04789}{{\tt 2407.04789}}

\bibitem{Jorstad:2024yzm}
E.~J\o{}rstad and S.~Pasterski, \emph{{A Comment on Boundary Correlators: Soft
  Omissions and the Massless S-Matrix}},
\href{http://www.arXiv.org/abs/2410.20296}{{\tt 2410.20296}}

\bibitem{Ruzziconi:2024kzo}
R.~Ruzziconi and A.~Saha, \emph{{Holographic Carrollian currents for massless
  scattering}}, JHEP {\bf 01} (2025) 169,
\href{http://www.arXiv.org/abs/2411.04902}{{\tt 2411.04902}}

\bibitem{Kraus:2024gso}
P.~Kraus and R.~M. Myers, \emph{{Carrollian Partition Functions and the Flat
  Limit of AdS}},
\href{http://www.arXiv.org/abs/2407.13668}{{\tt 2407.13668}}

\bibitem{Kraus:2025wgi}
P.~Kraus and R.~M. Myers, \emph{{Carrollian Partition Function for Bulk
  Yang-Mills Theory}},
\href{http://www.arXiv.org/abs/2503.00916}{{\tt 2503.00916}}

\bibitem{Nguyen:2025sqk}
K.~Nguyen and J.~Salzer, \emph{{Operator Product Expansion in Carrollian CFT}},
\href{http://www.arXiv.org/abs/2503.15607}{{\tt 2503.15607}}

\bibitem{Surubaru:2025fmg}
I.~Surubaru and B.~Zhu, \emph{{Carrollian Amplitudes and Holographic
  Correlators in AdS3/CFT2}},
\href{http://www.arXiv.org/abs/2504.07650}{{\tt 2504.07650}}

\bibitem{Duval:2014uva}
C.~Duval, G.~W. Gibbons  and P.~A. Horvathy, \emph{{Conformal Carroll groups
  and BMS symmetry}}, Class. Quant. Grav. {\bf 31} (2014) 092001,
\href{http://www.arXiv.org/abs/1402.5894}{{\tt 1402.5894}}

\bibitem{Barducci:2018thr}
A.~Barducci, R.~Casalbuoni  and J.~Gomis, \emph{{Vector SUSY models with
  Carroll or Galilei invariance}}, Phys. Rev. D {\bf 99} (2019), no.~4, 045016,
\href{http://www.arXiv.org/abs/1811.12672}{{\tt 1811.12672}}

\bibitem{Bagchi:2019xfx}
A.~Bagchi, A.~Mehra  and P.~Nandi, \emph{{Field Theories with Conformal
  Carrollian Symmetry}}, JHEP {\bf 05} (2019) 108,
\href{http://www.arXiv.org/abs/1901.10147}{{\tt 1901.10147}}

\bibitem{Bagchi:2019clu}
A.~Bagchi, R.~Basu, A.~Mehra  and P.~Nandi, \emph{{Field Theories on Null
  Manifolds}}, JHEP {\bf 02} (2020) 141,
\href{http://www.arXiv.org/abs/1912.09388}{{\tt 1912.09388}}

\bibitem{Grumiller:2020elf}
D.~Grumiller, J.~Hartong, S.~Prohazka  and J.~Salzer, \emph{{Limits of JT
  gravity}}, JHEP {\bf 02} (2021) 134,
\href{http://www.arXiv.org/abs/2011.13870}{{\tt 2011.13870}}

\bibitem{Gomis:2020wxp}
J.~Gomis, D.~Hidalgo  and P.~Salgado-Rebolledo, \emph{{Non-relativistic and
  Carrollian limits of Jackiw-Teitelboim gravity}}, JHEP {\bf 05} (2021) 162,
\href{http://www.arXiv.org/abs/2011.15053}{{\tt 2011.15053}}

\bibitem{deBoer:2021jej}
J.~de~Boer, J.~Hartong, N.~A. Obers, W.~Sybesma  and S.~Vandoren,
  \emph{{Carroll Symmetry, Dark Energy and Inflation}}, Front. in Phys. {\bf
  10} (2022) 810405,
\href{http://www.arXiv.org/abs/2110.02319}{{\tt 2110.02319}}

\bibitem{Henneaux:2021yzg}
M.~Henneaux and P.~Salgado-Rebolledo, \emph{{Carroll contractions of
  Lorentz-invariant theories}}, JHEP {\bf 11} (2021) 180,
\href{http://www.arXiv.org/abs/2109.06708}{{\tt 2109.06708}}

\bibitem{Gupta:2020dtl}
N.~Gupta and N.~V. Suryanarayana, \emph{{Constructing Carrollian CFTs}}, JHEP
  {\bf 03} (2021) 194,
\href{http://www.arXiv.org/abs/2001.03056}{{\tt 2001.03056}}

\bibitem{Baiguera:2022lsw}
S.~Baiguera, G.~Oling, W.~Sybesma  and B.~T. S\o{}gaard, \emph{{Conformal
  Carroll Scalars with Boosts}},
\href{http://www.arXiv.org/abs/2207.03468}{{\tt 2207.03468}}

\bibitem{Barnich:2022bni}
G.~Barnich, K.~Nguyen  and R.~Ruzziconi, \emph{{Geometric action for extended
  Bondi-Metzner-Sachs group in four dimensions}}, JHEP {\bf 12} (2022) 154,
\href{http://www.arXiv.org/abs/2211.07592}{{\tt 2211.07592}}

\bibitem{Rivera-Betancour:2022lkc}
D.~Rivera-Betancour and M.~Vilatte, \emph{{Revisiting the Carrollian scalar
  field}}, Phys. Rev. D {\bf 106} (2022), no.~8, 085004,
\href{http://www.arXiv.org/abs/2207.01647}{{\tt 2207.01647}}

\bibitem{Hao:2022xhq}
P.-X. Hao, W.~Song, Z.~Xiao  and X.~Xie, \emph{{A BMS-invariant free fermion
  model}},
\href{http://www.arXiv.org/abs/2211.06927}{{\tt 2211.06927}}

\bibitem{Bagchi:2022eui}
A.~Bagchi, A.~Banerjee, R.~Basu, M.~Islam  and S.~Mondal, \emph{{Magic
  fermions: Carroll and flat bands}}, JHEP {\bf 03} (2023) 227,
\href{http://www.arXiv.org/abs/2211.11640}{{\tt 2211.11640}}

\bibitem{Miskovic:2023zfz}
O.~Miskovic, R.~Olea, P.~M. Petropoulos, D.~Rivera-Betancour  and K.~Siampos,
  \emph{{Chern-Simons action and the Carrollian Cotton tensors}}, JHEP {\bf 12}
  (2023) 130,
\href{http://www.arXiv.org/abs/2310.19929}{{\tt 2310.19929}}

\bibitem{Ara:2024vbe}
N.~Ara, A.~Banerjee, R.~Basu  and B.~Krishnan, \emph{{Flat Bands and Compact
  Localised States: A Carrollian roadmap}},
\href{http://www.arXiv.org/abs/2412.18965}{{\tt 2412.18965}}

\bibitem{Banerjee:2024ldl}
A.~Banerjee, R.~Basu, A.~Bhattacharyya  and N.~Chakrabarti, \emph{{Symmetry
  resolution in non-Lorentzian field theories}}, JHEP {\bf 06} (2024) 121,
\href{http://www.arXiv.org/abs/2404.02206}{{\tt 2404.02206}}

\bibitem{deBoer:2023fnj}
J.~de~Boer, J.~Hartong, N.~A. Obers, W.~Sybesma  and S.~Vandoren,
  \emph{{Carroll stories}}, JHEP {\bf 09} (2023) 148,
\href{http://www.arXiv.org/abs/2307.06827}{{\tt 2307.06827}}

\bibitem{Chen:2021xkw}
B.~Chen, R.~Liu  and Y.-f. Zheng, \emph{{On Higher-dimensional Carrollian and
  Galilean Conformal Field Theories}},
\href{http://www.arXiv.org/abs/2112.10514}{{\tt 2112.10514}}

\bibitem{Chen:2023pqf}
B.~Chen, R.~Liu, H.~Sun  and Y.-f. Zheng, \emph{{Constructing Carrollian Field
  Theories from Null Reduction}},
\href{http://www.arXiv.org/abs/2301.06011}{{\tt 2301.06011}}

\bibitem{Chen:2024voz}
B.~Chen, H.~Sun  and Y.-f. Zheng, \emph{{Quantization of Carrollian conformal
  scalar theories}}, Phys. Rev. D {\bf 110} (2024), no.~12, 125010,
\href{http://www.arXiv.org/abs/2406.17451}{{\tt 2406.17451}}

\bibitem{Cotler:2024xhb}
J.~Cotler, K.~Jensen, S.~Prohazka, A.~Raz, M.~Riegler  and J.~Salzer,
  \emph{{Quantizing Carrollian field theories}}, JHEP {\bf 10} (2024) 049,
\href{http://www.arXiv.org/abs/2407.11971}{{\tt 2407.11971}}

\bibitem{Sharma:2025rug}
A.~Sharma, \emph{{Studies on Carrollian Quantum Field Theories}},
\href{http://www.arXiv.org/abs/2502.00487}{{\tt 2502.00487}}

\bibitem{Poulias:2025eck}
G.~Poulias and S.~Vandoren, \emph{{On Carroll partition functions and flat
  space holography}},
\href{http://www.arXiv.org/abs/2503.20615}{{\tt 2503.20615}}

\bibitem{Jorge-Diaz:2022dmy}
C.~Jorge-Diaz, S.~Pasterski  and A.~Sharma, \emph{{Celestial amplitudes in an
  ambidextrous basis}}, JHEP {\bf 02} (2023) 155,
\href{http://www.arXiv.org/abs/2212.00962}{{\tt 2212.00962}}

\bibitem{Banerjee:2018fgd}
S.~Banerjee, \emph{{Symmetries of free massless particles and soft theorems}},
  Gen. Rel. Grav. {\bf 51} (2019), no.~9, 128,
\href{http://www.arXiv.org/abs/1804.06646}{{\tt 1804.06646}}

\bibitem{1977asst.conf....1G}
R.~{Geroch}, \emph{{Asymptotic Structure of Space-Time}}, in {\em Asymptotic
  Structure of Space-Time}, F.~P. {Esposito} and L.~{Witten}, eds., p.~1.
\newblock
Jan., 1977.
\newblock

\bibitem{Henneaux:1979vn}
M.~Henneaux, \emph{{Geometry of Zero Signature Space-times}}, Bull. Soc. Math.
  Belg. {\bf 31} (1979)
47--63

\bibitem{Ashtekar:2014zsa}
A.~Ashtekar, \emph{{Geometry and Physics of Null Infinity}},
\href{http://www.arXiv.org/abs/1409.1800}{{\tt 1409.1800}}

\bibitem{Arutyunov:2002fh}
G.~Arutyunov, F.~A. Dolan, H.~Osborn  and E.~Sokatchev, \emph{{Correlation
  functions and massive Kaluza-Klein modes in the AdS / CFT correspondence}},
  Nucl. Phys. B {\bf 665} (2003) 273--324,
\href{http://www.arXiv.org/abs/hep-th/0212116}{{\tt hep-th/0212116}}

\bibitem{Gary:2009ae}
M.~Gary, S.~B. Giddings  and J.~Penedones, \emph{{Local bulk S-matrix elements
  and CFT singularities}}, Phys. Rev. D {\bf 80} (2009) 085005,
\href{http://www.arXiv.org/abs/0903.4437}{{\tt 0903.4437}}

\bibitem{Maldacena:2015iua}
J.~Maldacena, D.~Simmons-Duffin  and A.~Zhiboedov, \emph{{Looking for a bulk
  point}}, JHEP {\bf 01} (2017) 013,
\href{http://www.arXiv.org/abs/1509.03612}{{\tt 1509.03612}}

\bibitem{Bastianelli:1999vm}
F.~Bastianelli and R.~Zucchini, \emph{{Three point functions for a class of
  chiral operators in maximally supersymmetric CFT at large N}}, Nucl. Phys. B
  {\bf 574} (2000) 107--129,
\href{http://www.arXiv.org/abs/hep-th/9909179}{{\tt hep-th/9909179}}

\bibitem{Lee:1998bxa}
S.~Lee, S.~Minwalla, M.~Rangamani  and N.~Seiberg, \emph{{Three point functions
  of chiral operators in D = 4, N=4 SYM at large N}}, Adv. Theor. Math. Phys.
  {\bf 2} (1998) 697--718,
\href{http://www.arXiv.org/abs/hep-th/9806074}{{\tt hep-th/9806074}}

\bibitem{Bastianelli:1999en}
F.~Bastianelli and R.~Zucchini, \emph{{Three point functions of chiral primary
  operators in d = 3, N=8 and d = 6, N=(2,0) SCFT at large N}}, Phys. Lett. B
  {\bf 467} (1999) 61--66,
\href{http://www.arXiv.org/abs/hep-th/9907047}{{\tt hep-th/9907047}}

\bibitem{Dolan:2004mu}
F.~A. Dolan, L.~Gallot  and E.~Sokatchev, \emph{{On four-point functions of
  1/2-BPS operators in general dimensions}}, JHEP {\bf 09} (2004) 056,
\href{http://www.arXiv.org/abs/hep-th/0405180}{{\tt hep-th/0405180}}

\bibitem{Symanzik:1972wj}
K.~Symanzik, \emph{{On Calculations in conformal invariant field theories}},
  Lett. Nuovo Cim. {\bf 3} (1972)
734--738

\bibitem{Mack:2009gy}
G.~Mack, \emph{{D-dimensional Conformal Field Theories with anomalous
  dimensions as Dual Resonance Models}}, Bulg. J. Phys. {\bf 36} (2009)
  214--226,
\href{http://www.arXiv.org/abs/0909.1024}{{\tt 0909.1024}}

\bibitem{Penedones:2010ue}
J.~Penedones, \emph{{Writing CFT correlation functions as AdS scattering
  amplitudes}}, JHEP {\bf 03} (2011) 025,
\href{http://www.arXiv.org/abs/1011.1485}{{\tt 1011.1485}}

\bibitem{Sannan:1986tz}
S.~Sannan, \emph{{Gravity as the Limit of the Type {II} Superstring Theory}},
  Phys. Rev. D {\bf 34} (1986)
1749

\bibitem{Chowdhury:2019kaq}
S.~D. Chowdhury, A.~Gadde, T.~Gopalka, I.~Halder, L.~Janagal  and S.~Minwalla,
  \emph{{Classifying and constraining local four photon and four graviton
  S-matrices}}, JHEP {\bf 02} (2020) 114,
\href{http://www.arXiv.org/abs/1910.14392}{{\tt 1910.14392}}

\bibitem{Alday:2021odx}
L.~F. Alday, C.~Behan, P.~Ferrero  and X.~Zhou, \emph{{Gluon Scattering in AdS
  from CFT}}, JHEP {\bf 06} (2021) 020,
\href{http://www.arXiv.org/abs/2103.15830}{{\tt 2103.15830}}

\bibitem{Chen:2020ipe}
H.-Y. Chen and J.-i. Sakamoto, \emph{{Superconformal Block from Holographic
  Geometry}}, JHEP {\bf 07} (2020) 028,
\href{http://www.arXiv.org/abs/2003.13343}{{\tt 2003.13343}}

\bibitem{Bagchi:2022owq}
A.~Bagchi, D.~Grumiller  and P.~Nandi, \emph{{Carrollian superconformal
  theories and super BMS}},
\href{http://www.arXiv.org/abs/2202.01172}{{\tt 2202.01172}}

\bibitem{Zheng:2025cuw}
Y.-f. Zheng and B.~Chen, \emph{{Structure of Carrollian (conformal)
  superalgebra}},
\href{http://www.arXiv.org/abs/2503.22160}{{\tt 2503.22160}}

\bibitem{Park:1999cw}
J.-H. Park, \emph{{Superconformal symmetry in three-dimensions}}, J. Math.
  Phys. {\bf 41} (2000) 7129--7161,
\href{http://www.arXiv.org/abs/hep-th/9910199}{{\tt hep-th/9910199}}

\bibitem{Barnich:2009se}
G.~Barnich and C.~Troessaert, \emph{{Symmetries of asymptotically flat 4
  dimensional spacetimes at null infinity revisited}}, Phys. Rev. Lett. {\bf
  105} (2010) 111103,
\href{http://www.arXiv.org/abs/0909.2617}{{\tt 0909.2617}}

\bibitem{Barnich:2011ct}
G.~Barnich and C.~Troessaert, \emph{{Supertranslations call for
  superrotations}}, PoS {\bf CNCFG} (2010) 010,
  \href{http://www.arXiv.org/abs/1102.4632}{{\tt 1102.4632}},
[Ann. U. Craiova Phys.21,S11(2011)]

\bibitem{Fotopoulos:2020bqj}
A.~Fotopoulos, S.~Stieberger, T.~R. Taylor  and B.~Zhu, \emph{{Extended Super
  BMS Algebra of Celestial CFT}},
\href{http://www.arXiv.org/abs/2007.03785}{{\tt 2007.03785}}

\bibitem{Henneaux:2020ekh}
M.~Henneaux, J.~Matulich  and T.~Neogi, \emph{{Asymptotic realization of the
  super-BMS algebra at spatial infinity}}, Phys. Rev. D {\bf 101} (2020),
  no.~12, 126016,
\href{http://www.arXiv.org/abs/2004.07299}{{\tt 2004.07299}}

\bibitem{Fuentealba:2021xhn}
O.~Fuentealba, M.~Henneaux, S.~Majumdar, J.~Matulich  and T.~Neogi,
  \emph{{Local supersymmetry and the square roots of Bondi-Metzner-Sachs
  supertranslations}}, Phys. Rev. D {\bf 104} (2021), no.~12, L121702,
\href{http://www.arXiv.org/abs/2108.07825}{{\tt 2108.07825}}

\bibitem{Bagchi:2024efs}
A.~Bagchi, A.~Lipstein, M.~Mandlik  and A.~Mehra, \emph{{3d Carrollian
  Chern-Simons theory and 2d Yang-Mills}},
\href{http://www.arXiv.org/abs/2407.13574}{{\tt 2407.13574}}

\bibitem{Bagchi:2024gnn}
A.~Bagchi, P.~Dhivakar  and S.~Dutta, \emph{{3D Stress Tensor for Gravity in 4D
  Flat Spacetime}},
\href{http://www.arXiv.org/abs/2408.05494}{{\tt 2408.05494}}

\bibitem{Caron-Huot:2018kta}
S.~Caron-Huot and A.-K. Trinh, \emph{{All tree-level correlators in
  AdS$_{5}$\texttimes{}S$_{5}$ supergravity: hidden ten-dimensional conformal
  symmetry}}, JHEP {\bf 01} (2019) 196,
\href{http://www.arXiv.org/abs/1809.09173}{{\tt 1809.09173}}

\bibitem{Abl:2020dbx}
T.~Abl, P.~Heslop  and A.~E. Lipstein, \emph{{Towards the Virasoro-Shapiro
  amplitude in AdS$_{5} \times S^{5}$}}, JHEP {\bf 04} (2021) 237,
\href{http://www.arXiv.org/abs/2012.12091}{{\tt 2012.12091}}

\bibitem{Aprile:2020mus}
F.~Aprile, J.~M. Drummond, H.~Paul  and M.~Santagata, \emph{{The
  Virasoro-Shapiro amplitude in AdS$_{5}$ \texttimes{} S$^{5}$ and level
  splitting of 10d conformal symmetry}}, JHEP {\bf 11} (2021) 109,
\href{http://www.arXiv.org/abs/2012.12092}{{\tt 2012.12092}}

\bibitem{Caron-Huot:2021usw}
S.~Caron-Huot and F.~Coronado, \emph{{Ten dimensional symmetry of $ \mathcal{N}
  $ = 4 SYM correlators}}, JHEP {\bf 03} (2022) 151,
\href{http://www.arXiv.org/abs/2106.03892}{{\tt 2106.03892}}

\bibitem{Zhou:2017zaw}
X.~Zhou, \emph{{On Superconformal Four-Point Mellin Amplitudes in Dimension
  $d>2$}}, JHEP {\bf 08} (2018) 187,
\href{http://www.arXiv.org/abs/1712.02800}{{\tt 1712.02800}}

\bibitem{Fitzpatrick:2011hu}
A.~L. Fitzpatrick and J.~Kaplan, \emph{{Analyticity and the Holographic
  S-Matrix}}, JHEP {\bf 10} (2012) 127,
\href{http://www.arXiv.org/abs/1111.6972}{{\tt 1111.6972}}

\bibitem{Fitzpatrick:2011ia}
A.~L. Fitzpatrick, J.~Kaplan, J.~Penedones, S.~Raju  and B.~C. van Rees,
  \emph{{A Natural Language for AdS/CFT Correlators}}, JHEP {\bf 11} (2011)
  095,
\href{http://www.arXiv.org/abs/1107.1499}{{\tt 1107.1499}}

\bibitem{Binder:2018yvd}
D.~J. Binder, S.~M. Chester  and S.~S. Pufu, \emph{{Absence of $D^4 R^4$ in
  M-Theory From ABJM}}, JHEP {\bf 04} (2020) 052,
\href{http://www.arXiv.org/abs/1808.10554}{{\tt 1808.10554}}

\end{thebibliography}\endgroup

\end{document}